\documentclass[a4paper, 10pt]{article}

\usepackage{bm}
\usepackage[T1]{fontenc}
\usepackage[utf8]{inputenc}
\usepackage[italian, english]{babel}
\usepackage{graphicx}                    				
\usepackage{booktabs}                   				
\usepackage{textcomp}                   				
\usepackage{marvosym}                  					
\usepackage{eurosym}                    				
\usepackage{longtable}    				 				
\usepackage{fancyhdr}                    				
\usepackage{listings}                       			
\usepackage[]{color} 									
\usepackage{mathrsfs} 					 				
\usepackage{amsmath}									
\usepackage{amssymb}
\usepackage{float} 										
\usepackage{pdfpages} 									
\usepackage{lipsum}					 
\usepackage{empheq}                    					
\usepackage[svgnames, dvipsnames, table]{xcolor} 		
\usepackage{array} 										
\usepackage{subfig}		
\usepackage{placeins}
\usepackage{multirow}
\usepackage{caption}
\usepackage{rotating}
\usepackage{mathtools}
\usepackage{enumitem} 
\usepackage{pifont}
\usepackage[colorlinks=true,linkcolor=blue,urlcolor=blue]{hyperref}
\usepackage[normalem]{ulem}

\usepackage[top=3cm,bottom=3cm,left=2cm,right=2cm,bindingoffset=0cm]{geometry}
\usepackage[font=small,format=hang,labelfont={bf}]{caption}
\usepackage{palatino}

\usepackage{helvet}

\definecolor{greenSea}{RGB}{22, 160, 133}
\graphicspath{{immagini/}} 

\DeclareMathOperator{\tr}{tr}

\newcommand{\scalp}{\bm \cdot}

\newcommand{\Reals}{\mathbb{R}}

\newcommand{\ba}{\bm a}
\newcommand{\bb}{\bm b}

\newcommand{\be}{\bm e}
\newcommand{\bef}{\bm f}
\newcommand{\bg}{\bm g}

\newcommand{\bn}{\bm n}

\newcommand{\bq}{\bm q}

\newcommand{\bt}{\bm t}

\newcommand{\bv}{\bm v}

\newcommand{\bx}{\bm x}
\newcommand{\by}{\bm y}

\newcommand{\bA}{\bm A}

\newcommand{\bD}{\bm D}

\newcommand{\bG}{\bm G}

\newcommand{\bI}{\bm I}
\newcommand{\bJ}{\bm J}
\newcommand{\bK}{\bm K}
\newcommand{\bL}{\bm L}
\newcommand{\bM}{\bm M}

\newcommand{\bP}{\bm P}

\newcommand{\bS}{\bm S}

\newcommand{\bU}{\bm U}

\newcommand{\boldeta}{\bm \eta}

\newcommand{\bxi}{\bm \xi}

\newcommand{\bPhi}{\bm \Phi}

\newcommand{\bGamma}{\bm \Gamma}

\newcommand{\mH}{\mathcal H}

\newcommand{\mV}{\mathcal V}

\definecolor{alizarin}{RGB}{231, 76, 60}
\definecolor{pomegranate}{RGB}{192, 57, 43}
\definecolor{turquoise}{RGB}{26, 188, 156}
\definecolor{emerald}{RGB}{46, 204, 113}				
\definecolor{greenSea}{RGB}{22, 160, 133}					
\definecolor{nephritis}{RGB}{39, 174, 96}		
\definecolor{peterRivers}{RGB}{52, 152, 219}
\definecolor{belizeHole}{RGB}{41, 128, 185}
\definecolor{amethyst}{RGB}{155, 89, 182}				
\definecolor{wisteria}{RGB}{142, 68, 173}
\definecolor{wetAsphalt}{RGB}{52, 73, 94}				
\definecolor{midnightBlue}{RGB}{44, 62, 80}
\definecolor{sunflower}{RGB}{241, 196, 15}
\definecolor{orange}{RGB}{243, 156, 18}
\definecolor{carrot}{RGB}{230, 126, 34}			
\definecolor{pumpkin}{RGB}{211, 84, 0}
\definecolor{cloud}{RGB}{236, 240, 241}				
\definecolor{silver}{RGB}{189, 195, 199}				
\definecolor{concrete}{RGB}{149, 165, 166}				
\definecolor{asbestos}{RGB}{127, 140, 141}

\newcommand{\mycolorbar}[8]
{   
	\foreach \x [count=\c] in {#3}{ \xdef\numcolo{\c}}
	\pgfmathsetmacro{\pieceheight}{#1/(\numcolo-1)}
	\xdef\lowcolo{}
	\foreach \x [count=\c] in {#3}
	{ \ifthenelse{\c = 1}
		{}
		{   \fill[bottom color=\lowcolo,top color=\x] (0,{(\c-2)*\pieceheight}) rectangle (#2,{(\c-1)*\pieceheight});
		}
		\xdef\lowcolo{\x}
	}
	\draw[thin] (0,0) rectangle (#2,#1);
	\pgfmathsetmacro{\secondlabel}{#4+#6}
	\pgfmathsetmacro{\lastlabel}{#5+0.01}
	\pgfkeys{/pgf/number format/.cd,fixed,fixed zerofill,precision=#7}
	\foreach \x in {#4,\secondlabel,...,\lastlabel}
	{ \draw[thin] (#2,{(\x-#4)/(#5-#4)*#1}) -- ++ (0.15,0); 
		\node[left] at ({(#2+0.15)+(#8+4)*1ex},{(\x-#4)/(#5-#4)*#1}) {\scriptsize{\pgfmathprintnumber{\x}}}; 
	}
}

\usepackage[autostyle, italian=guillemets]{csquotes}
\usepackage[
    backend=biber,
    style=numeric,
    citestyle=numeric-comp,
    maxbibnames=99,
    minbibnames=99,
    maxcitenames=2,
    mincitenames=1,
    giveninits=true,
    sorting=none,
    sortlocale=auto,
    natbib=true,
    url=false,
    isbn=false,
    doi=true,
    eprint=true
]{biblatex}
\usepackage{tikz}
\usepackage{fp}
\usepackage{ifthen}
\usepackage{fullpage}
\usepackage{pgfmath}
\usepackage{pgfplots}
\usepackage{tikz-3dplot}
\usetikzlibrary{backgrounds}
\usetikzlibrary{decorations.pathmorphing,backgrounds,fit,calc,through}
\usetikzlibrary{arrows}
\usetikzlibrary{arrows.meta}
\usetikzlibrary{shapes,decorations,shadows}
\usetikzlibrary{fadings}
\usetikzlibrary{patterns}
\usetikzlibrary{mindmap}
\usetikzlibrary{decorations.text}
\usetikzlibrary{decorations.shapes}
\usetikzlibrary{3d}
\usetikzlibrary{hobby}
\usetikzlibrary{graphs}
\usetikzlibrary{math}
\usetikzlibrary{angles,positioning}
\usetikzlibrary{external}

\usetikzlibrary{decorations.pathreplacing, intersections}
\usetikzlibrary{decorations.markings}

\usepackage[affil-it,auth-sc]{authblk}
\usepackage{blindtext}
\usepackage{abstract}

\title{Fusion of two stable elastic structures resulting in an unstable system}

\author[1]{Marco Rossi}
\author[2]{Andrea Piccolroaz}
\author[2]{Davide Bigoni\footnote{Corresponding author:\,e-mail:\, bigoni@ing.unitn.it; phone:\,+39\,0461\,282507.}}

\affil[1]{Department of Engineering and Architecture, University of Trieste, Via Valerio 6/1, 34127 Trieste, Italy}
\affil[2]{Department of Civil, Environmental and Mechanical Engineering, University of Trento, via Mesiano 77, 38123 Trento, Italy}

\date{\today}

\pgfplotsset{compat=1.17}

\mathtoolsset{showonlyrefs=true}

\begin{document}

\maketitle

\begin{abstract}
    It is shown that a compound elastic structure, which displays a dynamic instability, may be designed as the union (or \lq fusion') of two structures which are stable when separately analyzed. The compound elastic structure has two degrees of freedom and is made up of a rigid rod connected with two springs to a smooth support, which evidences a jump in the  {\it curvature} at the equilibrium configuration. Instability is proven in a linearized context and is related to the application of a non-conservative load of the follower type so that the instability disappears under dead loads. In the fully nonlinear range, the instability is also confirmed through numerical simulations. The obtained results may be useful in the design of new mechanical sensors, devices for energy harvesting, or architected materials. In addition, our findings have conceptual implications on piecewise-linear theories of mechanics such as for instance plasticity or frictional contact.
\end{abstract}

\section{Introduction}

Consider two elastic structures loaded within their stable range and imagine that these represent two parts of a third elastic structure obtained through the \lq fusion' of the initial  structures, becoming the \lq constituents' of the new structure. The stability of the compound structure is expected when the latter is subject to the same load at which the two component structures are stable. This intuitive belief is true for dead loading and smooth systems, but it is shown in the present article  to be false for follower loads and non-smooth mechanical behaviour. 
From a purely mathematical point of view, the singularity of a non-smooth behaviour obtained as a combination of two stable dynamical systems may be expected, as was advocated through a purely mathematical example by Branicky \cite{Branicky1998} and Carmona et al. \cite{carmonaContinuousMatchingTwo2006a} for abstract systems of non-smooth differential equations. However, an elastic structure exhibiting this peculiar kind of instability has never been discovered so far. The purpose of the present article is to fill this gap through the invention and the theoretical and numerical analysis of a structure designed to demonstrate the instability of a simple two d.o.f. non-smooth mechanical system that is composed of two stable smooth subsystems. This finding brings a mathematical result into the realm of mechanics.

\begin{figure}[H]
	\centering
	\,\,\,{\includegraphics[width=0.32\linewidth]{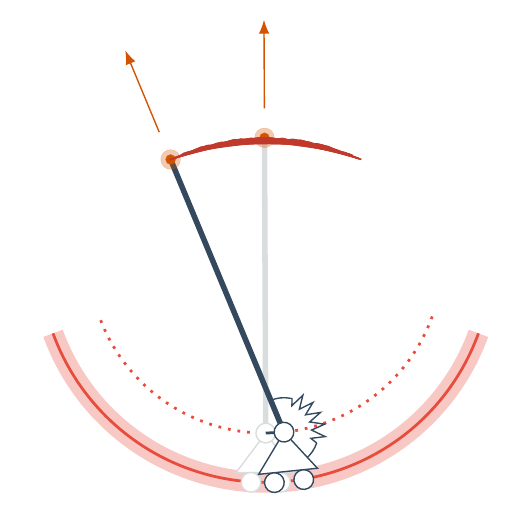}} \,\,\,
	{\includegraphics[width=0.32\linewidth]{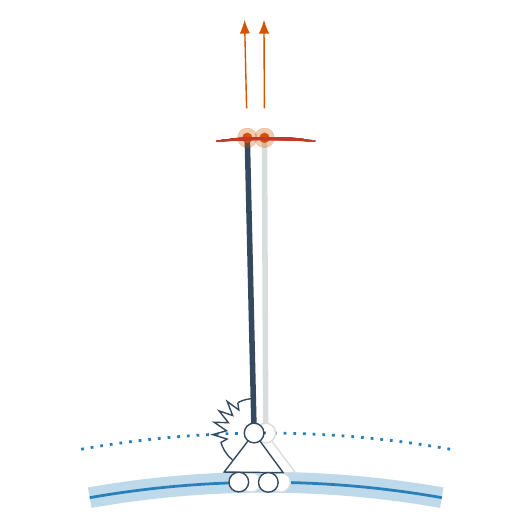}} 
	{\includegraphics[width=0.32\linewidth]{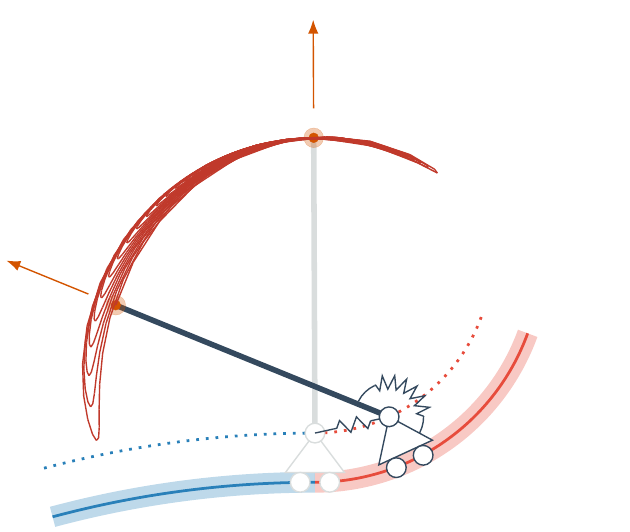}}\\
	{\includegraphics[width=0.32\linewidth]{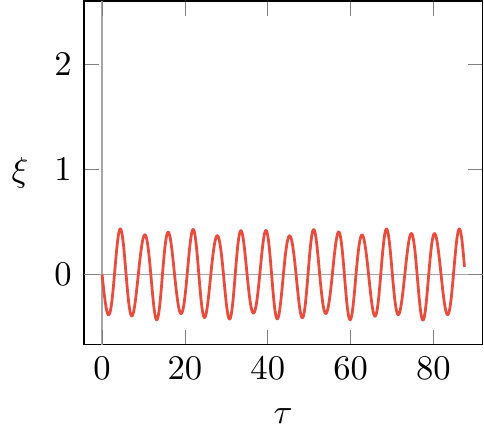}} 
	{\includegraphics[width=0.32\linewidth]{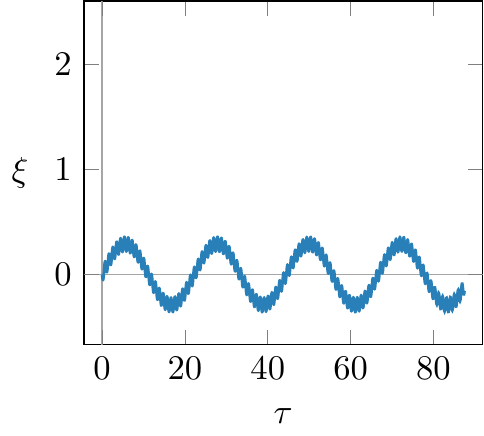}} 
	{\includegraphics[width=0.32\linewidth]{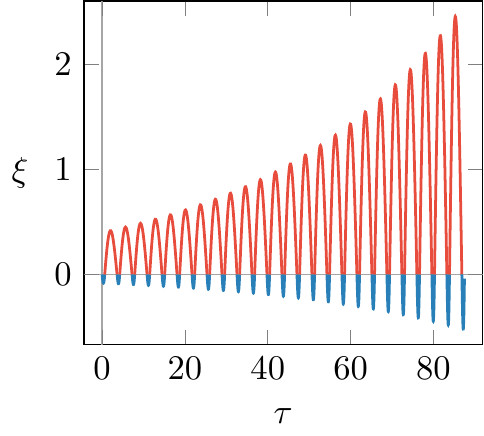}}\\
	\caption{Two stable smooth subsystems with positive and negative curvature of a sliding constraint (upper part: left and center) and the fusion of these two structures, namely, a compound non-smooth structure displaying instability (upper part: right), although the two \lq components' are stable. The trajectories of the end of the structures are also reported for vibrational motion, together with the corresponding arc-length $\xi$ vs time $\tau$ behaviours, showing sinusoidal (i.e. stable, lower part: left and center) and unstable (lower part: right) oscillations. The tensile force acting at the free end of the rods is tangentially follower and the same for all three structures, lying well below the critical load for instability in the case of the two smooth \lq component systems', both displaying  motions  confined in a neighborhood of the trivial equilibrium configuration. Differently, a flutter-like instability is observed for the composite structure (upper part: right), as evidenced by the unstable and exponentially growing oscillations of the loaded end. See the video file in the electronic supplementary material.}
	\label{fig:example1}
\end{figure}

The elastic structures considered here are of the type shown in Fig.~\ref{fig:example1}, consisting of a rigid rod with an end sliding on a smooth profile, while the other end is subject to a tangential follower load, remaining parallel to the bar. The deformed configuration of the structure is defined by two degrees of freedom, specifically the arc length distance characterizing the roller position along the profile and the angle of rotation of the bar with respect to the vertical direction. The structure is stiffened by a longitudinal spring connecting the roller to a fixed point and a rotational spring interposed between the rod and the roller. 

The first two structures (from left) shown in Fig.~\ref{fig:example1} are characterized by a circular profile (with positive and negative curvature, respectively). These structures are described by smooth dynamical systems and suffer flutter and divergence instabilities. Assuming that the magnitude of the follower force is well below the critical values, the vertical trivial equilibrium configuration is stable, so that a small perturbation in the initial conditions generates a motion which remains confined within a small neighborhood of the fixed point, as  shown in the lower part of Fig.~\ref{fig:example1} (obtained through numerical integration of the nonlinear equations of motion and representing the arc-length distance $\xi$ traveled by the roller as a function of the elapsed time $\tau$).

The third elastic structure shown on the right in Fig.~\ref{fig:example1} is also described by two degrees of freedom and is obtained as the \lq fusion' of the two structures sketched on the left and center of the same figure. This new structure is characterized by a smooth sliding profile, which evidences a jump in the curvature so that the dynamics is characterized by piecewise smooth differential equations. The tangent to the profile at the junction is horizontal and at this point, the longitudinal spring is unloaded, so that the vertical configuration of the rigid rod is the trivial equilibrium configuration. 

If this structure is subject to a load smaller than the critical loads of the two \lq generating structures', it might be expected that the structure would be stable. This becomes true when the load is  conservative, but now the load is follower, so that: 
\begin{quote}
    \emph{it is shown in this article that the non-smooth structure (Fig.~\ref{fig:example1} right), obtained as the fusion of two smooth structures (Fig.~\ref{fig:example1} left and center), may be dynamically unstable at a load well inside the stability domains of both the generating structures.}
\end{quote}

Indeed, at a load well below both critical loads of the smooth
constituent structures, the compound non-smooth structure exhibits the exponentially growing oscillation illustrated in Fig.~\ref{fig:example1} (lower part, obtained through numerical integration of the nonlinear equations of motion), resembling the flutter instability, occurring in smooth systems, for instance, the celebrated Ziegler double pendulum \cite{zieglerPrinciplesStructuralStability1977}, see the video file in the electronic supplementary material.

Note that the structures shown in Fig.~\ref{fig:example1} are similar to those investigated in \cite{bigoniEffectsConstraintCurvature2012a} and \cite{misseroniDeformationElasticRod2015a}, but now the load is non-conservative and follower, so that the instability landscape results completely changed and, in particular, the possibility arises of finding instabilities unrelated to the instabilities of the component structures. 

The above-stated result, referred to the structure shown in Fig.~\ref{fig:example1} on the right, follows from the combination of two features, namely, the presence of (i.) a non-conservative follower load and (ii.) a jump in curvature in the sliding constraint. The latter feature implies  that the acceleration is discontinuous at the junction between the two circular profiles at the basis of the structure and thus the system of governing equations becomes non-smooth. 

Piecewise-smooth dynamical systems are common in the mechanics of solids and structures. In fact, elastoplasticity and contact with friction are based on piece-wise linear equations of the rate type; bi-linear elasticity defines solids with different tensile and compressive elastic moduli; 
structures impacting against unilateral constraints involve two sets of equations of motion. 
These systems are known to exhibit peculiar forms of instability, such as for instance stick and slip motion for frictional contact \cite{krogerExperimentalInvestigationAvoidance2008}, or blowing-up vibrations for non-associative elastoplasticity \cite{burghardtNonuniquenessInstabilityClassical2015, pucikNonuniquenessInstabilityClassical2015}.
Every elastoplastic constitutive equation is always piecewise linear in the rate response and nonassociative flow rules lead to a lack of symmetry similar in essence to that induced by follower loads in structural elements. 
It is known that the combination of plasticity and non-associativity may lead to flutter instability in a continuum, an instability largely unknown, as the effects of piecewise linearity are still almost unexplored \cite{bigoni_note_2002}.

Therefore, the structures designed in the present paper are governed by equations sharing strong similarity with elastoplasticity, so that our results lead to important conjectures in that field. More in detail, stability analysis in elastoplasticity is performed on the so-called \lq comparison solids' \cite{Bigoni2012}, which are the exact counterpart of the component structures introduced here. Stability of the comparison solids is usually {\it assumed} to imply stability of the true piecewise linear behaviour, but our structural examples demonstrate that this may be false, an implication that would {\it completely revolutionize the stability theory for nonassociative elastoplasticity}. 

Accordingly to its interest in mechanics, the instability of piecewise-smooth dynamical systems has recently attracted growing interest. The first proof that a compound system generated as the fusion between two stable systems can be unstable is due to Carmona et al. \cite{carmonaContinuousMatchingTwo2006a}. 
For non-smooth dynamical systems, they have provided a sufficient condition for instability, based on the detection of a so-called `invariant cone'. This condition has been further developed in various directions \cite{ hoshamBifurcationsFourdimensionalSwitched2018, kupperReductionInvariantCones2011b, weissInvariantManifoldsNonsmooth2012a, weissInvariantManifoldsNonsmooth2015}. All these works provide a mathematical framework and open new directions for research in bifurcation theory. However, applications of the mathematical setting to mechanics are scarce and so far limited to simplified systems characterized by Coulomb friction \cite{kalmar-nagyNonlinearAnalysis2DOF2016, 
leineBifurcationPhenomenaNonsmooth2006a, zouGeneralizedHopfBifurcation2006}. Therefore, the objective of the present article is to develop the analysis of invariant cones to demonstrate the instability of the structure shown in Fig.~\ref{fig:example1} on the right, obtained as the fusion of two stable structures. 

Although a general concept, the invariant cone can practically be applied only to the \emph{linearized} equations of motion governing a  mechanical model, so that structures which are proven to be unstable on the basis of a linearized analytical treatment are also numerically investigated in this article, to provide a complete picture of their mechanical behaviour. In this way, it is proven that in all cases in which the linearized equations display instability, the latter persists also when the fully  nonlinear piecewise-smooth problem is examined.

It has to be highlighted that the mathematical background so far developed only consists in sufficient conditions for instability so that when these conditions are not fulfilled, nothing can be concluded concerning stability. This situation is reflected in the results presented in the present article, where only examples of piecewise-smooth {\it unstable} structures are obtained, while the behaviour of the same structures at different loads is usually unknown (though open to numerical investigation), as the sufficient condition fails. 

The paper is organized as follows. After the introduction of the class of the addressed elastic structures (Section \ref{sec:jdescriptionStructure}), the differential equations governing their dynamics are formulated in Section \ref{sec:dynamicInstabilityStructures}, where the concept of invariant cone is provided as a sufficient condition for instability. The properties of invariant cones are demonstrated in Section \ref{sec:attractvityCone}, where it is shown, under broad hypotheses, that an unstable cone is always attractive. An algorithm to detect invariant cones for two d.o.f. mechanical systems is presented in Section \ref{sec:numericalExample} and numerical results on the linearized analysis are developed demonstrating the instability, which is also finally confirmed through numerical solutions  where the nonlinear behaviour is fully kept into account.

\section{Elastic structure on a curved constraint with a jump in curvature}
\label{sec:jdescriptionStructure}

\subsection{Nonlinear dynamics}
\label{balena1}

\begin{figure}[H]
	\centering
	\includegraphics[]{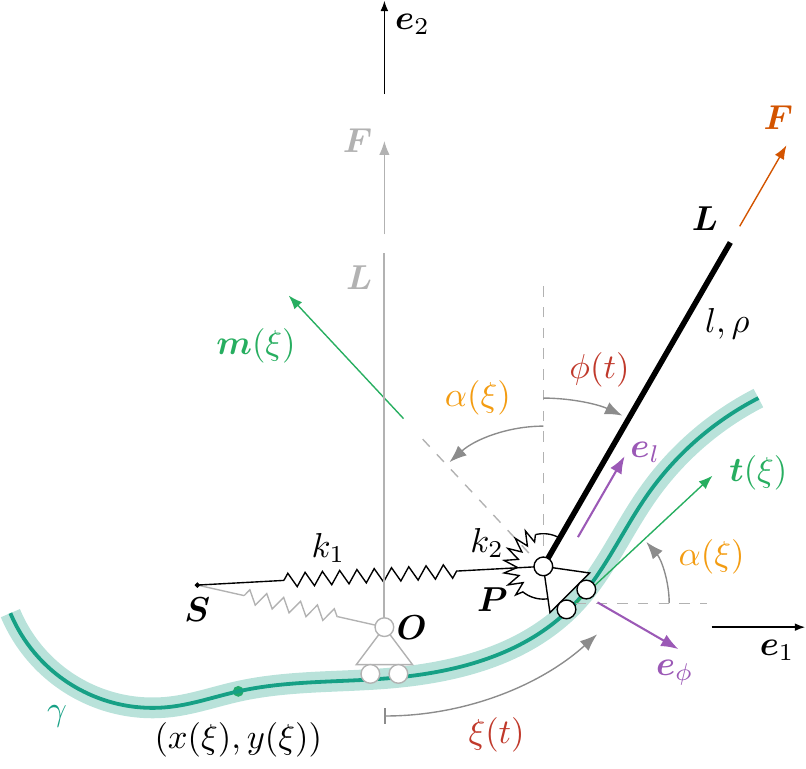}
	\caption{A 2 d.o.f. elastic structure made up of a rigid bar constrained to move with an elastic hinge on a curved profile and subject to a tensile follower force (remaining parallel to the bar).}
	\label{fig:structure}
\end{figure}

A two d.o.f. elastic structure is considered, Fig. \ref{fig:structure}, composed of a rigid bar, of mass density $\rho$ and length $l$, which is loaded at one end (point $\bL$) with a \emph{follower force}, positive when tensile, of constant modulus and parallel to the bar. At the other end (point $\bP$), the bar is connected to an elastic hinge of rotational stiffness $k_2$, which is constrained to move, without friction, along a smooth profile $\gamma$. The elastic hinge at the lower end of the rigid bar is linked to a fixed point $\bS$ (singled out by the coordinates $x_s$ and $y_s$), with a longitudinal linear spring of stiffness $k_1$. 

The rigid smooth profile, along which point $\bP$ is constrained to move, plays a fundamental role in the mechanics of the structure shown in Fig. \ref{fig:structure}. 
The formulation presented below is general enough to include profiles with discontinuous curvature, provided that higher-order derivatives are understood in the generalized sense. 

The profile may be described by parametric equations ($x(\xi)$, $y(\xi)$) in the plane $Oxy$ defined by the two unit vectors $\be_1$ and $\be_2$, so that the tangent, the unit tangent,  and the principal normal at the generic point $\bP(\xi)$ of the profile are
\begin{equation}
    \label{eq:defTangentNormal}
    \boldsymbol{P}' = x'(\xi) \boldsymbol{e}_1 + y'(\xi) \boldsymbol{e}_2, \qquad
    \boldsymbol{t} = \frac{\boldsymbol{P}'}{|\boldsymbol{P}'|}, \qquad
    \boldsymbol{n} = \frac{\boldsymbol{t}'}{|\boldsymbol{t}'|}, 
\end{equation}
in which a dash $(~)'$ denotes differentiation with respect to the parameter $\xi$. The signed curvature of the profile is defined as 
\begin{equation}
    \kappa = \frac{\alpha'}{|\boldsymbol{P}'|}, 
\end{equation}
where $\alpha$ is the angle between $\bP'$ and the $x$-axis, 
\begin{equation}
    \label{eq:defAlpha}
    \alpha(\xi) = \arctan \frac{y'(\xi)}{x'(\xi)},
\end{equation}
defined in such a way that $\bt$ and the unit vector $\bm$ obtained through an anticlockwise rotation of $\bt$ by $\pi/2$ may be represented as 
\begin{equation}
    \label{eq:defTangent}
    \boldsymbol{t} = \cos \alpha(\xi) \, \boldsymbol{e}_1 + \sin \alpha(\xi) \, \boldsymbol{e}_2,  \qquad
    \boldsymbol{m} = -\sin \alpha(\xi) \, \boldsymbol{e}_1 + \cos \alpha(\xi) \, \boldsymbol{e}_2.
\end{equation}
The derivative of equation \eqref{eq:defAlpha} with respect to $\xi$ leads to
\begin{equation}
    \alpha'  =  \frac{x'y'' - x''y'}{x'^2 + y'^2} =  \boldsymbol{m} \cdot \boldsymbol{t}'.
\end{equation}	

Note that the parameter $\xi$ may be identified with the arc length of the profile and in such case $|\bP'|=1$ so that $\alpha'$ coincides with the signed curvature $\kappa$. 

The deformation of the structure is described by two generalized coordinates: the arc length $\xi$ of the curve describing the profile and the angle $\phi$ between the rigid bar and the $y$-axis, positive if clockwise, which are assumed to be continuous functions of time, namely, $\xi=\xi(t)$ and $\phi=\phi(t)$. 

Two unit vectors  $\boldsymbol{e}_l$ and $\boldsymbol{e}_\phi$ are defined, attached to the rigid bar and aligned parallel and transverse to it respectively, as 
\begin{equation}
    \label{eq:defNewRefSys}
    \boldsymbol{e}_l = \sin\phi \, \boldsymbol{e}_1 + \cos\phi \, \boldsymbol{e}_2, \qquad 
    \boldsymbol{e}_\phi = \cos\phi \, \boldsymbol{e}_1 - \sin\phi \, \boldsymbol{e}_2. 
\end{equation}
Differentiation of equations \eqref{eq:defNewRefSys} yields (denoting with a superimposed dot the derivative with respect to time)
\begin{equation}
    \dot{\boldsymbol{e}}_l = \dot{\phi}\,\boldsymbol{e}_\phi, ~~~~ \dot{\boldsymbol{e}}_\phi = -\dot{\phi}\,\boldsymbol{e}_l, 
\end{equation}
and therefore the positions of the end points $\boldsymbol{P}$ and $\boldsymbol{L}$ and of the generic point $\boldsymbol{R}$ of the bar (at distance $r$ from $\boldsymbol{P}$) can be written as 
\begin{equation}
    \boldsymbol{P} = x(\xi)\boldsymbol{e}_1 + y(\xi)\boldsymbol{e}_2 + \boldsymbol{O} , \qquad
    \boldsymbol{L}   = l \boldsymbol{e}_l +  \boldsymbol{P} , \qquad
    \boldsymbol{R}  = r\, \boldsymbol{e}_l +  \boldsymbol{P} ,
\end{equation}
where it can be noted that $\boldsymbol{R}(l)=\boldsymbol{L}$. The velocities of points $\boldsymbol{P}$, $\boldsymbol{L}$ and $\boldsymbol{R}$ are
\begin{equation}
    \dot{\boldsymbol{P}} = \dot{\xi}\boldsymbol{P}',   \qquad        
    \dot{\boldsymbol{L}}  = l \dot{\phi} \boldsymbol{e}_\phi +  \dot{\boldsymbol{P}}, \qquad
    \dot{\boldsymbol{R}}  = r\, \dot{\phi} \boldsymbol{e}_\phi +  \dot{\boldsymbol{P}},
\end{equation}
where $\boldsymbol{P}'$ provides the tangent to the rigid profile at $\boldsymbol{P}$, equation \eqref{eq:defTangentNormal}$_1$. The accelerations can be calculated as 
\begin{equation}
    \ddot{\boldsymbol{P}} = \ddot{\xi}\boldsymbol{P}'+\dot{\xi}^2\boldsymbol{P}'', 	\qquad 
    \ddot{\boldsymbol{L}} = l\ddot{\phi} \boldsymbol{e}_\phi - l \dot{\phi}^2 \boldsymbol{e}_l +\ddot{\boldsymbol{P}}, \qquad
    \ddot{\boldsymbol{R}} = r\,\ddot{\phi} \boldsymbol{e}_\phi - r\, \dot{\phi}^2 \boldsymbol{e}_l +\ddot{\boldsymbol{P}}. 	
\end{equation}
The follower force has a constant modulus $F$ and remains always parallel to the rigid bar,
\begin{equation}
    \boldsymbol{F} =  F\,\boldsymbol{e}_l.
\end{equation}
The longitudinal spring with stiffness $k_1$ produces an elastic force proportional to the vector $\boldsymbol{P}-\boldsymbol{S}$
\begin{equation}
    \boldsymbol{F}_s = -k_1 (\boldsymbol{P} - \boldsymbol{S}) ,
\end{equation}
while the rotational spring of stiffness $k_2$ applies a moment (positive when anticlockwise) to the end $\boldsymbol{P}$ of the rigid bar, which is given by 
\begin{equation}
    M = -k_2 (\phi + \alpha), 
\end{equation}
where $\alpha$ has been defined in formula \eqref{eq:defAlpha}.

The equations of motion governing the dynamics of the mechanical system under analysis can be found using the principle of virtual work, which can be written as
\begin{equation}
    \label{eq:plv}
    \boldsymbol{F} \cdot \delta\boldsymbol{L} - k_1 (\boldsymbol{P} - \boldsymbol{S}) \cdot \delta\boldsymbol{P} - k_2 (\phi + \alpha) (\delta\phi + \delta\alpha) - \rho  
    \int_0^l \ddot{\boldsymbol{R}} \cdot \delta\boldsymbol{R}  \, dr = 0 .
\end{equation}
The external work due to the follower force is
\begin{equation}
    \boldsymbol{F} \cdot \delta\boldsymbol{L} = F \boldsymbol{e}_l \cdot \delta \boldsymbol{P},
\end{equation}
while
\begin{equation}
    \int_0^l \ddot{\boldsymbol{R}} \cdot \delta\boldsymbol{R}  \, dr = \frac{l^2}{2}\left(\frac{2l}{3}\ddot{\phi} + \ddot{\boldsymbol{P}}\cdot \boldsymbol{e}_\phi\right)\delta\phi
    +l\left(\frac{l}{2} \ddot{\phi}\boldsymbol{e}_\phi - \frac{l}{2}\dot{\phi}^2 \boldsymbol{e}_l + \ddot{\boldsymbol{P}}\right)\cdot\delta \boldsymbol{P} ,
\end{equation}
so that equation \eqref{eq:plv} can be rewritten as 
\begin{multline}
    \left[
    \left( F + \rho \frac{l^2}{2}\dot{\phi}^2\right)\boldsymbol{e}_l -\rho\frac{l^2}{2} \ddot{\phi} \boldsymbol{e}_\phi 
    - k_1 (\boldsymbol{P} - \boldsymbol{S}) - \rho l \ddot{\boldsymbol{P}} 
    \right] 
    \cdot \boldsymbol{P}' \delta \xi - k_2 (\phi + \alpha) \alpha' \delta\xi \\
    - \left[ 
    k_2 (\phi + \alpha) + \rho \frac{l^2}{2}\left(\frac{2l}{3}\ddot{\phi} +\ddot{\boldsymbol{P}} \cdot \boldsymbol{e}_\phi\right) 
    \right] \delta\phi = 0,
\end{multline}
which, invoking the arbitrariness of $\delta\xi$ and $\delta\phi$, can be split into the two equations governing the dynamics of the structure
\begin{equation}
    \label{eq:nonlinearComplete}
    \begin{split}
        & \left( F + \rho \frac{l^2}{2}\dot{\phi}^2\right)\left(x'\sin\phi+y'\cos\phi \right) -\rho\frac{l^2}{2} \ddot{\phi}\left(x'\cos\phi -y'\sin\phi \right)  
        - k_1 \left[x'(x-x_S)+y'(y-y_S)\right] \\
        & - \rho l \left[\ddot{\xi}\left(x'^2+y'^2  \right)  + \dot{\xi}^2\left(x'x''+y'y''  \right) \right] - k_2 (\phi + \alpha) \alpha' = 0, \\[3mm]
        & k_2 (\phi + \alpha) + \rho \frac{l^3}{3}\ddot{\phi} + \rho \frac{l^2}{2} 
        \left[\ddot{\xi}\left(x'\cos\phi-y'\sin\phi \right) + \dot{\xi}^2\left(x''\cos\phi -y''\sin\phi\right)\right] = 0.
    \end{split}
\end{equation}
The nonlinear system \eqref{eq:nonlinearComplete} can be solved for $\xi$ and $\phi$, so that it can be equivalently written as
\begin{equation}
    \label{eq:lagrange}
    \ddot{\bq}(t) = \bg(\bq(t),\dot{\bq}(t)),
\end{equation}
where $\bq(t) = [\xi(t),\phi(t)]^T$ is a vector collecting the Lagrangian coordinates.

Alternatively to the above, the Hamiltonian formulation can be used, so that the system \eqref{eq:lagrange} becomes a first-order differential nonlinear system
\begin{equation}
    \label{eq:hamilton}
    \dot{\by}(t) = \bef(\by(t)),
\end{equation}
where the phase vector $\boldsymbol{y}(t) = [\boldsymbol{q}(t), \dot{\boldsymbol{q}}(t)]^{\text{T}} = [\xi, \phi, \dot{\xi}, \dot{\phi}]^T$ contains the vector of Lagrangian generalized coordinates and its first derivative in time, so defining a \emph{4-dimensional phase space}.

\subsection{Linearized dynamics}
\label{balena2}

The nonlinear differential system \eqref{eq:nonlinearComplete} can be linearized near $\xi=\phi =0$ as %
\begin{equation}
    \label{eq:linearisedEquationsMotion}
    \begin{split}
        & \rho\frac{l^2}{2} \ddot{\phi} x'(0)   +\left[k_2 \alpha' - F x'\right]_{\xi=0} \phi +\rho l \ddot{\xi}\left[x'^2+y'^2 \right]_{\xi=0}  \\
        & + \left[k_1 \left(x''(x-x_S)+x'^2+y''(y-y_S) +y'^2\right) - Fy'' + k_2\left(\alpha'^2 +\alpha \alpha''\right) \right]_{\xi=0}\xi \\
        & + \left[k_1 \left(x'(x-x_S)+y'(y-y_S)\right) - Fy' + k_2 \alpha' \alpha \right]_{\xi=0} = 0, \\[3mm]
        & \rho \frac{l^3}{3}\ddot{\phi} + k_2 \phi  + \rho \frac{l^2}{2} x'(0) \ddot{\xi} + k_2\alpha'(0) \xi + k_2\alpha(0) = 0 . 
    \end{split}
\end{equation}
Furthermore, with the introduction of the vector collecting the Lagrangian generalized coordinates,  $\boldsymbol{q} = \left[\xi,\phi \right]^T$, equations \eqref{eq:linearisedEquationsMotion} can be written in matrix form as 
\begin{equation}
    \label{eq:matrixGeneral}
    \boldsymbol{M} \ddot{\boldsymbol{q}}(t) +  \boldsymbol{K}\boldsymbol{q}(t) = \boldsymbol{f}(t), 
\end{equation}
where $\boldsymbol{M}$ is the mass matrix, $\boldsymbol{K}$ the stiffness matrix and $\boldsymbol{f}$ the vector of generalized forces, respectively
\begin{equation}
    \boldsymbol{M} = 
    \rho l \, 
    \begin{bmatrix}
    \displaystyle	x'^2 + y'^2  & \displaystyle \frac{l}{2}x'\\[2ex]
    \displaystyle	\frac{l}{2}x' & \displaystyle \frac{l^2}{3}
    \end{bmatrix}_{\xi=0},
\end{equation}
\begin{equation}
    \label{eq:stiffMatrix}
    \boldsymbol{K} = 
    \begin{bmatrix}
    k_1 \left(x''(x-x_S) + x'^2 + y''(y-y_S) + y'^2\right)-Fy'' + k_2\left(\alpha'^2 +\alpha \alpha''\right) & k_2 \alpha' - F x' \\[5 mm]
    k_2\alpha' & k_2
    \end{bmatrix}_{\xi=0} ,
\end{equation}
and 
\begin{equation}
    \boldsymbol{f} = \left[Fy' - k_1 \left(x'(x-x_S)+y'(y-y_S)\right) - k_2 \alpha' \alpha, ~~~ -k_2 \alpha \right]_{\xi=0}^T.
\end{equation}

The trivial solution $\xi=\phi=0$ is an equilibrium configuration only when $\boldsymbol{f} = \boldsymbol{0}$, which implies
\begin{equation}
    \label{eq:conditions}
    y'(0)=0, \qquad x(0) = x_S,
\end{equation}
so that the tangent to the profile has to be horizontal at $\xi=0$, and the fixed point $\boldsymbol{S}$ of the linear spring must be aligned vertically with the point of the curve at $\xi=0$.

\subsection{Piecewise-smooth structure: doubly circular profile}

All  equations obtained in the previous Sections \ref{balena1} and \ref{balena2} can be applied to a profile with discontinuous curvature, and hold for both branches of the profile. However, the discontinuity has to be made explicit.
The introduced structure can be particularised through the implementation of a specific curve for the constraint, given as a parametric function of the arc length $\xi$. 

\begin{figure}[H]
	\centering
	{\includegraphics[]{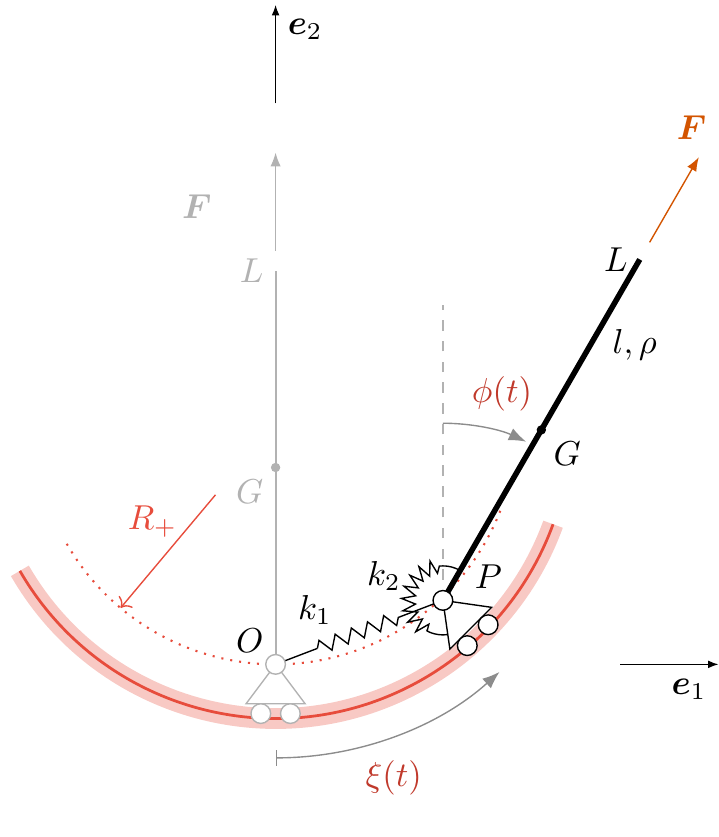}} \quad
	{\includegraphics[]{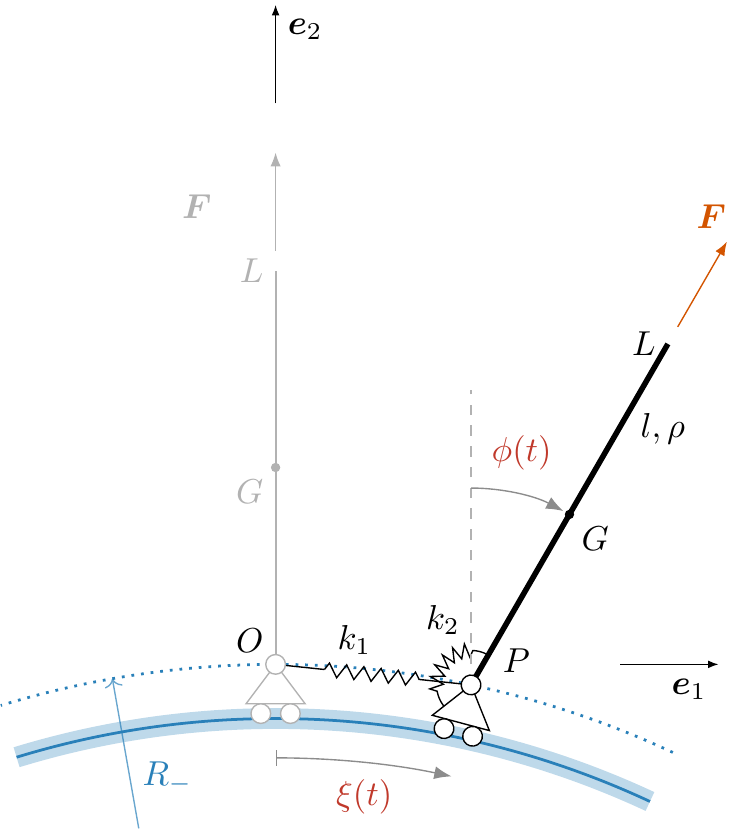}} \\
	\caption{Two elastic structures of the type shown in Fig.~\ref{fig:structure} with circular sliding profiles, having positive (on the left) and negative (on the right) curvatures.
	}
	\label{fig:singleStructures}
\end{figure}

Circular curves will be addressed with positive and negative curvatures, see Fig.~\ref{fig:singleStructures}, so that the coordinates of the point $\bP$ along the profile singled out by the arc length $\xi$ are 
\begin{equation}
    \label{eq:circular}
    x(\xi) = R_\pm \sin \frac{\xi}{R_\pm}, \qquad
    y(\xi) = \pm R_\pm \left(1 -  \cos \frac{\xi}{R_\pm}\right),
\end{equation}
where $R_\pm > 0$ is the radius of curvature and where the \lq $+$' sign (the \lq$-$' sign) applies for positive (for negative) curvature. For a circular curve described by the parametric representation \eqref{eq:circular}, the nonlinear governing equations are obtained from \eqref{eq:nonlinearComplete} by substituting
\begin{equation}
    \label{eq:derivatives}
    \begin{gathered}
        x'(\xi) = \cos \frac{\xi}{R_\pm}, \quad y'(\xi) = \pm \sin \frac{\xi}{R_\pm}, \quad x''(\xi) = - \frac{1}{R_\pm} \sin \frac{\xi}{R_\pm}, \quad y''(\xi) = \pm \frac{1}{R_\pm} \cos \frac{\xi}{R_\pm} \\
        \alpha(\xi) = \pm \frac{\xi}{R_\pm}, \quad \alpha'(\xi) = \pm \frac{1}{R_\pm}.
    \end{gathered}
\end{equation}
%

Moreover, conditions \eqref{eq:conditions} for a trivial equilibrium solution are fulfilled provided that $x_S = 0$. Referring to the linearized equations \eqref{eq:matrixGeneral}, the mass matrix remains the same for both positive and negative curvatures, while the stiffness matrix is different in the two cases, namely,
\begin{equation}
    \label{eq:massCirc}
    \boldsymbol{M} = \rho l 
    \begin{bmatrix}
        1   & l/2 \\[2ex]
        l/2  &  l^2/3 \\
    \end{bmatrix}, 
    \qquad
    \boldsymbol{K}^{\pm} = 
    \begin{bmatrix}
        \displaystyle k_1 + \frac{k_2}{R_\pm^2} \mp \frac{k_1 y_s}{R_\pm} \mp \frac{F}{R_\pm} \quad & \displaystyle \pm \frac{k_2}{R_\pm} - F \\[2ex]
        \displaystyle \pm \frac{k_2}{R_\pm}  & k_2\\
    \end{bmatrix}.
\end{equation}

\begin{figure}[H]
	\centering
	\includegraphics[]{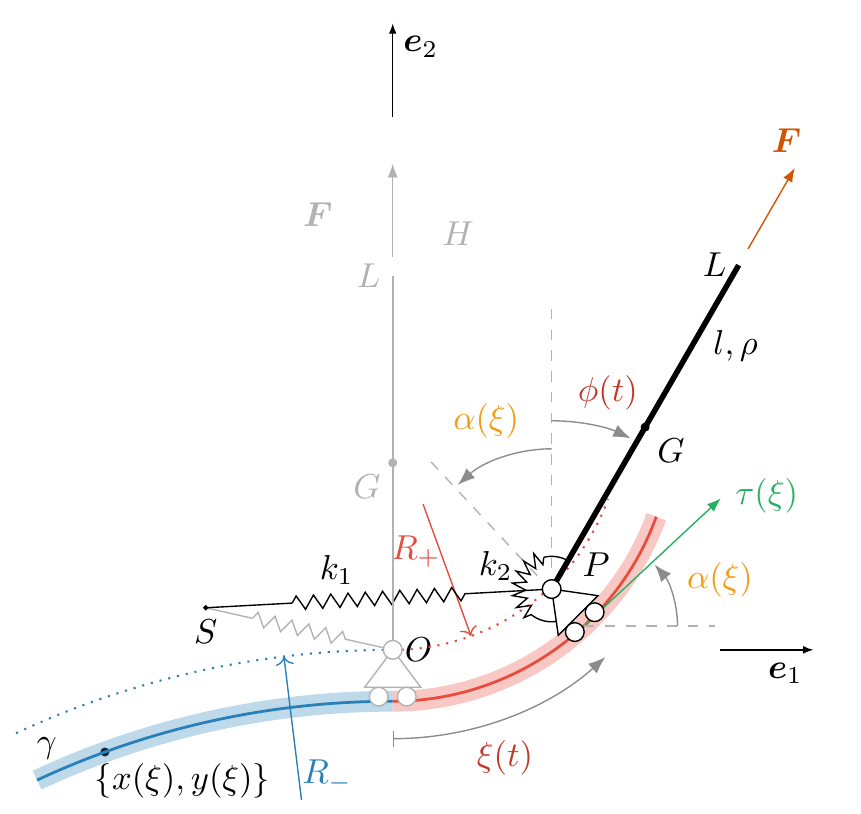}
	\caption{An elastic structure is obtained as the \lq fusion' of the two elastic structures shown in Fig. \ref{fig:singleStructures}. The profile, on which one end of the structure is forced to slide without friction, is composed of two circular paths, having negative and positive curvatures on the left and the right, respectively.}
	\label{fig:structureCircNegPos}
\end{figure}

A third elastic structure shown in Fig. \ref{fig:structureCircNegPos} is now considered, described by two degrees of freedom and obtained as the \lq fusion' of the two previously described subsystems with positive and negative curvatures. Specifically, the profile on the left (colored blue in the figure) is a circular path with negative curvature, while the profile on the right (colored red in the figure) is a circular path with positive curvature.
The tangent to the profile at the junction is horizontal and at this point, the longitudinal spring is unloaded, so that the vertical configuration of the rigid rod is the trivial equilibrium configuration.

This new structure is characterized by a smooth sliding profile, which however evidences a jump in the curvature at $\xi = 0$, so that the dynamics is characterized by piecewise smooth differential equations. Specifically, the equations of motion are obtained through a substitution of equation  \eqref{eq:derivatives} into equation \eqref{eq:nonlinearComplete} and considering the equations referred to the $+$ system ($-$ system) for $\xi> 0$ ($\xi < 0$).

Correspondingly, the small amplitude vibrations of the system are described by the following piecewise linear equations of motion
\begin{equation}
    \label{eq:equationPiecewiseLagrange}
    \left\lbrace
    \begin{aligned}
        & \boldsymbol{M} \ddot{\boldsymbol{q}}(t) +  \boldsymbol{K}^{-}\boldsymbol{q}(t) = \boldsymbol{0},  &\xi < 0, \\
        & \boldsymbol{M} \ddot{\boldsymbol{q}}(t) +  \boldsymbol{K}^{+}\boldsymbol{q}(t) = \boldsymbol{0},  &\xi > 0. \\
    \end{aligned}
    \right.
\end{equation}
Although composed of two linear differential problems, the piecewise system \eqref{eq:equationPiecewiseLagrange} is globally nonlinear, due to switching between two different sets of equations of motion.

Using the notation so far introduced, it will be shown that, due to the presence of the follower (non-conservative) load, the stability properties of this structure are non-trivial and, in particular, an unstable structure may result from the union of two structures which are stable when considered alone.

\section{Dynamics and instability for smooth and non-smooth structures} 
\label{sec:dynamicInstabilityStructures}

The analysis of a piecewise-smooth dynamical system described in the Lagrangian formalism by the equation of motions \eqref{eq:equationPiecewiseLagrange} is not trivial due to its nonlinear nature. In particular, as for stability analysis, the classical Lyapunov theorem on linear analysis cannot be applied, because the Jacobian matrix calculated at equilibrium is not unique, so all the standard criteria based on the nature of the eigenvalues of the Jacobian matrix are impracticable. Moreover, the complexity of the analysis increases due to the fact that the time intervals in which the solution is associated with a specific subdomain ($\xi > 0$ or $\xi < 0$) are \emph{a priori} unknown since they depend on the initial conditions applied to the structure. 

The aim of the present article is the definition of some general criteria that allow the design of a mechanical structure with given stability properties. In particular, the unusual unstable behaviour related to the coupling of two stable systems is investigated. In this Section, some extensions of the classical stability theory are introduced to deal with piecewise linear dynamical systems, with the aim of linking the stability properties of each single smooth system to those of the entire non-smooth structure.

\subsection{Linearized behaviour for a smooth structure}
\label{sec:smooth}

Before analyzing the non-smooth mechanical system of Fig.~\ref{fig:structureCircNegPos}, the stability of the smooth subsystems reported in Fig.~\ref{fig:singleStructures} is analyzed.
The treatment applies to both smooth mechanical systems with positive or negative curvature so that for simplicity the symbols \lq $\pm$' will be dropped in the following. The linearized dynamics is governed by the following system of two linear second-order differential equations
\begin{equation}
    \label{eq:linearSystemEquationMatrix}
    \boldsymbol{M} \ddot{\boldsymbol{q}} + \boldsymbol{K} \boldsymbol{q} = \boldsymbol{0},
\end{equation}
to be complemented by initial conditions on position and velocity, $\boldsymbol{q}(0) = \boldsymbol{q}_0$ and $\dot{\boldsymbol{q}}(0) = \dot{\boldsymbol{q}}_0$, respectively.

The solution to equation \eqref{eq:linearSystemEquationMatrix} can be expressed using exponential functions as
\begin{equation}
    \boldsymbol{q}(t) = \boldsymbol{\psi}^{(j)} e^{\lambda_j t},
\end{equation}	
leading to the eigenvalue problem
\begin{equation}
    \label{eq:eigenProblemLinearSystem}
    \left(\lambda_j^2 \boldsymbol{M} + \boldsymbol{K} \right) \boldsymbol{\psi}^{(j)} = \boldsymbol{0} , 
\end{equation}
which provides two values for $\lambda_j^2$. The eigenvalues $\lambda_j$ are related to the natural frequencies $\omega_j$ of the mechanical system through  $\lambda_j = i \omega_j$, where $i = \sqrt{-1}$ is the imaginary unit. 

The generalized coordinates $\boldsymbol{q}(t)$ that satisfy equation \eqref{eq:linearSystemEquationMatrix} can now be written as the linear combination of four exponential terms
\begin{equation}
    \label{eq:soutionLinSystemLagrange}
    \boldsymbol{q}(t) = \sum_{j=1}^2 \boldsymbol{\psi}^{(j)} \left( A_j e^{\lambda_j t}+ B_j e^{-\lambda_j t} \right),
\end{equation}
where $A_j$ and $B_j$ are four arbitrary constants that can be obtained from the initial conditions $\boldsymbol{q}_0$ and $\dot{\boldsymbol{q}}_0$. The response of the dynamical system to initial conditions near the trivial equilibrium configuration is now determined.

The linear equation \eqref{eq:linearSystemEquationMatrix} can be rewritten in the Hamiltonian form, namely, as a system of four linear first-order differential equations
\begin{equation}
    \label{eq:hamiltonianODEs}
    \dot{\boldsymbol{y}}(t) = \boldsymbol{A}\, \boldsymbol{y}(t),
\end{equation}
where 
\begin{equation}
    \boldsymbol{A} = 
    \begin{bmatrix} 
        \boldsymbol{0} & \boldsymbol{I} \\[2ex]
        -\boldsymbol{M}^{-1}\boldsymbol{K} & \boldsymbol{0}\\
    \end{bmatrix}
    = 
    \begin{bmatrix}
        \boldsymbol{0} & \boldsymbol{I} \\[2ex]
        \boldsymbol{\Gamma} & \boldsymbol{0}\\
    \end{bmatrix}
\end{equation}
is a $4\times4$ matrix, $\boldsymbol{I}$ is the $2\times2$ identity matrix, and the phase vector $\boldsymbol{y}(t) = [\boldsymbol{q}(t), \dot{\boldsymbol{q}}(t)]^{\text{T}} = [\xi, \phi, \dot{\xi}, \dot{\phi}]^T$ contains the vector of Lagrangian generalized coordinates and its first derivative in time, so defining a \emph{4-dimensional phase space}. 
The differential equation \eqref{eq:hamiltonianODEs} can be solved as for the Lagrangian formulation using an exponential ansatz $\boldsymbol{y}(t) = \boldsymbol{v}^{(j)} e^{\lambda_j t}$, leading to the eigenvalue problem
\begin{equation}
    \label{eq:eigProblemHamilt}
    \boldsymbol{A}\, \boldsymbol{v}^{(j)} = \lambda_j \boldsymbol{v}^{(j)} ,
\end{equation}
whose eigenvalues $\lambda_j$ \emph{are the same} appearing in equation \eqref{eq:eigenProblemLinearSystem}, while the eigenvectors $\boldsymbol{v}^{(j)}$ are related to the eigenvectors $\boldsymbol{\psi}^{(j)}$ through %
\begin{equation}
    \label{eq:eigenvectorsHamiltonian}
    \begin{aligned}
        \bv^{(1,2)} = &[\psi^{(1)}_1,  \psi^{(1)}_2,   \pm \lambda_1 \psi^{(1)}_1,   \pm \lambda_1 \psi^{(1)}_2 ]^{\text{T}},  \\ 
        \bv^{(3,4)} = &[\psi^{(2)}_1,  \psi^{(2)}_2,   \pm \lambda_2 \psi^{(2)}_1,    \pm \lambda_2 \psi^{(2)}_2 ]^{\text{T}}.
    \end{aligned}
\end{equation}
The solution of the initial value problem with assigned initial conditions $\boldsymbol{y}(0) = \boldsymbol{y}_0$ is unique and can be related to $\boldsymbol{y}_0$ through the so-called \emph{fundamental solution matrix}, which is the matrix exponential $e^{\boldsymbol{A} t}$ defined in such a way that
\begin{equation}
    \label{eq:solutionLinearSystemHamiltonian}
    \boldsymbol{y}(t) = e^{\boldsymbol{A} t}\, \boldsymbol{y}_0 ,
\end{equation}
which can also be written \emph{in extenso} for the case under analysis as
\begin{equation}
    \begin{bmatrix}
        \xi(t)\\
        \phi(t)\\
        \dot{\xi}(t)\\
        \dot{\phi}(t)\\
    \end{bmatrix}
    =
    e^{\boldsymbol{A} t}
    \begin{bmatrix}
        \xi(0)\\
        \phi(0)\\
        \dot{\xi}(0)\\
        \dot{\phi}(0)\\
    \end{bmatrix}.
\end{equation}
The \emph{stability analysis} of the equilibrium configuration for a given smooth mechanical system can be performed using the Lyapunov theorem. In particular, this analysis is based on the nature of the eigenvalues $\lambda_j$ of the matrix $\boldsymbol{A}$,
\begin{equation}
    \label{eq:structureEigenvalues}
    \lambda_j = \pm \sqrt{\frac{I_1 \pm \sqrt{I_1^2 - 4 I_2}}{2}},
\end{equation} 
where 
\begin{equation}
    \label{eq:invariants}
    I_1 = \tr \bGamma \quad \text{and} \quad I_2 = \det\bGamma
\end{equation}
are the first and second invariants of the matrix $\bGamma = -\bM^{-1} \bK$. 

\begin{figure}[H]
	\centering
	\includegraphics{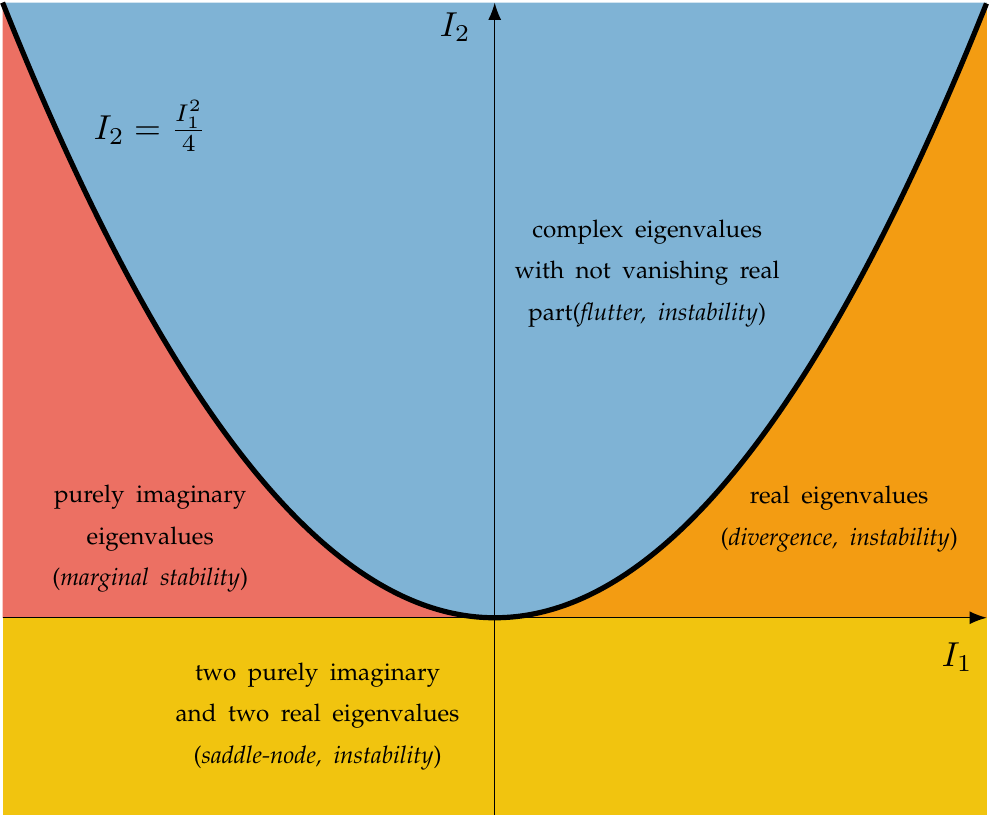}
	\caption{Representation in the plane $I_1 - I_2$ of the stability domains for a dynamical system characterized by 2 degrees of freedom}
	\label{fig:parabolaEigs}
\end{figure}

The Lyapunov theorem  \cite{kirillovNonconservativeStabilityProblems2013, zieglerPrinciplesStructuralStability1977} states that the equilibrium configuration of a nonlinear dynamical system is stable when the real parts of all eigenvalues of the Jacobian matrix are negative, whereas it is unstable when the real part of at least one eigenvalue is positive. The case in which the real part of one or more eigenvalues is zero and all the others have a negative real part is not covered by the Lyapunov theorem and is referred to as {\it critical case}, in the context of mechanics \cite{zieglerPrinciplesStructuralStability1977}, or as \emph{marginally stable}, in the context of abstract dynamical systems  \cite{kirillovNonconservativeStabilityProblems2013}. Due to the symmetry of the eigenvalues with respect to the imaginary axis given by the structure of the formula \eqref{eq:structureEigenvalues}, only three cases can be distinguished with reference to Fig.~\ref{fig:parabolaEigs}:
\begin{itemize}
    \item (Marginal) stability: the two squared eigenvalues $\lambda_j^2$ are real and negative, so that the corresponding $\lambda_j \in i\mathbb{R}$ are two purely imaginary conjugate pairs, say,
    $\lambda_{1,2} = \pm i \omega_1$ and $\lambda_{3,4} = \pm i \omega_2$. To be more specific, this is the critical case mentioned above. 
    Note that the sub-structures originating the piecewise-smooth structure analyzed in this article are stable in the sense considered here and belong to the zone marked red in the figure.
    
    \item Divergence instability: at least one of $\lambda_j^2$ is real and positive, that produces $\lambda_{1,2} = \pm \omega_1$. This situation corresponds to the zones marked orange and yellow in Fig. \ref{fig:parabolaEigs}.
    
    \item Flutter instability: $\lambda_j^2$ are complex conjugate pairs, so that $\lambda_{1,2,3,4} = \pm (\alpha \pm i\beta$), $\alpha \neq 0$. The behaviour is unstable and the presence of a non-null imaginary part produces oscillations in the solution, zone marked blue in the figure.
\end{itemize}

The aim of the present article is to show that two smooth elastic structures, which are stable when taken separately, may lead to an unstable structure when combined together. Therefore, our interest is in the case when both smooth subsystems are stable, for which $\lambda_{1,2}=\pm i\omega_1$ and $\lambda_{3,4}=\pm i\omega_2$. Assuming $\omega_1\neq \omega_2$
\footnote{The coalescence $\omega_1=\omega_2$ would denote grazing of an unstable boundary, an occurrence which is excluded here.}
, matrix $\bA$ is diagonalizable as $\boldsymbol{A} = \boldsymbol{U} \boldsymbol{J} \boldsymbol{U}^{-1}$ and the matrix exponential fulfils the relation
\begin{equation}
    \label{eq:jordanForm}
    e^{\bA} = \bU e^{\bJ} \bU^{-1},
\end{equation}
where 
\begin{equation}
    \boldsymbol{J} = 
    \begin{bmatrix}
        i \omega_1  		& 0 			& 0 			& 0 		\\
        0  				& -i \omega_1  	& 0				& 0 		\\
        0  				& 0				& i \omega_2  	& 0 		\\
        0			  	& 0 			& 0 			& -i \omega_2      \\
    \end{bmatrix}, 
    \qquad 
    \boldsymbol{U} = 
    \begin{bmatrix}
        \psi^{(1)}_1 & \psi^{(1)}_1 & \psi^{(2)}_1 & \psi^{(2)}_1 \\[1mm]
        \psi^{(1)}_2 & \psi^{(1)}_2 & \psi^{(2)}_2 & \psi^{(2)}_2 \\[1mm]
        i \omega_1 \psi^{(1)}_1 & -i \omega_1  \psi^{(1)}_1 & i \omega_2  \psi^{(2)}_1 & -i \omega_2  \psi^{(2)}_1 \\[1mm]
        i \omega_1 \psi^{(1)}_2 & -i \omega_1  \psi^{(1)}_2 & i \omega_2  \psi^{(2)}_2 & -i \omega_2  \psi^{(2)}_2 
    \end{bmatrix},
\end{equation}
and the vectors $\boldsymbol{\psi}^{(1)}$ and $\boldsymbol{\psi}^{(2)}$ are solutions of equation  \eqref{eq:eigenProblemLinearSystem}, while the inverse of $\boldsymbol{U}$ is given by
\begin{equation}
    \boldsymbol{U}^{-1} = \frac{1}{2(\psi^{(1)}_1\psi^{(2)}_2 - \psi^{(1)}_2\psi^{(2)}_1)}
    \begin{bmatrix}
        \psi^{(2)}_2 & -\psi^{(2)}_1 & -i \psi^{(2)}_2/\omega_1 & i \psi^{(2)}_1/\omega_1 \\[1mm]
        \psi^{(2)}_2 & -\psi^{(2)}_1 & i \psi^{(2)}_2/\omega_1 & -i \psi^{(2)}_1/\omega_1 \\[1mm]
        -\psi^{(1)}_2 & \psi^{(1)}_1 & i \psi^{(1)}_2/\omega_2 & -i \psi^{(1)}_1/\omega_2 \\[1mm]
        -\psi^{(1)}_2 & \psi^{(1)}_1 & -i \psi^{(1)}_2/\omega_2 & i \psi^{(1)}_1/\omega_2 
    \end{bmatrix}.
\end{equation}
Therefore, the matrix exponential $e^{\boldsymbol{J} t}$ becomes 
\begin{equation}
    e^{\boldsymbol{J} t} =
    \begin{bmatrix}
        \cos \omega_1 t + i \sin \omega_1 t & 0 & 0 & 0 \\
        0 & \cos \omega_1 t - i \sin \omega_1 t & 0 & 0 \\
        0 & 0 & \cos \omega_2 t + i \sin \omega_2 t & 0 \\
        0 & 0 & 0 &\cos \omega_2 t - i \sin \omega_2 t\\
    \end{bmatrix} 
\end{equation}
and finally, the matrix exponential $e^{\boldsymbol{A}t}$ for the four-dimensional smooth mechanical system considered here is
\begin{equation}
    \label{eq:exponentialMatrix}
    e^{\boldsymbol{A}t} =
    \begin{bmatrix}
        \frac{a_1 \cos  \omega_1  t -a_2  \cos  \omega_2  t }{a_1 -a_2 } & \frac{a_1  a_2  (\cos  \omega_2  t -\cos  \omega_1  t )}{a_1 -a_2 } & \frac{a_1  \omega_2  \sin  \omega_1  t -a_2  \omega_1  \sin  \omega_2  t }{a_1  \omega_1  \omega_2 -a_2  \omega_1  \omega_2 } & \frac{a_1  a_2  (\omega_1  \sin  \omega_2  t - \omega_2  \sin  \omega_1  t) }{a_1  \omega_1  \omega_2 -a_2  \omega_1  \omega_2 } \\[2ex]
        \frac{\cos  \omega_1  t -\cos  \omega_2  t }{a_1 -a_2 } & \frac{a_1  \cos  \omega_2  t -a_2  \cos  \omega_1  t }{a_1 -a_2 } & \frac{\omega_2  \sin  \omega_1  t -\omega_1  \sin  \omega_2  t }{a_1  \omega_1  \omega_2 -a_2  \omega_1  \omega_2 } & \frac{a_1  \omega_1  \sin  \omega_2  t -a_2  \omega_2  \sin  \omega_1  t }{a_1  \omega_1  \omega_2 -a_2  \omega_1  \omega_2 } \\[2ex]
        \frac{a_2  \omega_2  \sin  \omega_2  t -a_1  \omega_1  \sin  \omega_1  t }{a_1 -a_2 } & \frac{a_1  a_2  (\omega_1  \sin  \omega_1  t -\omega_2  \sin  \omega_2  t )}{a_1 -a_2 } & \frac{a_1  \cos  \omega_1  t -a_2  \cos  \omega_2  t }{a_1 -a_2 } & \frac{a_1  a_2  (\cos  \omega_2  t -\cos  \omega_1  t )}{a_1 -a_2 } \\[2ex]
        \frac{\omega_2  \sin  \omega_2  t -\omega_1  \sin  \omega_1  t }{a_1 -a_2 } & \frac{a_2  \omega_1  \sin  \omega_1  t -a_1  \omega_2  \sin  \omega_2  t }{a_1 -a_2 } & \frac{\cos  \omega_1  t -\cos  \omega_2  t }{a_1 -a_2 } & \frac{a_1  \cos  \omega_2  t -a_2  \cos  \omega_1  t }{a_1 -a_2 } \\
    \end{bmatrix},
\end{equation}
where 
\begin{equation}
    a_1 = \psi^{(1)}_1/\psi^{(1)}_2 =  -\frac{\omega_1^2+\Gamma_{22}}{\Gamma_{21}}, \qquad
    a_2 = \psi^{(2)}_1/\psi^{(2)}_2 =  -\frac{\omega_2^2+\Gamma_{22}}{\Gamma_{21}} . 
\end{equation}
It will be instrumental later to calculate the inverse of the matrix given by equation \eqref{eq:exponentialMatrix}. Because of the property 
\begin{equation}
    \boldsymbol{A} \boldsymbol{B} = \boldsymbol{B} \boldsymbol{A} \Longrightarrow e^{\boldsymbol{A}} e^{\boldsymbol{B}} = e^{\boldsymbol{A} + \boldsymbol{B}}
\end{equation}
the inverse of the exponential matrix can be obtained as 
\begin{equation}
    \label{eq:inverseMatrix}
    (e^{\boldsymbol{A} t} )^{-1} = e^{-\boldsymbol{A} t}, 
\end{equation}
corresponding to a change in the sign of $t$ in equation \eqref{eq:exponentialMatrix}. 

As described above, for a smooth mechanical system, the stability can simply be judged through the calculation of the eigenvalues $\lambda^2_j$ of the jacobian matrix. However, this is not enough for piecewise smooth structures, for which a complete understanding of the dynamics of the system is required, by a direct solution of the equations of motion. This analysis will be performed in the next paragraph through the computation of the solution of the piecewise linear system \eqref{eq:equationPiecewiseLagrange}.

\subsection{Linearized behaviour for a piecewise-smooth structure}

When a piecewise smooth dynamical system is considered, equation \eqref{eq:equationPiecewiseLagrange} can be rewritten in the Hamiltonian form as
\begin{equation}
    \label{eq:piecewiseHamilton}
    \dot{\boldsymbol{y}}(t) = 
    \left\lbrace
    \begin{aligned}
        & \boldsymbol{A}^- \boldsymbol{y}(t), & y_1 <0 ,\\
        & \boldsymbol{A}^+ \boldsymbol{y}(t), & y_1 > 0 ,\\
    \end{aligned}
    \right.
\end{equation}
where $y_1=\xi$. A dynamical system with a discontinuous right-hand side such as that 
expressed by equation \eqref{eq:piecewiseHamilton} is referred to as \lq Filippov system' \cite{filippovDifferentialEquationsDiscontinuous1988a}.

Note that, although the accelerations $\dot y_3 = \ddot \xi$ and $\dot y_4 = \ddot \phi$ suffer a jump across the discontinuity in the curvature of the profile, $y_1=0$, the velocities $\dot{y}_1 = y_3 = \dot{\xi}$ and $\dot{y}_2 = y_4 = \dot{\phi}$ remain continuous, so that
\begin{equation}
    \label{eq:continuityOnSwitchingManifold}
    \be_1 \scalp \boldsymbol{A}^- \boldsymbol{y}(t) = \be_1 \scalp \boldsymbol{A}^+ \boldsymbol{y}(t),
    \text{ and } 
    \be_2 \scalp \boldsymbol{A}^- \boldsymbol{y}(t) = \be_2 \scalp \boldsymbol{A}^+ \boldsymbol{y}(t),
     \text{ when } y_1=0.
\end{equation}
Equation \eqref{eq:piecewiseHamilton} shows that the piecewise smooth structure under consideration defines a 4-dimensional {\it phase space}, with canonical basis denoted by $\{\be_1,\be_2,\be_3,\be_4\}$, which can be divided into two subdomains $\mathcal{V}^\pm$ within which the system can be considered smooth. The manifold separating the two subdomains, called {\it switching manifold}, is the hyperplane $\Sigma = \{\by \in \mathbb{R}^4 : y_1 = 0\}$. The negative part of equations \eqref{eq:piecewiseHamilton} applies to the subdomain defined by $\mathcal{V}^- = \left\lbrace \boldsymbol{y}\in \mathbb{R}^4 : y_1 < 0\right\rbrace$, while the positive part applies to $\mathcal{V}^+ = \left\lbrace \boldsymbol{y} \in \mathbb{R}^4 : y_1 > 0\right\rbrace$. Note that the origin of the configuration space $(\xi,\phi)=(y_1,y_2)=(0,0)$ represents the trivial equilibrium configuration of the structure and belongs to the switching manifold $\Sigma$.

The discontinuous system \eqref{eq:piecewiseHamilton} does not define the time derivative $\dot{\by}(t)$ when the configuration of the system $\by(t)$ is on the switching boundary $\Sigma$. 
To overcome this difficulty, Filippov has developed a technique, known as {\it Filippov's convex method}, that extends the discontinuous system \eqref{eq:piecewiseHamilton} to a {\it differential inclusion} of the form
\begin{equation}
    \label{eq:differentialInclusion}
    \dot{\by}(t) \in 
    \left\lbrace
    \begin{aligned}
        & \boldsymbol{A}^- \boldsymbol{y}(t), & y_1 < 0 , \\
        & \overline{\operatorname{co}} \lbrace \boldsymbol{A}^- \boldsymbol{y}(t), \boldsymbol{A}^+ \boldsymbol{y}(t) \rbrace, & y_1 = 0 , \\
        & \boldsymbol{A}^+ \boldsymbol{y}(t), & y_1 > 0 , \\
    \end{aligned}
    \right.
\end{equation}
where the closed convex hull $\overline{\operatorname{co}}$ of the two right-hand sides $\bef^-$ and $\bef^+$ is defined by
\begin{equation}
    \overline{\operatorname{co}} \lbrace \bef^-, \bef^+ \rbrace = 
    \lbrace (1 - \eta) \bef^- + \eta \bef^+, \forall \eta \in [0,1] \rbrace .
\end{equation}
A \emph{solution in the Filippov sense} of the discontinuous system \eqref{eq:piecewiseHamilton} is a solution of the differential inclusion \eqref{eq:differentialInclusion}. In this sense, the mechanical system admits among the possible solutions, the so-called {\it sliding modes}, namely, motions in which the system remains on the switching manifold $\Sigma$.

More precisely, there are three possible ways in which the mechanical system behave around the switching boundary $\Sigma$, namely:
\begin{itemize}
    \item Transverse intersection: both vector fields $\bA^+ \by(t)$ and $\bA^- \by(t)$ point on the same side of $\Sigma$
    \begin{equation}
        \label{eq:transversal}
        \big[\be_1 \scalp \bA^+ \by(t)\big] \big[\be_1 \scalp \bA^- \by(t)\big] > 0,
    \end{equation}
    so that a solution, that evolves in one subdomain and at some instant of time hits $\Sigma$, will cross it transversely and proceed in the other subdomain. In this case, the solution is locally unique.
    \item Attractive sliding modes: both vector fields $\bA^+ \by(t)$ and $\bA^- \by(t)$ point to $\Sigma$
    \begin{equation}
        \label{eq:attractive}
        \be_1 \scalp \bA^+ \by(t) < 0 \quad \text{and} \quad \be_1 \scalp \bA^- \by(t)] > 0,
    \end{equation}
    hence a solution that hits the switching boundary $\Sigma$ will not leave it and will therefore move along $\Sigma$. Also in this case the solution is locally unique.
    \item Repulsive sliding modes: both vector fields $\bA^+ \by(t)$ and $\bA^- \by(t)$ point in the opposite direction to $\Sigma$
    \begin{equation}
        \label{eq:repulsive}
        \be_1 \scalp \bA^+ \by(t) > 0 \quad \text{and} \quad \be_1 \scalp \bA^- \by(t)] < 0,
    \end{equation}
    which implies that a solution emanating from $\Sigma$ can remain in $\Sigma$ or leave it by entering either subdomain. Consequently, the solution, in this case, is not unique.
\end{itemize}

For the two d.o.f. elastic structure under consideration, the condition of transverse  intersection, equation  \eqref{eq:transversal}, is satisfied almost everywhere, that is for any configuration belonging to $\Sigma$ and such that $y_3 = \dot{\xi} \neq 0$, since the velocity $\dot{\xi}$ is continuous across the switching manifold $\Sigma$. Sliding modes are only possible in a zero-measure subset of $\Sigma$, namely the set $\lbrace \by \in \Reals^4 : \by \scalp \be_1 = \by \scalp \be_3 = 0 \rbrace$, and will not be investigated further.

Assuming that transverse intersection always prevails at the intersections of the orbits with the switching manifold $\Sigma$, a solution that evolves in one subdomain and hits $\Sigma$ necessarily crosses the hyperplane and enters the other subdomain. The intersection point between the orbit and the hyperplane $\Sigma$ can then be used as the initial condition for the subsequent evolution in the subdomain in which the orbit is entering. Therefore, the solution of the piecewise-linear system \eqref{eq:piecewiseHamilton} can be obtained as the composition of exponential matrices as follows
\begin{equation}
    \label{eq:solution}
    \by(t) = 
    \left\{
    \begin{array}{ll}
        e^{\bA^-(t-t_0)} \by_0, & t_0 \leq t < t_1, \\
        e^{\bA^+(t-t_1)} e^{\bA^-(t_1-t_0)} \by_0, & t_1 \leq t < t_2, \\
        \cdots & \cdots \\
        e^{\bA^-(t-t_{k-1})} e^{\bA^+(t_{k-1}-t_{k-2})} \cdots e^{\bA^+(t_2-t_1)} e^{\bA^-(t_1-t_0)} \by_0, & t_{k-1} \leq t < t_{k}, \\
        e^{\bA^+(t-t_{k})} e^{\bA^-(t_{k}-t_{k-1})} \cdots e^{\bA^+(t_2-t_1)} e^{\bA^-(t_1-t_0)} \by_0, & t_{k} \leq t < t_{k+1}, \\
        \cdots & \cdots \\
    \end{array}
    \right.
\end{equation}
where, without loss of generality, it is assumed that the initial condition belongs to the negative subdomain, $\by(0) = \by_0 \in \mV^-$, and $\{t_1, t_2, \cdots, t_{k-1}, t_{k}, \cdots\}$ is the sequence of intersection times at which the orbit crosses the switching boundary and changes subdomain. However, it should be pointed out that the piecewise solution \eqref{eq:solution} is not completely determined, because the intersection times $\{t_1, t_2, \cdots, t_{k-1}, t_{k}, \cdots\}$ are \emph{a priori} unknown and depend on the initial condition $\by_0$. To completely define the solution \eqref{eq:solution}, the intersection times have to be determined by tracing the evolution of the orbit from the initial condition $\by_0$ and numerically detecting the roots of the crossing condition $\xi = 0$.

A classical expedient to simplify the description of a dynamical system is the introduction of a discrete map, known as \emph{Poincar\'e map}. A Poincar\'e map transforms a $n$-dimensional continuous-time system into a $(n-1)$-dimensional discrete-time system, with the introduction of a hyperplane embedded in the $n$-dimensional phase space, called \emph{Poincar\'e section}. 

For the two d.o.f. elastic system under consideration, described by the 4-dimensional non-smooth dynamical system \eqref{eq:piecewiseHamilton}, a natural and convenient choice for the 3-dimensional Poincar\'e section is the switching manifold $\Sigma$. In this case, the Poincar\'e map $\bP: \Sigma \to \Sigma$ links points on $\Sigma$ through the orbits defined by the solution \eqref{eq:solution}. 

More precisely, considering a point $\bx \in \Sigma$ and assuming that the orbit starting from the initial condition $\bx$ enters the negative subspace $\mV^-$, the time evolution in the negative subsystem is described by the first expression in \eqref{eq:solution}
\begin{equation}
    \by(t) = e^{\bA^- (t - t_0)} \bx,
\end{equation}
until the orbit reaches $\Sigma$ at point $\bxi = \by(t_1) = \exp{(\bA^- \Delta{t^-})}\, \bx$ in a given time interval, namely $\Delta{t^-} = t_1 - t_0$ measured from the initial time $t_0$. Then the orbit crosses the switching manifold and enters the positive subsystem $\mV^+$, following the orbit described by
\begin{equation}
    \by(t) = e^{\bA^+ (t - t_1)} e^{\bA^- \Delta{t^-}} \bx = e^{\bA^+ (t - t_1)} \bxi, 
\end{equation}
so that the point $\bxi$ can be interpreted as a new initial condition for the second part of the orbit, evolving within the positive subdomain $\mV^+$. The orbit remains inside $\mV^+$ until it hits $\Sigma$ for a second time at point $\boldeta = \by(t_2) = \exp{(\bA^+ \Delta{t^+})}\, \bxi$ in a time interval $\Delta{t^+} = t_2 - t_1$, measured from $t = t_1$.

The just described sequence  defines two Poincar\'e half-maps
\begin{equation}
    \label{eq:halfmaps}
    \bxi = \bP^-(\bx) = e^{\bA^- \Delta{t^-}(\bx)} \bx, \qquad \boldeta = \bP^+(\bxi) = e^{\bA^+ \Delta{t^+}(\bxi)} \bxi,
\end{equation}
and that the complete Poincar\'e map $\bP: \Sigma \to \Sigma$ may be obtained through the composition of the two half-maps, $\bP = \bP^+ \circ \bP^-$, such that
\begin{equation}
   \label{eq:map}
   \boldeta = \bP(\bx) = \bP^+(\bP^-(\bx)) = e^{\bA^+ \Delta{t^+}(\bxi(\bx))} e^{\bA^- \Delta{t^-}(\bx)} \bx. 
\end{equation}
It should be noted that the time intervals $\Delta{t^-}(\bx)$ and $\Delta{t^+(\bxi)}$ are non linear functions of the initial conditions $\bx$ and $\bxi$, respectively, and can be defined as
\begin{equation}
    \label{eq:intervals}
    \begin{aligned}
        \Delta{t^-}(\bx) &= \inf \left\{ \Delta{t} > 0: \be_1 \scalp e^{\bA^- \Delta{t}} \bx = 0 \right\}, \\
        \Delta{t^+}(\bxi) &= \inf \left\{ \Delta{t} > 0: \be_1 \scalp e^{\bA^+ \Delta{t}} \bxi = 0 \right\}, 
    \end{aligned}
\end{equation}
where $\be_1$ is the normal to the switching manifold $\Sigma$. These definitions simply represent the fact that the points $\bxi$ and $\boldeta$ must be on the switching manifold and that they identify the first two intersections between the considered trajectory and the hyperplane $\Sigma$.

The use of the Poincar\'e map will be crucial in the next section for the analysis of the stability of the piecewise smooth elastic structure.

\begin{figure}[h]
	\centering
	\includegraphics{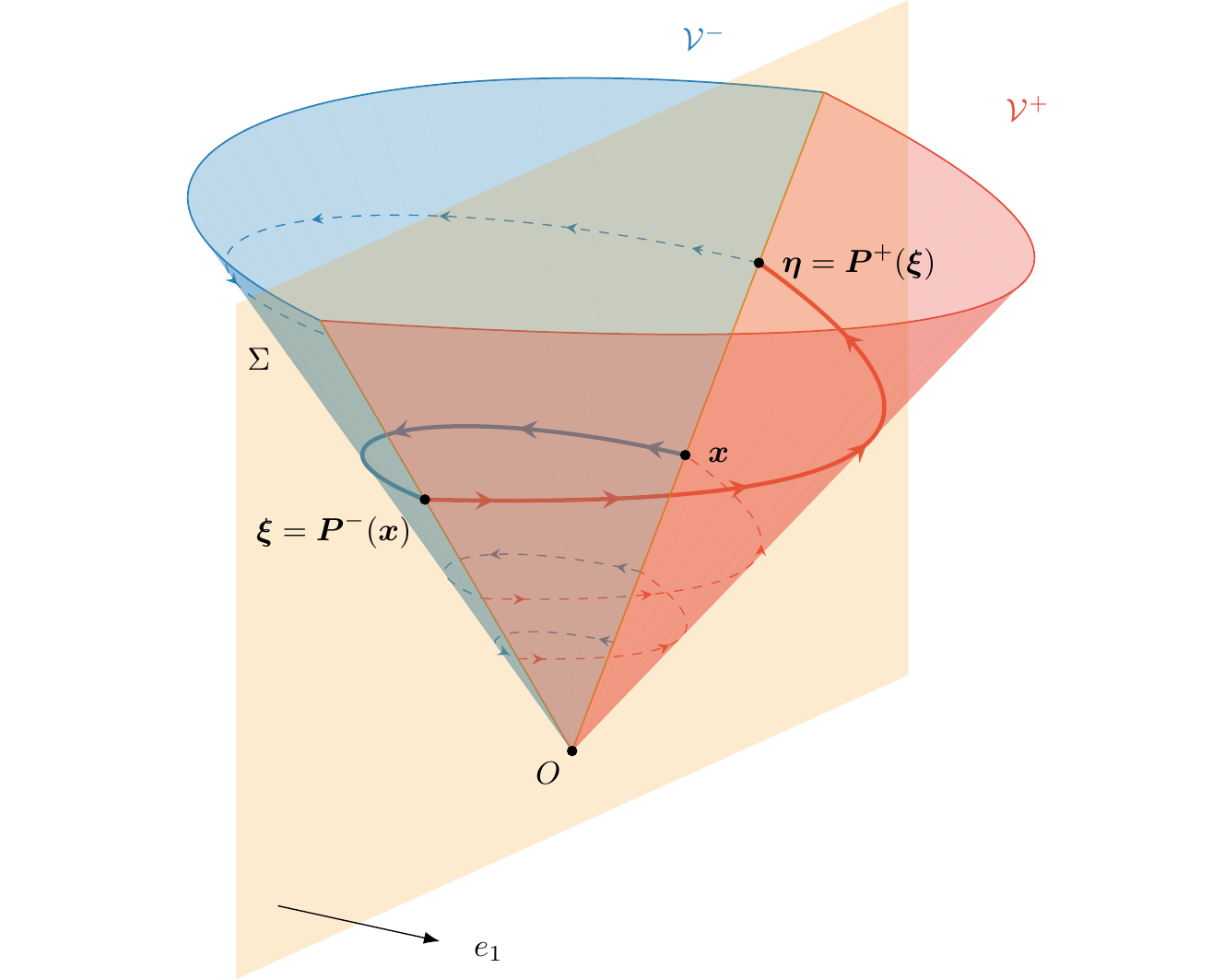}
	\caption{A pictorial view (in which a 4-dimensional space is reduced to a 3D sketch) of an unstable cone with the Poincaré half-maps $\bP^-$ and $\bP^+$, separated by the hyperplane $\Sigma$. The three points $\bx,\bxi,\boldeta$ identify a solution belonging to the invariant cone.}
	\label{fig:pictorialViewInvariantCone}
\end{figure}

\subsection{Invariant cones: instability of piecewise-linear mechanical systems}

The stability analysis of the piecewise-linear system \eqref{eq:piecewiseHamilton} cannot be pursued using  standard methods for smooth systems, such as the analysis of eigenvalues at the equilibrium point. In this section, a tool will be developed, based on the existence of special invariant sets, called invariant cones.
An \emph{invariant set} of an autonomous dynamical system, such that described by equation \eqref{eq:hamiltonianODEs} for smooth structures or by equation \eqref{eq:piecewiseHamilton} for piecewise smooth structures, 
is a subset $\mathcal{S}$ of the phase space such that every initial condition 
belonging to the set leads to a solution remaining within the same set  $\mathcal{S}$. From a  
mathematical point of view, this {\it invariance} can be written as \cite{guckenheimerNonlinearOscillationsDynamical1983, kuznetsovElementsAppliedBifurcation, parkerPracticalNumericalAlgorithms1989}:
$$
\boldsymbol{y}(t_0) \in \mathcal{S} \mbox{ implies } \boldsymbol{y}(t) \in \mathcal{S}, 
~~\mbox{ for all times }
t > t_0.
$$

An \emph{invariant cone} is a special type of invariant set  \cite{carmonaBIFURCATIONINVARIANTCONES2005, carmonaContinuousMatchingTwo2006a, kupperInvariantConesNonsmooth2008a, weissInvariantManifoldsNonsmooth2012a}, defined by the condition that a vector $\bn$ belonging to a Poincaré section $\Sigma$ exists, for which the Poincaré map $\bP$ describing the evolution of the dynamical system fulfills the condition
\begin{equation}
    \label{eq:conditionCone}
    \bP(\bn) = \mu\, \bn,
\end{equation} 
where $\mu \in \mathbb{R}^{+}$ is a positive real number and $\bn \in \Sigma$. 

In the specific case of the piecewise-linear system \eqref{eq:piecewiseHamilton}, the Poincar\'e map is given by the composition of two half-maps \eqref{eq:halfmaps}. 
In this case, an {\it invariant cone} exists, equation \eqref{eq:conditionCone}, for the elastic system under investigation if and only if there exist a scalar $\mu > 0$ and a vector $\bx \in \Sigma$ such that 
\begin{equation}
    \label{eq:eigenvalue}
    \boldeta = e^{\bA^+ \Delta{t^+}(\bxi(\bx))} e^{\bA^- \Delta{t^-}(\bx)} \bx = \mu\, \bx.
\end{equation}
Equation \eqref{eq:eigenvalue} has the structure of an eigenvalue problem: the multiplier $\mu$ can be interpreted as a generalized eigenvalue and the vector $\bx$ as a generalized eigenvector of the matrix $e^{\bA^+ \Delta{t^+}(\bxi(\bx))} e^{\bA^- \Delta{t^-}(\bx)}$, which defines the Poincar\'e map. However, the  eigenvalue problem is  nonlinear, because the matrix itself depends on the eigenvector, hence the solution of \eqref{eq:eigenvalue} in terms of $\mu$ and $\bx$ is not trivial and the usual procedures for linear eigenvalue problems cannot be adopted.

A solution of eq. \eqref{eq:eigenvalue}, provided it exists, identifies an invariant cone and is given by the list
\begin{equation}
    \label{eq:invariantConeSol}
    \{\Delta{t^-}, \Delta{t^+}, \bx, \mu\},
\end{equation}
comprising the time intervals $\Delta{t^-}$ and $\Delta{t^+}$, expended by the mechanical system in passing through the subdomains $\mV^-$ and $\mV^+$, respectively, an eigenvector $\bx$ belonging to the invariant cone (and also to the Poincar\'e section), and finally the corresponding eigenvalue $\mu$.

Before proceeding, it is instrumental to establish some fundamental properties of invariant cones. 

First of all, it is noted that the assumption that the system enters first in the negative subdomain $\mV^-$, used in eq. \eqref{eq:eigenvalue}, is not restrictive, as the \emph{same} invariant cone may be identified by assuming instead that the system enters first the positive subdomain $\mV^+$ and, thus, by solving the eigenvalue problem
\begin{equation}
    \label{eq:eigenvalue2}
    e^{\bA^- \Delta{t^-}} e^{\bA^+ \Delta{t^+}} \bxi = \gamma\, \bxi,
\end{equation}
where $\Delta{t^-}$ and $\Delta{t^+}$ are the same as in eq. \eqref{eq:eigenvalue}. It is easy to check that $\bxi = e^{\bA^- \Delta{t^-}} \bx$ solves the problem \eqref{eq:eigenvalue2} with $\gamma = \mu$. This means that the solution $\{\Delta{t^+}, \Delta{t^-}, \bxi, \mu\}$ of eq. \eqref{eq:eigenvalue2} identifies the same invariant cone as the solution $\{\Delta{t^-}, \Delta{t^+}, \bx, \mu\}$ of eq. \eqref{eq:eigenvalue}. To simplify notations, here and in the sequel, the three points $\bx,\bxi,\boldeta$ identify a solution belonging to the invariant cone.

Next, it is noted that the time intervals $\Delta{t^-}(\bx)$ and $\Delta{t^+}(\bxi)$ are homogeneous functions of degree zero, 
\begin{equation}
    \label{eq:homozero}
    \Delta{t^-}(\alpha \bx) = \Delta{t^-}(\bx), \qquad \Delta{t^+}(\alpha \bxi) = \Delta{t^+}(\bxi), 
    \qquad \forall \alpha > 0,
\end{equation}
as can be easily checked from the definitions \eqref{eq:intervals}. These conditions are a direct consequence of the linearity of the dynamical system within each individual subdomain, $\mV^-$ and $\mV^+$, and define the structure of the invariant cone. In fact, when the point $\boldeta$ is assumed as an initial condition for the motion of the system after the two initial half-maps, $\bxi = \bP^-(\bx)$ and $\boldeta = \bP^+(\bxi)$, the new intersection time intervals become
\begin{equation}
    \Delta{t^-}(\boldeta) = \Delta{t^-}(\mu \bx) = \Delta{t^-}(\bx), \qquad
    \Delta{t^+}(\mu\bxi) = \Delta{t^+}(\bxi), 
\end{equation}
hence the time intervals $\Delta{t^\pm}$ are constants and do not change in the application of further half-maps.

The properties \eqref{eq:homozero} immediately imply that both the Poincar\'e half-maps $\bP^-(\bx)$ and  $\bP^+(\bxi)$, defined according to equation \eqref{eq:halfmaps} as well as the complete Poincar\'e map $\bP(\bx)$, defined by equation  \eqref{eq:map}, are homogeneous functions of degree one
\begin{equation}
\label{wlf}
    \bP^-(\alpha \bx) = \alpha \bP^-(\bx), \qquad \bP^+(\alpha \bxi) = \alpha \bP^+(\bxi), \qquad
    \bP(\alpha \bx) = \alpha \bP(\bx), \qquad \forall \alpha > 0.
\end{equation}
The latter equations have a clear geometrical interpretation, in that the Poincar\'e half-maps $\bP^-(\bx)$ and $\bP^+(\bxi)$, plus the Poincar\'e map $\bP(\bx)$, transform straight half-lines belonging to the switching boundary $\Sigma$ and intersecting the origin into straight half-lines also belonging to $\Sigma$ and intersecting the origin. 
This is the reason why the invariant set defined with the property expressed by equation \eqref{wlf} is a {\it cone}.

The last observation is crucial for the stability analysis of a dynamical system when an invariant cone is present. In fact, when an initial condition $\by_0$ belongs to the Poincar\'e section and to the invariant cone, the evolution of the system can be described by the recursive applications of Poincar\'e maps such that
\begin{equation}
    \label{eq:invariantCone}
    \begin{aligned}
        y(\Delta{t^-} + \Delta{t^+}) &= e^{A^+ \Delta{t^+}} e^{A^- \Delta{t^-}} \by_0 = \mu\, \by_0, \\
        y(2\Delta{t^-} + 2\Delta{t^+}) &= \Big(e^{A^+ \Delta{t^+}} e^{A^- \Delta{t^-}}\Big)^2 \by_0 = \mu^2\, \by_0, \\
        y(3\Delta{t^-} + 3\Delta{t^+}) &= \Big(e^{A^+ \Delta{t^+}} e^{A^- \Delta{t^-}}\Big)^3 \by_0 = \mu^3\, \by_0, \\
        \cdots & \\
        y(k\Delta{t^-} + k\Delta{t^+}) &= \Big(e^{A^+ \Delta{t^+}} e^{A^- \Delta{t^-}}\Big)^k \by_0 = \mu^k\, \by_0, \\
        \cdots & 
    \end{aligned}
\end{equation}
defining a discrete exponential relation. 

Accordingly, when an invariant cone is present, the dynamics of the mechanical system on the invariant cone can easily be understood. This is because the orbits may spiral in or out along the cone, depending on the value 
of the eigenvalue $\mu$, and may evolve either towards the vertex at the equilibrium point or away from it. 
In particular, the following conclusions can be drawn:
\begin{itemize}
    \item If $\mu > 1$, a family of \lq spiraling out' trajectories, belonging to the invariant cone, exists,  moving away for $t > 0$ from the vertex of the cone, which represents the equilibrium configuration of the system. When an initial condition is selected, belonging to the invariant cone, the motion of the structure evolves in time diverging from the fixed point, hence the equilibrium configuration at the origin is \emph{unstable}.
    
    \item If $\mu < 1$, a family of \lq spiraling in' trajectories belonging to the invariant cone exists, moving towards the vertex of the cone. Although this corresponds to a stable behaviour, other trajectories different from those on the invariant cone may exist being divergent, so that stability of the fixed point cannot be guaranteed.
    
    \item If $\mu = 1$, a family of periodic trajectories, belonging to the invariant cone, exists. In this case, nothing can be concluded about the stability of the equilibrium configuration.
\end{itemize}
Summarising the above statements, for the dynamical system represented by equation  \eqref{eq:piecewiseHamilton}, the existence of an invariant cone, eq.~\eqref{eq:eigenvalue},  with $\mu > 1$ is a \emph{sufficient condition for instability}. 
Vice-versa, assuming the existence of an invariant cone, fulfillment of condition \eqref{eq:eigenvalue} with $\mu \leq 1$ is a \emph{necessary condition for stability}. 

It is important to emphasise that these conditions hold true, regardless of the stability behaviour of each single subsystem. Therefore, an equilibrium configuration of a piecewise-linear mechanical system, which results stable when analysed separately for each constituting subsystem, can become unstable when the composed structure is considered.

A proposition is now proven which establishes a general property of invariant cones for autonomous non-dissipative mechanical systems. Although this proposition is proven for the 2 d.o.f. mechanical system under investigation, it can be easily extended to a system with any number of degrees of freedom.

\paragraph{A proposition on reciprocal eigenvalues.} 
If the autonomous non-dissipative mechanical system \eqref{eq:piecewiseHamilton} admits an invariant cone, that is a solution $\{\Delta{t^-}, \Delta{t^+}, \bx, \mu\}$ of the eigenvalue problem \eqref{eq:eigenvalue} with $\mu \neq 1$, then another invariant cone $\{\Delta{t^-}, \Delta{t^+}, \bJ\bxi, 1/\mu\}$ exists, associated to the eigenvalue $1/\mu$, reciprocal of $\mu$, where
\begin{equation}
    \bxi = e^{A^- \Delta{t^-}} \bx, \qquad 
    \bJ = 
    \begin{bmatrix}
    1 & 0 & 0 & 0 \\
    0 & 1 & 0 & 0 \\
    0 & 0 & -1 & 0 \\
    0 & 0 & 0 & -1
    \end{bmatrix}.
\end{equation}
Consequently, 
\begin{quote}
    \emph{when the problem \eqref{eq:eigenvalue} admits a real eigenvalue  $\mu\neq 1$, it admits also the eigenvalue $1/\mu$. Therefore, when a solution of equation \eqref{eq:eigenvalue} is found with $\mu \neq 1$, the trivial equilibrium configuration is certainly unstable.}
\end{quote}

The above proposition can be proven by preliminarily noting that the application of matrix $\bJ$ to a state vector $\by = [\xi, \phi, \dot{\xi}, \dot{\phi}]^T$ has the effect of changing the sign of the velocities
\begin{equation}
    \bJ 
    \begin{bmatrix}
    \xi \\
    \phi \\
    \dot{\xi} \\
    \dot{\phi}
    \end{bmatrix} = 
    \begin{bmatrix}
    \xi \\
    \phi \\
    -\dot{\xi} \\
    -\dot{\phi}
    \end{bmatrix}.
\end{equation}
Note also that $\bJ = \bJ^{-1}$, $e^{-\bA^\pm t^\pm} = \bJ e^{\bA^\pm t^\pm} \bJ$ and that vector $\bJ \bxi$ enters in the negative subdomain, as the vector $\bxi$, without the minus signs, enters in the positive subdomain, being a solution of the half-map \eqref{eq:halfmaps}$_1$ by assumption.

The above statement follows from the two properties
\begin{equation}
    \label{eq:inversion}
    \bJ \bxi = e^{-\bA^-\Delta{t^-}} \bJ \bx, \qquad 
    \bJ \boldeta = e^{-\bA^+\Delta{t^+}} \bJ \bxi,
\end{equation}
equations that can be directly checked considering the form of matrix $e^{\bA t}$, equation \eqref{eq:exponentialMatrix}. The above properties have a clear mechanical meaning, as they correspond to an inversion of the motion: the time is inverted by changing the sign to the variable $t$ and, correspondingly, the velocities in the vectors $\bx$, $\bxi$ and $\boldeta$ are inverted through multiplication by matrix $\bJ$.

A multiplication of equation \eqref{eq:inversion}$_1$ by $e^{\bA^+\Delta{t^+}}e^{\bA^-\Delta{t^-}}$ and subsequent use of $\bx = \boldeta/\mu$, as well as consideration of equation \eqref{eq:inversion}$_2$ leads to
\begin{equation}
    e^{\bA^+\Delta{t^+}}e^{\bA^-\Delta{t^-}} \bJ \bxi = e^{\bA^+\Delta{t^+}} \bJ \bx = 
    \frac{1}{\mu} e^{\bA^+\Delta{t^+}} \bJ \boldeta = \frac{1}{\mu} \bJ \bxi,
\end{equation}
which proves the proposition.

\subsection{Mechanical energy for the piecewise smooth system}
\label{sec:energy}

The unusual unstable behaviour of the piecewise smooth structure, composed of two stable substructures, can be qualitatively explained from a mechanical point of view by analyzing the mechanical energy characterizing the non-smooth system.
In particular, the analysis which follows shows that the presence of the follower laod is essential, in the sense that the instability does not occur when the load is dead.

The stiffness matrix $\boldsymbol{K}^\pm$, defined in equation \eqref{eq:massCirc}, collects the effect of the elastic springs, $k_1$ and $k_2$, and of the follower force $F$ and can be decomposed into the sum of a symmetric and an unsymmetric components, $\bK^\pm = \widehat{\bK}^\pm + \bG^\pm$, 
\begin{equation}
    \label{eq:matrixKGenergy}
    \widehat{\boldsymbol{K}}^{\pm}
    = 
    \begin{bmatrix}
        \displaystyle k_1 \left(1\mp \frac{y_s}{R_\pm}\right) +\frac{k_2}{R_\pm^2} \quad & \displaystyle \pm \frac{k_2}{R_\pm} \\[3ex]
        \displaystyle \pm \frac{k_2}{R_\pm} & k_2 
    \end{bmatrix},
    \qquad
    \boldsymbol{G}^{\pm} = F
    \begin{bmatrix}
        \displaystyle \mp \frac{1}{R_\pm} \quad & \displaystyle - 1 \\[3ex]
        \displaystyle 0 & 0 
    \end{bmatrix},
\end{equation}
where $\widehat{\boldsymbol{K}}^\pm$ is a symmetric matrix collecting only the spring stiffnesses and, possibly, other conservative loads applied to the structure, while $\boldsymbol{G}^\pm$ contains only the nonconservative forces.

A scalar product of equation \eqref{eq:linearSystemEquationMatrix} by $\dot{\boldsymbol{q}}$ and a factorization of the time derivative yield
\begin{equation}
    \label{eq:conservationMechEnergyPrinciple}
    \frac{d}{dt} \left(\frac{1}{2} \dot{\boldsymbol{q}}\cdot\boldsymbol{M} \dot{\boldsymbol{q}} + \frac{1}{2} \boldsymbol{q} \cdot \widehat{\boldsymbol{K}}\boldsymbol{q} \right) = 
    -\dot{\boldsymbol{q}} \cdot \boldsymbol{G}\boldsymbol{q}.
\end{equation}
The mechanical energy 
$\mathcal{H}$ 
is defined as the sum of the kinetic and the total potential energy 
\begin{equation}
    \label{eq:defMechEnergy}
    \mathcal{H} =  \underbrace{\frac{1}{2} \dot{\boldsymbol{q}}\cdot\boldsymbol{M} \dot{\boldsymbol{q}}}_{\text{Kinetic energy}} + 
    \underbrace{\frac{1}{2} \boldsymbol{q} \cdot \widehat{\boldsymbol{K}}\boldsymbol{q}}_{\text{Total potential energy}}, 
\end{equation}
where the total potential energy is the sum of the elastic energy and the potential energy of  conservative external loads, when present. 

Equation \eqref{eq:defMechEnergy} can be rewritten in matrix notation as a function of the state vector $\boldsymbol{y}(t)$ as
\begin{equation}
    \label{eq:mechanicalEnergyHamiltonian}
    \mathcal{H}(t) = \frac{1}{2} \boldsymbol{y}(t) \cdot \boldsymbol{D} \boldsymbol{y}(t),
\end{equation}
where
\begin{equation}
    \boldsymbol{D}
    = 
    \begin{bmatrix}
        \boldsymbol{\widehat{K}} & \boldsymbol{0}\\[2ex]
        \boldsymbol{0} & \boldsymbol{M}\\
    \end{bmatrix},
\end{equation}
is a symmetric matrix. An integration of equation \eqref{eq:conservationMechEnergyPrinciple} yields 
\begin{equation}
    \label{eq:mechEnergyIntegral}
    \mathcal{H}(t_f) - \mathcal{H}(t_0)  = 	\underbrace{-\int_{t_0}^{t_f} \dot{\boldsymbol{q}} \cdot \boldsymbol{G}\boldsymbol{q}\,dt}_{\text{Work done by the nonconservative loads}} ,
\end{equation}
therefore,  a change in the mechanical energy equals a corresponding work done on the structure by the nonconservative loads. 

When in a system, smooth or not, only the symmetric part $\widehat{\boldsymbol{K}}$ of the stiffness matrix is present, as in the case of a dead load not explicitly evidenced here, the mechanical energy remains  constant and flutter instability is excluded. On the contrary, when an unsymmetric $\boldsymbol{G}$ is present, the mechanical energy  $\mathcal{H}(t)$ varies in time. Therefore, the presence of follower load breaking the symmetry is essential to the instability that is analyzed here.

The mechanical energy for a solution near a stable equilibrium configuration of a smooth system with follower loads can be studied by substituting the solution \eqref{eq:solutionLinearSystemHamiltonian} with the matrix exponential \eqref{eq:exponentialMatrix} into equation \eqref{eq:mechanicalEnergyHamiltonian}, hence
\begin{equation}
    \label{eq:mechanicalEnergyInitialConditions}
    \mathcal{H} = \frac{1}{2} \by_0 \scalp (e^{\bA t})^T \bD e^{\bA t}\by_0.
\end{equation}
When a solution is stable, the mechanical energy \eqref{eq:mechanicalEnergyInitialConditions} results as a sum of trigonometric functions, which in general is not a periodic function, but it is bounded, so that in a stable system the mechanical energy cannot indefinitely increase. Formally, this can be shown by considering that the mechanical energy \eqref{eq:mechanicalEnergyInitialConditions} can be bounded as 
\begin{equation}
    \by_0\scalp (e^{\bA t})^T \bD e^{\bA t} \by_0 
    \leq  ||(e^{\bA t})^T \bD e^{\bA t}|| \, \by_0^2  
    \leq ||e^{\bA t}||^2 \, ||\bD|| \, \by_0^2,
\end{equation}
where, in the case of stability, matrix $e^{\bA t}$ is given by equation \eqref{eq:exponentialMatrix}, so that its 
norm is bounded in time for every value of $t \in [0,\infty)$. 

On the contrary, in case of a piecewise linear system described by equation \eqref{eq:piecewiseHamilton}, the mechanical energy may become unbounded, although the subsystems forming the non smooth structure are both stable. 

In fact, the mechanical energy at time $t = 0$, equation \eqref{eq:mechanicalEnergyInitialConditions},  for initial conditions $\boldsymbol{y}_0$ is
\begin{equation}
    \label{eq:mechEnergyNC_0}
    \mathcal{H}_0 = \mathcal{H}(0) = \frac{1}{2} \boldsymbol{y}_0 \cdot \boldsymbol{D} \boldsymbol{y}_0,
\end{equation}
where $\bD$ can be identified with either $\bD^-$ or $\bD^+$, because both matrices provide the same result, being the energy continuous at the switching manifold $\xi=0$. 

During the motion, the variation of mechanical energy with time can be computed by substituting the piecewise solution \eqref{eq:solution} into eq.~\eqref{eq:mechanicalEnergyHamiltonian}, thus obtaining 
\begin{equation}
    \label{eq:solution2}
    \mH(t) = 
    \left\{
    \begin{array}{ll}
        \frac{1}{2} e^{\bA^-(t-t_0)} \by_0 \scalp \bD^- e^{\bA^-(t-t_0)} \by_0, & t_0 \leq t < t_1, \\[3mm]
        \frac{1}{2} e^{\bA^+(t-t_1)} e^{\bA^-(t_1-t_0)} \by_0 \scalp \bD^+ e^{\bA^+(t-t_1)} e^{\bA^-(t_1-t_0)} \by_0, & t_1 \leq t < t_2, \\[3mm]
        \cdots & \cdots \\[3mm]
        \frac{1}{2} e^{\bA^-(t-t_{k-1})} \cdots e^{\bA^-(t_1-t_0)} \by_0 \scalp \bD^- e^{\bA^-(t-t_{k-1})} \cdots e^{\bA^-(t_1-t_0)} \by_0, & t_{k-1} \leq t < t_{k}, \\[3mm]
        \frac{1}{2} e^{\bA^+(t-t_{k})} \cdots e^{\bA^-(t_1-t_0)} \by_0 \scalp \bD^+ e^{\bA^+(t-t_{k})} \cdots e^{\bA^-(t_1-t_0)} \by_0, & t_{k} \leq t < t_{k+1}, \\[3mm]
        \cdots & \cdots 
    \end{array}
    \right.
\end{equation}
Assuming that an invariant cone exists, eq.~\eqref{eq:invariantCone}, the mechanical energy at the intersection time $k\Delta{t^-}+k\Delta{t^+}$ 
(where $k$ is a positive integer) 
can be computed as
\begin{equation}
    \mH(k\Delta{t^-}+k\Delta{t^+}) = \frac{1}{2} \mu^k \by_0 \scalp \bD \mu^k \by_0 = \mu^{2k} \mH_0. 
\end{equation}
Therefore, when $\mu > 1$ \emph{the mechanical energy suffers an unbounded exponential growth in time}, revealing an unstable behaviour of the system.
This exponential growth is accompanied by oscillations, a situation similar to flutter instability in smooth systems loaded by follower forces so that the instability behaviour under investigation can be interpreted as a condition of flutter, but for non-smooth mechanical systems. 

\begin{figure}[H]
	\centering
	\includegraphics[]{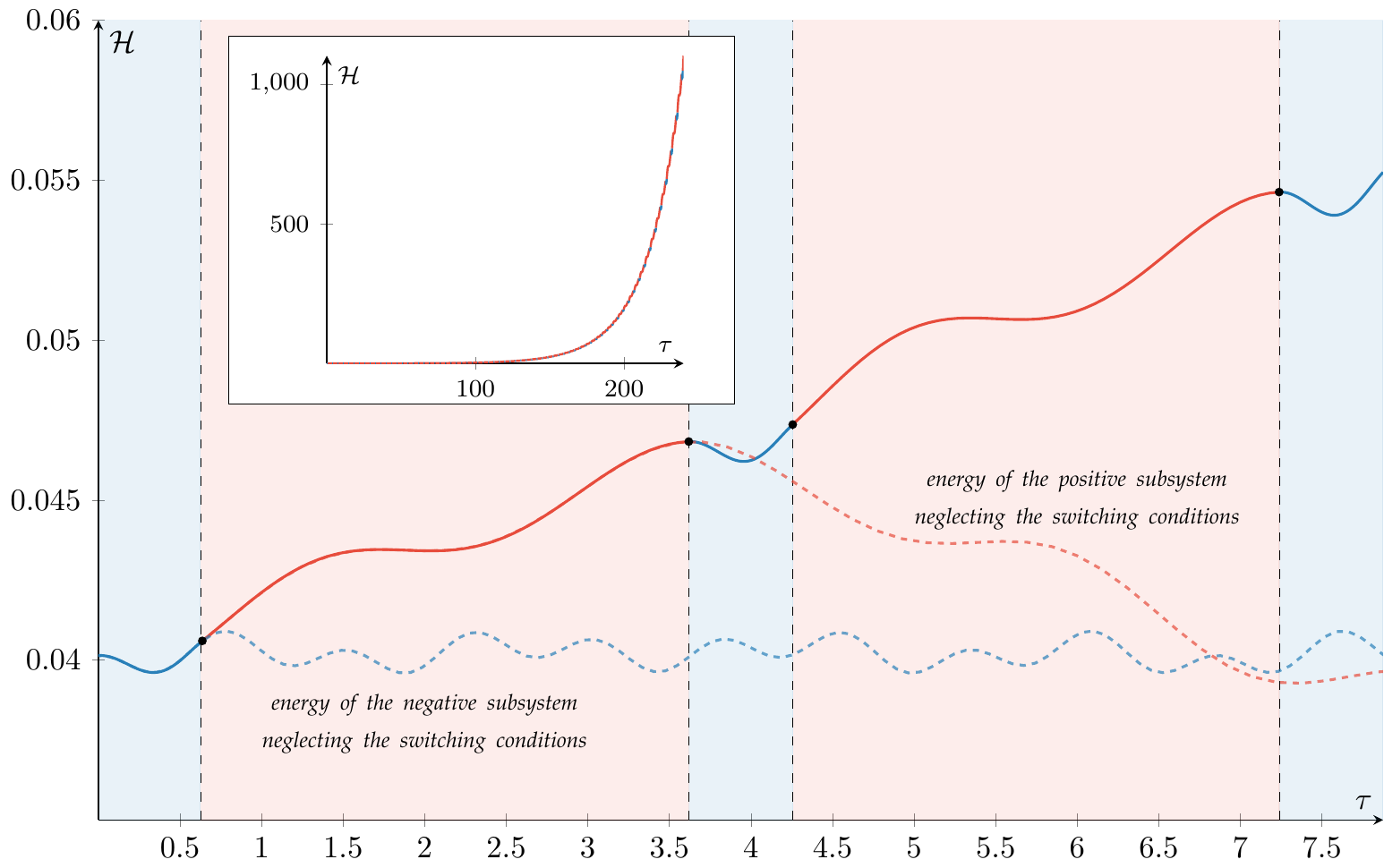}
	\caption{
	Evolution in time 
	of the mechanical energy for the non-smooth structure 
	[shown in Fig. \ref{fig:structureCircNegPos} and characterized by the parameters in the list
	(\ref{eq:parameterNumRef})].  
	The behaviours of the two \lq component' smooth mechanical structures (shown in Fig. \ref{fig:singleStructures}) are also reported with dotted lines. The energy in the latter case remains bounded, denoting stability. 
	The exponential growth of the mechanical energy corresponding to a flutter instability is visible in the main graph, but may be vividly observed in the inset, referred to a longer time interval. 
	}
	\label{fig:mechEnergy}
\end{figure}

The evolution in time (made dimensionless through division by a reference time $T$) of the mechanical energy $\mH$ is reported in 
Fig. \ref{fig:mechEnergy}
for the non-smooth structure reported in Fig. \ref{fig:structureCircNegPos}, characterized by the values of parameters listed in 
(\ref{eq:parameterNumRef}), 
together with the two \lq component' smooth structures 
shown in Fig. \ref{fig:singleStructures} (responses reported dashed).

The instability of the non-smooth system corresponds to a blow-up of the mechanical energy (note the vivid representation in the inset, showing exponential growth), while the two stable substructures evidence a bounded evolution. 

%

\section{Attractivity of the cone: the effects of perturbations}
\label{sec:attractvityCone}

It has been demonstrated in the previous sections that the existence of an invariant cone  with eigenvalue $\mu \neq 1$ in the phase space of a non-smooth dynamical system of the type  \eqref{eq:piecewiseHamilton} is a sufficient condition for instability. 
Now the effect of perturbations in the initial conditions has to be analyzed. 
In particular, the question arises whether a motion generated from an initial condition  slightly outside an unstable cone will still display unstable unbounded growth. When the latter behaviour is found, the cone is called {\it attractive}, so that it will be asymptotically approached by a trajectory originated from initial conditions sufficiently close to it. 

The answer to the above question is provided in the following, showing that for the structure under consideration, the unstable cone (associated to $\mu > 1$) is always attractive. In other words, the instability detected with $\mu > 1$ can be considered \lq genuine'.

\paragraph{A proposition on the attractivity of the cone.} 
A cone with eigenvector $\bx = \by(0) = \by_0$ and its associated eigenvalue $\mu$ are assumed to exist, solution of equation \eqref{eq:eigenvalue}, together with the corresponding vectors $\bxi = \by(\Delta{t}^-) = e^{\bA^-\Delta{t}^-} \by_0$ and $\boldeta = \by(\Delta{t}^- + \Delta{t}^+) = e^{\bA^+\Delta{t}^+} e^{\bA^-\Delta{t}^-} \by_0$, defining the state of the system at the first crossing (from $\mV^-$ to $\mV^+$) and the second crossing (from $\mV^+$ to $\mV^-$) of the switching manifold $\Sigma$. These vectors describe a solution $\by(t)$ belonging to the invariant cone that is considered as a reference solution and that is perturbed in order to study the attractivity of the cone. The reference solution $\by(t)$ crosses the switching manifold $\Sigma$ separating the negative and the positive subspaces the first time at the instant $t^- = \Delta{t}^-$ and a second time at the instant $t^+ = \Delta{t}^- + \Delta{t}^+$.

\begin{figure}[H]
	\centering
	\includegraphics[]{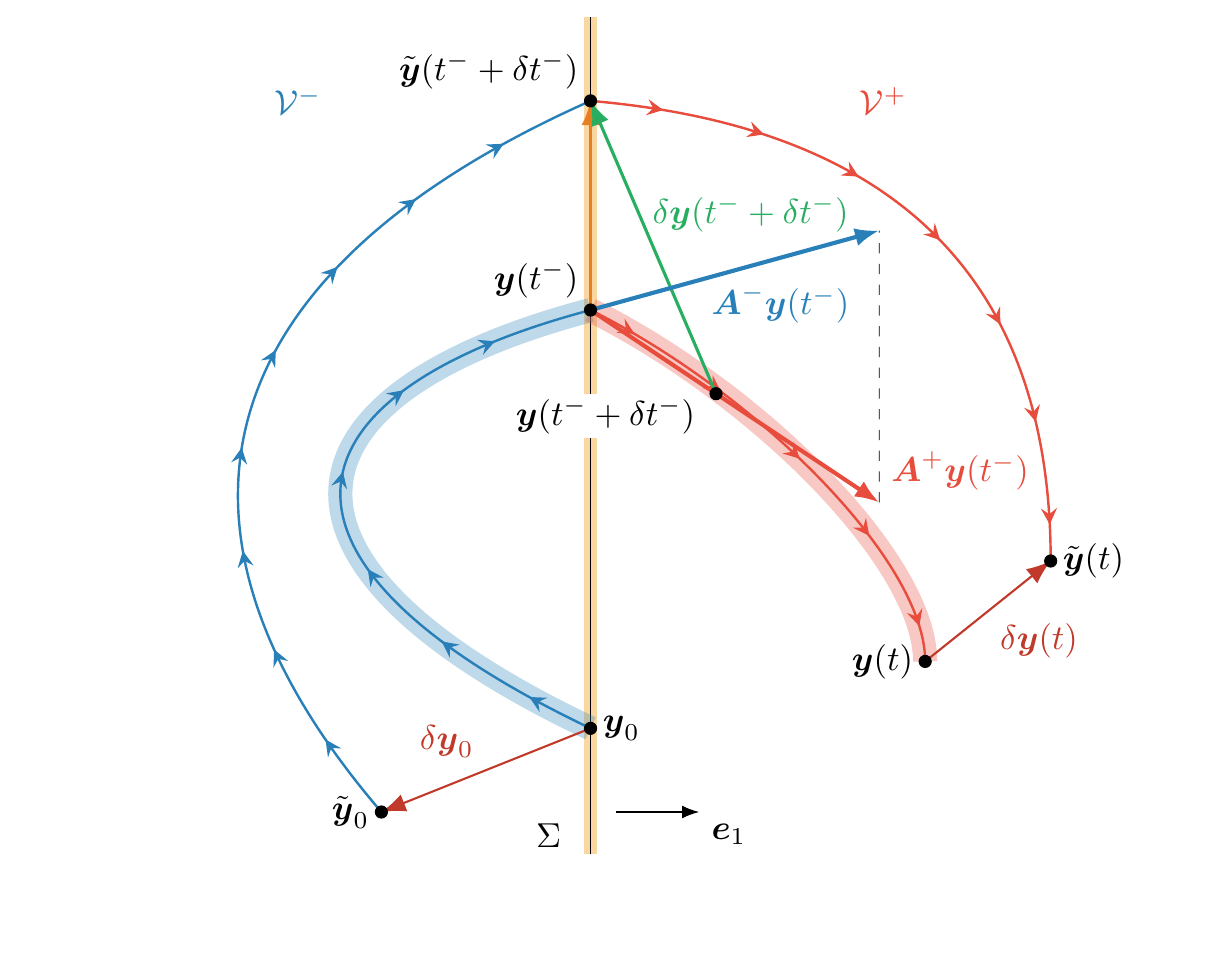}
	\caption{Sketch of the effect of a perturbation $\delta \by_0$ of a given piecewise solution (developing from $\tilde{\by}_0$) for a non-smooth dynamical system.
	The perturbed motion, starting at $\tilde{\by}_0 + \delta \by_0$, is assumed to develop in a close neighborhood of the unperturbed motion, starting at $\tilde{\by}_0$
	}
	\label{fig:saltation}
\end{figure}

A generic small perturbation $\delta \by_0$ in the initial conditions is assumed, with the restriction that $\tilde{\by}_0 = \by_0 + \delta \by_0$ belongs to the negative subspace $\mV^-$, as sketched in Fig. \ref{fig:saltation}. The perturbed initial condition $\tilde{\by}_0 = \by_0 + \delta \by_0$ gives rise to a perturbed solution $\tilde{\by}(t) = \by(t) + \delta \by(t)$ which crosses for the first time the switching manifold $\Sigma$ at a different instant of time $t^- + \delta t^-$.
Therefore, equation  \eqref{eq:solutionLinearSystemHamiltonian} can be applied to obtain
\begin{equation}
	\tilde{\by}(t^- + \delta t^-) = e^{\boldsymbol{A}^- (t^- +\delta t^-)} (\by_0 + \delta \by_0),
\end{equation}
where, due to the smallness of $\delta \by_0$, it is assumed that $\delta t^-$ is also sufficiently small to justify the following Taylor series approximation
\begin{equation}
	\label{eq:perturbedSolutionAttractivity}
	\tilde{\by}(t^- + \delta t^-) = e^{\bA^- t^-}  \by_0 + e^{\bA^- t^-} \delta \by_0 + 
	\delta t^- \bA^- e^{ \bA^- t^-} \by_0 .
\end{equation}
The reference solution at time $t^- + \delta t^-$ belongs to the positive subspace $\mV^+$ and can be calculated with an analogous Taylor expansion (assuming that $\delta t^-$ is sufficiently small) as
\begin{equation}
	\label{melone3}
	\by(t^- + \delta t^-) = e^{\bA^+ \delta t^-} e^{\bA^-   t^-} \by_0 = 
	e^{\bA^- t^-} \by_0 + \delta t^- \bA^+ e^{\bA^- t^-} \by_0,
\end{equation}
where $\bA^+ e^{\bA^- t^-} \by_0 = \bA^+ \by(t^-) = \dot{\by}(t^-)$ is the orbital velocity of the reference solution entering at the time instant $t^-$ into the positive subspace.

The evolution of the perturbation after the crossing of the switching manifold is described by the quantity $\delta \by(t^- + \delta t^-) = \tilde{\by}(t^- + \delta t^-) - \by(t^- + \delta t^-)$, that can be computed as
\begin{equation}
	\label{melone5}
	\delta \by(t^- + \delta t^-) =
	e^{\bA^- t^-} \delta \by_0 +
	\delta t^- (\bA^- - \bA^+) e^{\bA^- t^-} \by_0.
\end{equation}
The small time interval $\delta t^-$ can be calculated noting that the vector $\tilde{\by}(t^- + \delta t^-) - \by(t^-)$ belongs to the switching manifold and is orthogonal to the vector $\be_1$, hence
\begin{equation}
	\left[\tilde{\by}(t^- + \delta t^-) - \by(t^-)\right] \scalp \be_1 = 
	\left[e^{\bA^- t^-} \delta \by_0 + \delta t^- \bA^- e^{\bA^- t^-} \by_0 \right] \scalp \be_1 = 0,
\end{equation}
an equation that can be solved in $\delta t^-$, leading to
\begin{equation}
	\label{eq:deltatMinusAttractivity}
	\delta t^- = -\frac{1}{\dot{\xi}(t^-)} \be_1 \scalp e^{\bA^- t^-} \delta \by_0,
\end{equation}
where $\dot{\xi}(t^-) = \be_1 \scalp \bA^- \by(t^-)$. Note also that $\dot{\xi}(t^-) = \be_1 \scalp  \bA^- \by(t^-) = \be_1 \scalp \bA^+ \by(t^-)$, because the velocities $\dot{\xi}$ and $\dot{\phi}$ remain continuous crossing the switching manifold.

Substituting eq.  \eqref{eq:deltatMinusAttractivity} into eq. \eqref{melone5}, the evolution of the perturbation after the crossing can be written as
\begin{equation}
	\delta \by (t^- + \delta t^-) = \bS^- e^{\bA^- t^-} \delta \by_0 ,
\end{equation}
where the {\it saltation matrix} $\bS^-$ is defined as
\begin{equation}
	\label{eq:defSaltationMatrix}
	 \bS^- = \bI + \frac{1}{\dot{\xi}(t^-)}
	\left[
	\bA^+ \by(t^-) - \bA^- \by(t^-)
	\right]
	\otimes \be_1. 
\end{equation}
The saltation matrix defines the perturbation $\delta \by (t^- + \delta t^-)$ after the crossing of the switching manifold $\Sigma$, for a given  perturbation $\delta \by(t^-) = e^{\bA^- t^-} \delta \by_0$ defined just before the crossing.

Due to the structure of the saltation matrix and because  $\det(\bI + \ba \otimes \bb) = 1 + \ba \scalp \bb$, it follows that $\det(\bS^-) = 1$.

A similar calculation can be performed when the unperturbed solution crosses the switching manifold from the positive to the negative subspace at time $t^+ = \Delta{t^-} + \Delta{t^+}$ and the perturbed solution at time $t^+ + \delta t^+$, so that a new saltation matrix can be defined as
\begin{equation}
	\label{eq:defSaltationMatrixPlus}
	\bS^+ = \bI + \frac{1}{\dot{\xi}(t^+)}
	\left[ \bA^- \by(t^+) - \bA^+ \by(t^+) \right] \otimes \be_1, 
\end{equation}
in which $\dot{\xi}(t^+) = \be_1 \scalp \bA^+ \by(t^+)$ and $\by(t^+) = e^{\bA^+ \Delta{t}^+}e^{\bA^- \Delta{t}^-} \by_0$. 

Therefore, the initial perturbation $\delta \by_0$ evolves in time in a way that after two crossings of the switching manifold, the difference between the reference and the perturbed solutions is governed by the relation
\begin{equation}
    \delta \by (t^+ + \delta t^+) = \bS^+ e^{\bA^+ \Delta{t}^+} \bS^- e^{\bA^- \Delta{t}^-} \delta \by_0,
\end{equation}
where the matrix 
\begin{equation}
    \bPhi_T = \bS^+ e^{\bA^+ \Delta{t}^+} \bS^- e^{\bA^- \Delta{t}^-}
\end{equation}
is referred to as the \emph{monodromy matrix} in the context of stability of periodic solutions. Consequently, the attractivity of the invariant cone is related to the four eigenvalues of the monodromy matrix, the so-called {\it Floquet multipliers}.

The monodromy matrix for the 2 d.o.f. mechanical system under consideration possesses the following properties:
\begin{itemize}
	\item The determinant is equal to the unit,
	\begin{equation}
		\det \left(\bS^+ e^{\bA^+ \Delta{t}^+} \bS^- e^{\bA^- \Delta{t}^-}\right) = 1,
	\end{equation}
	because $\det e^{\bA t } = e^{\tr \bA t } = 1$ and $\det \bS^- = \det \bS^+ =1$;
	\item Two eigenvectors and two eigenvalues coincide with those of the eigenvalue problem \eqref{eq:eigenvalue} defining the invariant cone
	\begin{equation}
		\left(\bS^+ e^{\bA^+ \Delta{t}^+} \bS^- e^{\bA^- \Delta{t}^-}\right) \by_0 = \mu\, \by_0, \qquad
		\left(\bS^+ e^{\bA^+ \Delta{t}^+} \bS^- e^{\bA^- \Delta{t}^-}\right) \bJ \by(t^-) = \frac{1}{\mu} \bJ \by(t^-), 
	\end{equation}
	two identities following directly  from $\bS^- \by(t^-) = \bS^+ \by(t^-) = \by(t^-)$ and 
	$\bS^- \bJ\by_0 = \bS^+ \bJ\by_0 = \bJ\by_0$ (the saltation matrices $\bS^-$ and $\bS^+$ leave unchanged any vector belonging to the switching manifold $\Sigma$). 
	\item A third eigenvector is equal to $\bA^- \by_0$, with corresponding eigenvalue $\mu$, 
	\begin{equation}
	    \left(\bS^+ e^{\bA^+ \Delta{t}^+} \bS^- e^{\bA^- \Delta{t}^-}\right) \bA^-\by_0 = \mu\, \bA^-\by_0,
	\end{equation}
	which follow from the commutativity, $e^{\bA^\pm \Delta{t}^\pm} \bA^\pm = \bA^\pm e^{\bA^\pm \Delta{t}^\pm}$, and from the two identities 
	\begin{equation}
	    \bS^- \bA^- \by(t^-) = \bA^+ \by(t^-) \quad \text{ and } \quad 
	    \bS^+ \bA^+ \by(t^+) = \bA^- \by(t^+).
	\end{equation}
\end{itemize}

It follows from all the above that the monodromy matrix $\bS^+ e^{\bA^+ \Delta{t}^+} \bS^- e^{\bA^- \Delta{t}^-}$ possesses the four eigenvalues $\{\mu, \mu, 1/\mu, 1/\mu\}$. The first two eigenvalues $\{\mu, \mu\}$ are associated with initial perturbations $\delta \by_0$ along the directions $\by_0$ and $\bA^- \by_0$, i.e. belonging to the invariant cone. The other two $\{1/\mu, 1/\mu\}$ are associated to perturbations outside the invariant cone. 

In conclusion, 
\begin{quote}
	\emph{since a generic perturbation can always be decomposed along the eigenvectors of the monodromy matrix $\bS^+ e^{\bA^+ \Delta{t}^+} \bS^- e^{\bA^- \Delta{t}^-}$, the unstable cone associated to the eigenvalue $\mu > 1$ is always attractive. Therefore, when an eigenvalue $\mu \neq 1$ is found as the solution of problem \eqref{eq:eigenvalue}, the structure admits a stable (non attractive) cone, associated to $\mu < 1$, and an unstable (attractive) cone, associated to $\mu > 1$.}
\end{quote}

Note that in the case $\mu = 1$ the motion is periodic and the cone is not attractive, hence conclusions about instability of the mechanical system cannot be reached. This case is therefore not further considered.

\section{Numerical examples} 
\label{sec:numericalExample}

\subsection{Numerical algorithm for the identification of invariant cones for piecewise linear systems}
\label{sec:algorithmInvariantCone}

Assume that all the mechanical parameters of the system, together with the applied follower force, are given. A numerical procedure is proposed in this section for the identification of possible invariant cones, i.e. solutions of the generalized nonlinear eigenvalue problem \eqref{eq:eigenvalue}, recalled  here for convenience
\begin{equation}
    \label{eq:eigenvalueal}
    e^{\bA^+ \Delta{t}^+} e^{\bA^- \Delta{t}^-} \bx = \mu\, \bx,
\end{equation}
where the initial condition vector $\bx$ belongs to the switching manifold $\Sigma$, so that it has the following form
\begin{equation}
    \bx = [0, x_2, x_3, x_4]^T.
\end{equation}
Note that the unknowns of the problem are the eigenvalue $\mu$, the eigenvector $\bx$ and the two time intervals $\Delta{t^-}$ and $\Delta{t^+}$. Note also that the modulus of the vector $\bx$ is arbitrary, given the structure of the equation. Thus eq.~ \eqref{eq:eigenvalue} can be scaled as follows
\begin{equation}
    \label{eq:scaled}
    e^{\bA^+ \Delta{t}^+} e^{\bA^- \Delta{t}^-} 
    \begin{bmatrix}
    0 \\ x_2 \\x_3 \\ 1
    \end{bmatrix} 
    = 
    \mu\, 
    \begin{bmatrix}
    0 \\ x_2 \\x_3 \\ 1
    \end{bmatrix},
\end{equation}
from which it is clear that there are five scalar unknowns to be determined $\Delta{t}^-$,$\Delta{t}^+$, $x_2$, $x_3$ and $\mu$. However, the system \eqref{eq:scaled} provides only four scalar equations, and thus it has to be complemented by an additional scalar equation, which is provided by the condition that also the intermediate point $\bxi$ belongs to the switching manifold, 
\begin{equation}
    \label{eq:additional}
    \xi_1 = [e^{A^- \Delta{t}^-} \bx]_1 = 0.
\end{equation}
The system of equations \eqref{eq:scaled} and \eqref{eq:additional} is nonlinear, so that an algorithm is proposed below to partially decouple the system and reduce it to two equations for the unknowns $\Delta{t^-}$ and $\Delta{t^+}$.

The starting point $\bx$ and the final point $\boldeta$ of the Poincar\'e map \eqref{eq:map} can be expressed in terms of the intermediate point $\bxi$ using the Poincar\'e half maps \eqref{eq:halfmaps} as follows
\begin{equation}
    \label{eq:halfmapsAlgorithm}
    \bx = e^{-\bA^- \Delta{t}^-} \bxi, 
    \quad
    \boldeta = e^{\bA^+ \Delta{t}^+} \bxi.
\end{equation}
The condition that both $\bx$ and $\boldeta$ belong to the switching manifold provides two equations, 
\begin{equation}
    \label{eq:linear}
    x_1 = \left[ e^{-\bA^- \Delta{t}^-} \bxi \right]_1 =0, \quad \eta_1 = \left[ e^{\bA^+ \Delta{t}^+} \bxi \right]_1 = 0,
\end{equation}
that are linear in $\xi_2$, $\xi_3$, and $\xi_4$, and hence can be solved for $\xi_2$ and $\xi_3$ as
\begin{equation}
    \label{eq:firstRowSolution}
    \xi_2 = \xi_4\, h(\Delta{t}^-,\Delta{t}^+), 
    \quad 
    \xi_3 = \xi_4\, k(\Delta{t}^-,\Delta{t}^+), 
\end{equation}
where the coefficients $h(\Delta{t}^-,\Delta{t}^+)$ and $k(\Delta{t}^-,\Delta{t}^+)$ are the following functions of $\Delta{t}^-$ and $\Delta{t}^+$
\begin{equation}
    \begin{aligned}
        \displaystyle h(\Delta{t}^-,\Delta{t}^+) &=  
        \frac{\left[e^{\bA^+\Delta{t}^+} \right]_{14}  \left[e^{-\bA^-\Delta{t}^-}\right]_{13} 
        - \left[e^{\bA^+\Delta{t}^+}\right]_{13}\left[e^{-\bA^-\Delta{t}^-}\right]_{14} }
        {\left[e^{\bA^+\Delta{t}^+}\right]_{13}\left[e^{-\bA^-\Delta{t}^-}\right]_{12} -\left[e^{\bA^+\Delta{t}^+}\right]_{12}\left[e^{-\bA^-\Delta{t}^-}\right]_{13} },  \\[7 mm]
        \displaystyle k(\Delta{t}^-,\Delta{t}^+) &=  
        \frac{\left[e^{\bA^+\Delta{t}^+} \right]_{14}  \left[e^{-\bA^-\Delta{t}^-}\right]_{12} 
        - \left[e^{\bA^+\Delta{t}^+}\right]_{12}\left[e^{-\bA^-\Delta{t}^-}\right]_{14} }
        {\left[e^{\bA^+\Delta{t}^+}\right]_{12}\left[e^{-\bA^-\Delta{t}^-}\right]_{13} -\left[e^{\bA^+\Delta{t}^+}\right]_{13}\left[e^{-\bA^-\Delta{t}^-}\right]_{12} }. 
    \end{aligned}
\end{equation}

Turning back the attention to equations \eqref{eq:halfmapsAlgorithm}, these can be rewritten, using equations \eqref{eq:firstRowSolution}, as
\begin{equation}
    \label{eq:halfmapsAlgorithmReduced}
    \begin{bmatrix}
        x_2 \\
        x_3 \\
        x_4
    \end{bmatrix}
    =  \xi_4\, \widehat{e^{-\bA^-\Delta{t}^-}}
    \begin{bmatrix}
        h(\Delta{t}^-, \Delta{t}^+) \\
        k(\Delta{t}^-, \Delta{t}^+)\\
        1
    \end{bmatrix},
    \qquad
    \begin{bmatrix}
        \eta_2 \\
        \eta_3 \\
        \eta_4
    \end{bmatrix}
    = \xi_4\, \widehat{e^{\bA^+\Delta{t}^+}}
    \begin{bmatrix}
        h(\Delta{t}^-, \Delta{t}^+) \\
        k(\Delta{t}^-, \Delta{t}^+) \\
        1
    \end{bmatrix},
\end{equation}
where $\widehat{e^{-\bA^-\Delta{t}^-}}$ and $\widehat{e^{\bA^+\Delta{t}^+}}$ are the submatrices obtained from the original matrices through elimination of the first row and column. 

Now, to be solutions of the eigenvalue problem \eqref{eq:eigenvalue}, the two vectors given by equation \eqref{eq:halfmapsAlgorithmReduced} have to be parallel, namely,
\begin{equation}
    \label{eq:parallel}
    \frac{\eta_2}{x_2} = \frac{\eta_3}{x_3} = \frac{\eta_4}{x_4} = \mu,
\end{equation}
providing two equations for the unknowns $\Delta{t}^-$ and $\Delta{t}^+$, as follows
\begin{equation}
    \label{eq:finalEquationsAlgorithm}
    \begin{aligned}
        &\frac{\left[e^{\bA^+\Delta{t}^+}\right]_{22}h(\Delta{t}^-,\Delta{t}^+) + \left[e^{\bA^+\Delta{t}^+}\right]_{23}k(\Delta{t}^-,\Delta{t}^+) + \left[e^{\bA^+\Delta{t}^+}\right]_{24}}{\left[e^{-\bA^-\Delta{t}^-}\right]_{22}h(\Delta{t}^-,\Delta{t}^+) + \left[e^{-\bA^-\Delta{t}^-}\right]_{23}k(\Delta{t}^-,\Delta{t}^+) + \left[e^{-\bA^-\Delta{t}^-}\right]_{24}} \\[3mm]
        &= \frac{\left[e^{\bA^+\Delta{t}^+}\right]_{32}h(\Delta{t}^-,\Delta{t}^+) + \left[e^{\bA^+\Delta{t}^+}\right]_{33}k(\Delta{t}^-,\Delta{t}^+) + \left[e^{\bA^+\Delta{t}^+}\right]_{34}}{\left[e^{-\bA^-\Delta{t}^-}\right]_{32}h(\Delta{t}^-,\Delta{t}^+) + \left[e^{-\bA^-\Delta{t}^-}\right]_{33}k(\Delta{t}^-,\Delta{t}^+) + \left[e^{-\bA^-\Delta{t}^-}\right]_{34}}  \\[3mm]
        &= \frac{\left[e^{\bA^+\Delta{t}^+}\right]_{42}h(\Delta{t}^-,\Delta{t}^+) + \left[e^{\bA^+\Delta{t}^+}\right]_{43}k(\Delta{t}^-,\Delta{t}^+) + \left[e^{\bA^+\Delta{t}^+}\right]_{44}}{\left[e^{-\bA^-\Delta{t}^-}\right]_{42}h(\Delta{t}^-,\Delta{t}^+) + \left[e^{-\bA^-\Delta{t}^-}\right]_{43}k(\Delta{t}^-,\Delta{t}^+) + \left[e^{-\bA^-\Delta{t}^-}\right]_{44}}.
    \end{aligned}
\end{equation}
The system \eqref{eq:finalEquationsAlgorithm} is nonlinear and must be solved numerically. Once the time intervals $\Delta{t}^-$ and $\Delta{t}^+$ are known, the eigenvector $\bx$ can be computed from \eqref{eq:halfmapsAlgorithmReduced}$_1$, whereas the eigenvalue $\mu$ is obtained from equation \eqref{eq:parallel}.

\begin{figure}[H]
	\centering
	\includegraphics[]{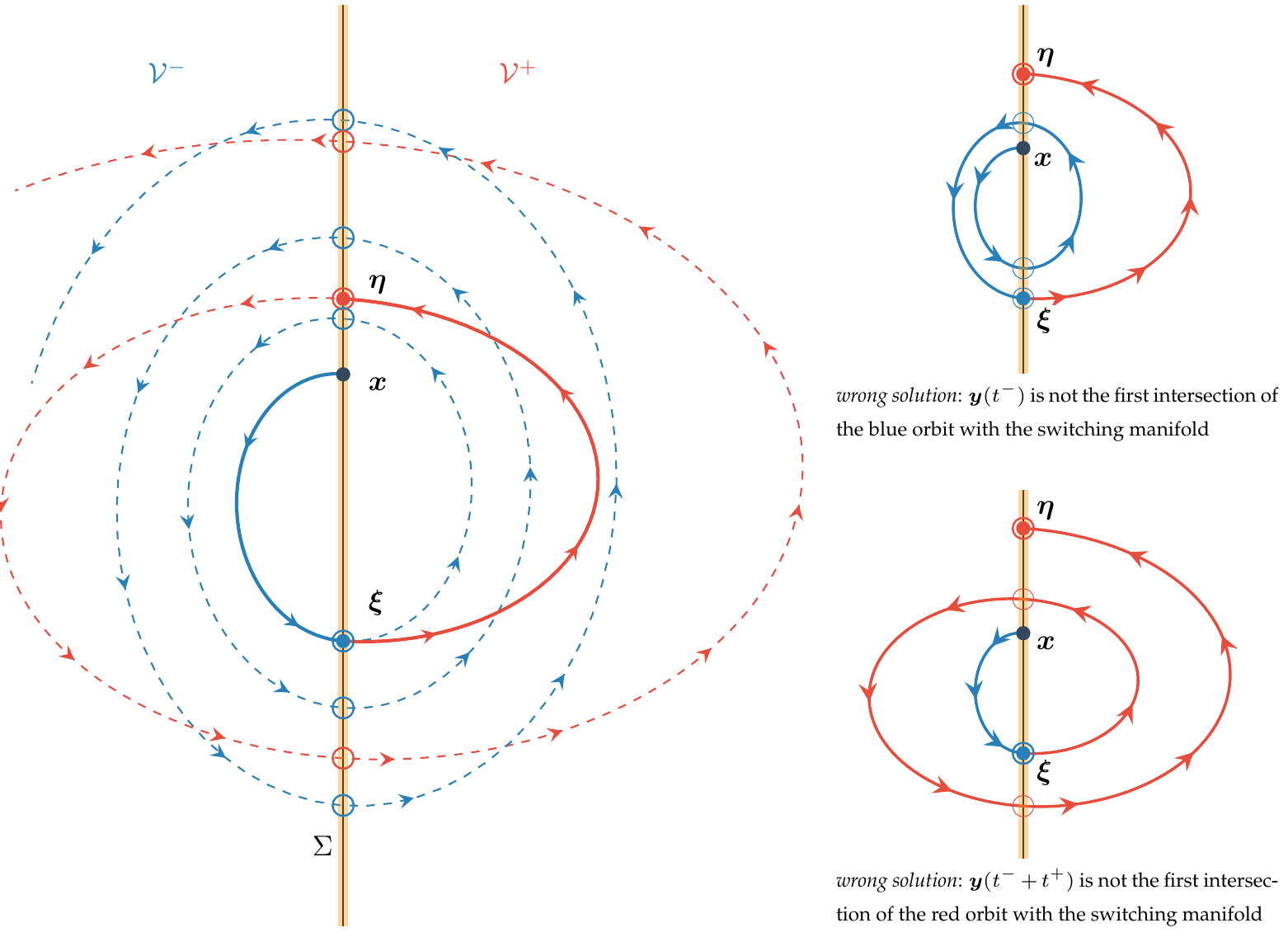}
	\caption{Sketch of a motion starting at point $\bx$ and initially developing inside the negative branch of the mechanical system (blue solid line). 
		At point $\bxi$ the motion crosses the switching manifold and further develops inside the positive branch (red solid line). However, from a purely mathematical point of view, the continuation with wrong equations referred to the negative branch (blue dashed line) are still solutions of equations \eqref{eq:finalEquationsAlgorithm} and their  orbits cross the switching manifold at several subsequent points. All the latter points have to be disregarded because they represent the solution of a smooth system, different from that under analysis. Something analogous happens with the solution continuation of the red line (represented dashed).  Therefore, there are infinite solutions on the switching manifold, and the selection of the correct points is the harder problem to be solved in finding instability cones. 
	}
	\label{fig:pictureWrongSolutionsAlgorithm}
\end{figure}

The calculation of the time intervals $\Delta{t}^-$ and $\Delta{t}^+$ is far from trivial, as the determining equations involve transcendental trigonometric functions, so that there are infinite values of $\Delta{t}^-$ and $\Delta{t}^+$ satisfying them. Among these infinite values, only those corresponding to motions crossing the switching manifold the first time at $\Delta{t}^-$ and the second at $\Delta{t}^- + \Delta{t}^+$ have to be retained, while the other disregarded, because they refer to orbits that cross the switching manifold multiple times, but erroneously remain in the same subdomain. This situation, sketched in Fig. \ref{fig:pictureWrongSolutionsAlgorithm}, has been solved through: (i) an estimation of the maximum time intervals $\Delta{t}^\pm_\text{max} = \max\{3\pi/(2\omega_1^\pm),3\pi/(2\omega_2^\pm)\}$ within which the first intersections of the orbit with the switching manifold $\Sigma$ occur, so that the solution is searched for in the bounded time domain $[0,\Delta{t}^-_\text{max}] \times [0,\Delta{t}^+_\text{max}]$ (details are reported in \cite{mrossi2021}); (ii) a systematic elimination of the solutions lacking mechanical meaning. In fact, an obtained solution is meaningful if and only if the following conditions are met:
\begin{itemize}
    \item \textbf{First crossings.} The time instants $\Delta{t}^-$ and $\Delta{t}^- + \Delta{t}^+$ correspond respectively to the first and second intersection time of an orbit on the invariant cone with the switching manifold $\Sigma$. Therefore, given the initial vector $\bx$ 
    (obtained from the above algorithm), the conditions have to be checked: (i.) that the value $\Delta{t}^-$ is the smallest solution of equation \eqref{eq:linear}$_1$, and (ii.) that the value $\Delta{t}^+$ is the smallest solution of equation \eqref{eq:linear}$_2$.
    \item \textbf{Switching conditions.} Any solution representing an invariant cone must fulfill
    \begin{equation}
        \label{eq:conditionFeasible1}
        x_3 = \dot{\xi}(0) < 0, \qquad \xi_3 = \dot{\xi}(\Delta{t}^-) > 0, \qquad \eta_3 = \dot{\xi}(\Delta{t}^- + \Delta{t}^+) < 0,
    \end{equation}
    denoting the fact that the orbit on the invariant cone must initially cross the switching manifold and enter in the negative subdomain (at $t=0$), while at $t = \Delta{t}^-$ has to enter into the positive one, and finally, at $\Delta{t}^- + \Delta{t}^+$ the negative subdomain has to be entered again. However, a solution to the problem \eqref{eq:finalEquationsAlgorithm} yields a vector $\bx$ defined except for the sign, which can always be adjusted, so that instead of conditions \eqref{eq:conditionFeasible1} for a solution to be meaningful, the time intervals $\Delta{t}^-$ and $\Delta{t}^+$ have to satisfy
    \begin{equation}
        \label{eq:conditionFeasibleFinal}
        \dot{\xi}(0)\, \dot{\xi}(\Delta{t}^- + \Delta{t}^+) > 0, \qquad \dot{\xi}(0)\, \dot{\xi}(\Delta{t}^-) < 0,
    \end{equation}
    meaning that the assumption made on the first entered subdomain (the negative) is arbitrary and the opposite assumption could be equally made.
\end{itemize}

\subsection{A non-smooth structure with an unstable invariant cone}
\label{subsec:refenenceNumericalExample}

\subsubsection{Existence of an unstable invariant cone}

It is instrumental now to introduce the following non-dimensional parameters
\begin{equation}
    \label{eq:nonDimensionalParameter}
    \zeta_\pm = \frac{R_\pm}{l} \qquad k = \frac{k_1 l^2}{k_2} \qquad \gamma = \frac{F l}{k_2} \qquad \sigma = \frac{y_s}{l},
\end{equation}
so that equations \eqref{eq:equationPiecewiseLagrange} governing the dynamics of the piecewise-smooth structure become 
\begin{equation}
    \label{eq:nondimensionalEquation}
    \Theta 
    \begin{bmatrix}
        \displaystyle 1 & \displaystyle \frac{1}{2} \\[2ex]
        \displaystyle \frac{1}{2} & \displaystyle \frac{1}{3}\\
    \end{bmatrix}
    \begin{bmatrix} 
        \displaystyle \tilde{\xi}_{,\tau\tau}(\tau) \\[2ex]
        \displaystyle \tilde{\phi}_{,\tau\tau}(\tau) \\
    \end{bmatrix}
    +
    \begin{bmatrix}
        \displaystyle \frac{k}{\zeta_\pm}(\zeta_\pm \mp \sigma) + \frac{1 \mp \gamma\zeta_\pm}{\zeta_\pm^2} \quad & \displaystyle \frac{-\gamma\zeta_\pm \pm 1}{\zeta_\pm} \\[2ex]
        \displaystyle \pm \frac{1}{\zeta_\pm}  & 1\\
    \end{bmatrix}
    \begin{bmatrix} 
        \displaystyle \tilde{\xi}(\tau) \\[2ex]
        \displaystyle \tilde{\phi}(\tau) \\
    \end{bmatrix}
    = \boldsymbol{0}
\end{equation}
where the non-dimensional time $\tau = t / T$ and the non-dimensional Lagrangian coordinates $\tilde{\xi}(\tau) = \xi(t) / l$, $\tilde{\phi}(\tau) = \phi(t)$ are introduced, together with the dimensionless mass density 
\begin{equation}
    \Theta = \frac{\rho l^3}{T^2 k_2}.
\end{equation}
Moreover, the non-dimensional parameter singling out the ratio between the two radii of the two constraints is defined as
\begin{equation}
    \label{eq:nonDimensionalParameter2}
    \chi = \frac{\zeta_-}{\zeta_+} .
\end{equation}
The presence of an unstable invariant cone has been numerically detected for a broad range of values of the above parameters. As a paradigmatic example, the following values for the parameters are considered in this section
\begin{equation}
    \label{eq:parameterNumRef}
    \begin{aligned}
    & \zeta_+ = \frac{R_+}{l} = 0.6, & k = \frac{k_1 l^2}{k_2} = 0.3, \quad \gamma &= \frac{F l}{k_2}= 0.06, \\ 
    & \sigma  = \frac{y_s}{l} = 0, & \Theta = \frac{\rho l^3}{T^2 k_2} = 1, \quad \chi &= \frac{\zeta_-}{\zeta_+} = 6
    \end{aligned}
\end{equation}
This reference numerical example contains all the most relevant features of the new kind of unstable structural behaviour disclosed in the present article. In the sequel, all quantities are dimensionless, according to the normalization \eqref{eq:nonDimensionalParameter}, and a superimposed dot stands for the derivative with respect to the dimensionless time $\tau$. However, the dimensionless Lagrangian coordinates are denoted $\xi$ and $\phi$ (without tilde) to ease the notation.

For the above geometry and loading, a piecewise invariant cone described by the initial condition
\begin{equation}
	\bx = \by_0 = \left[\xi(0), \phi(0), \dot{\xi}(0), \dot{\phi}(0) \right] = \left[ 0, -0.00838564, -0.372424, 0.928025 \right] 
\end{equation} 
is present, with a multiplier $\mu = 1.079995$. The intersection time intervals, calculated using the algorithm presented in the previous section for the detection of the invariant cone (and later confirmed by the numerical simulation of the mechanical system), are $\Delta{\tau}^- = 0.637108$ and $\Delta{\tau}^+ = 2.981694$.

\begin{figure}[H]
	\centering
	{\includegraphics[]{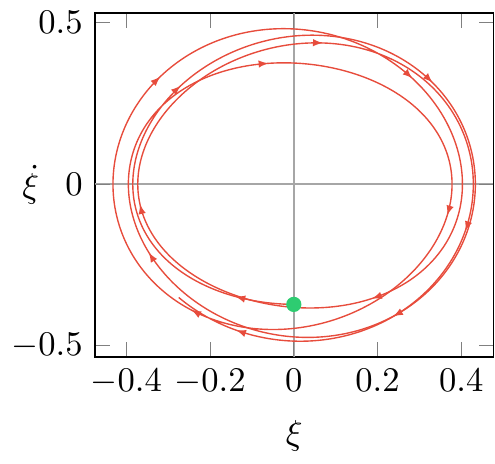}} \,
	{\includegraphics[]{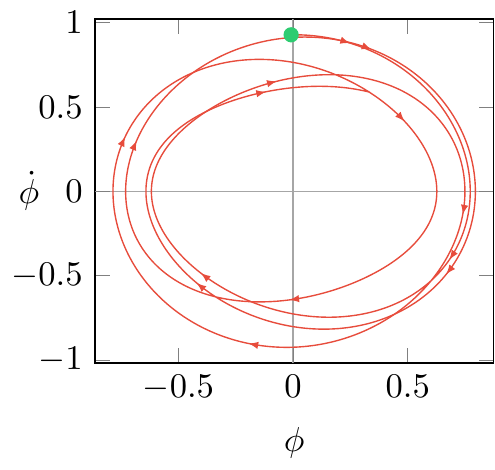}} \,
	{\includegraphics[]{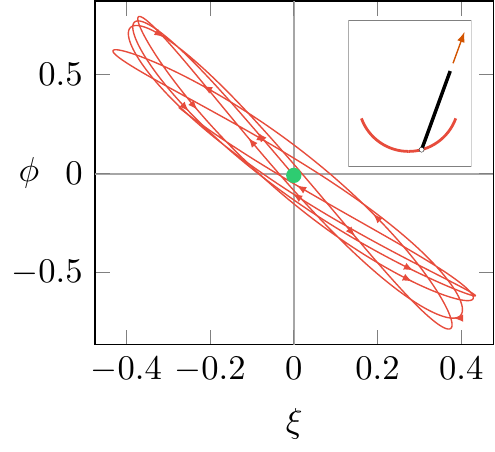}} \\
	\caption{Phase portraits, showing a stable response, for one (with positive curvature, see the inset on the left) of the two smooth systems forming the piecewise linear structure analyzed in Figs.
	\ref{fig:example1timePOSNEG}
	and
	\ref{fig:example1phasePOSNEG}}
	\label{fig:exemple1POS}
\end{figure}

\begin{figure}[H]
	\centering
	{\includegraphics[]{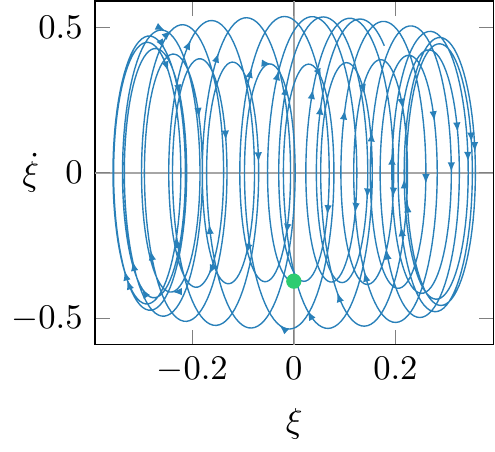}} \,
	{\includegraphics[]{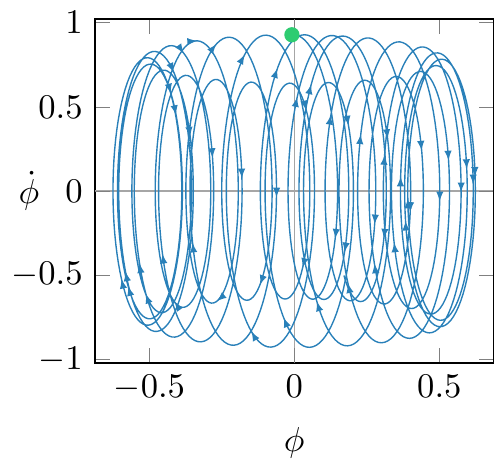}} \,
	{\includegraphics[]{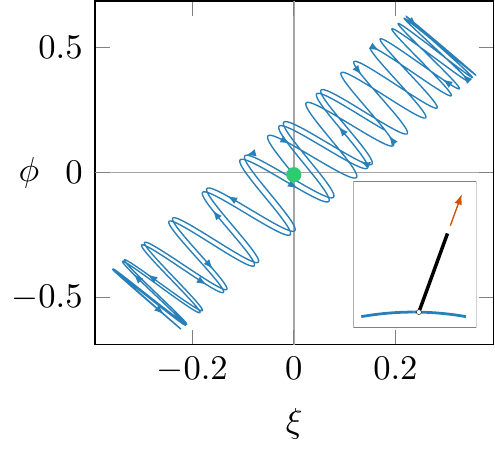}} \\
	\caption{Phase portraits, showing a stable response, for one (with negative curvature, see the inset on the left) of the two smooth systems forming the piecewise linear structure analyzed in Figs. 
	\ref{fig:example1timePOSNEG}
	and
	\ref{fig:example1phasePOSNEG}}
	\label{fig:exemple1NEG}
\end{figure}

The critical loads for flutter and divergence instability for the smooth substructures that compose the non-smooth mechanical structure can be analytically determined as (see Appendix)
\begin{equation}
    \label{eq:formulaFlutterDivergence}
    \begin{aligned}
        \gamma_{\text{flu},\text{div}}^+ = \frac{-2 \zeta_+  (3 \zeta_+ -2) (\zeta_+  (\zeta_+  (k+3)-k \sigma -3)+1) \pm 2 \sqrt{3} \sqrt{\zeta_+ ^5 (3 \zeta_+ -2)^2 k (\zeta_+ -\sigma )} }{(2-3 \zeta_+ )^2 \zeta_+ ^2},\\
        \gamma_{\text{flu},\text{div}}^- =\frac{-2 \zeta_-  (3 \zeta_- +2) (\zeta_-  (\zeta_-  (k+3)+k \sigma +3)+1) \pm 2 \sqrt{3} \sqrt{\zeta_- ^5 (3 \zeta_- +2)^2 k (\zeta_- +\sigma )}}{\zeta_- ^2 (3 \zeta_- +2)^2},
    \end{aligned}
\end{equation}
where $\gamma_{\text{flu},\text{div}}^+$ ($\gamma_{\text{flu},\text{div}}^-$) refers to the structure with positive (negative) curvature, while the critical loads for flutter correspond to the minimum absolute values. Therefore, a substitution of the values \eqref{eq:parameterNumRef} into \eqref{eq:formulaFlutterDivergence} leads to the conclusion that  the substructures are both stable when considered separately for an assumed {\it tensile} load $\gamma = 0.06$, because their critical loads are much higher in absolute value (one is negative and therefore compressive): 
\begin{equation}
    \gamma_{\text{flu}}^- = -1.83477, \qquad \text{and} \qquad \gamma_{\text{flu}}^+ = 0.774567.
\end{equation}
The stability of each smooth subsystem can be observed in the phase portraits reported in Figs.~\ref{fig:exemple1POS} and \ref{fig:exemple1NEG}, for the system with negative and positive curvature, respectively. The portraits refer to the linearized solution so that the orbits evolve remaining confined within the neighbourhood of the origin, representing the equilibrium point.


\begin{figure}[H]
	\centering
	{\includegraphics[]{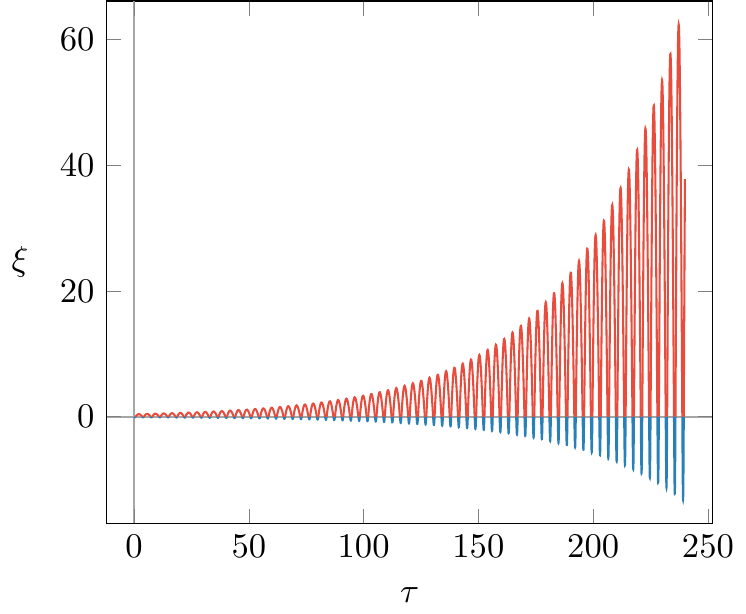}} \,
	{\includegraphics[]{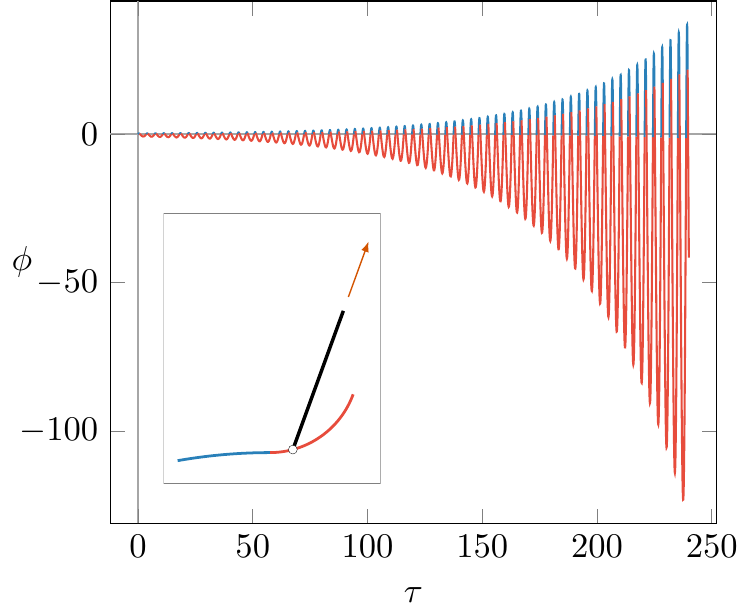}} \\
	\caption{Evolution of the Lagrangian parameters in  (dimensionless) time, for a non-smooth elastic structure (shown in the inset on the right) composed of two stable substructures. Exponential blow-up demonstrates instability. 
	}
	\label{fig:example1timePOSNEG}
\end{figure}

\begin{figure}[H]
	\centering
	{\includegraphics[]{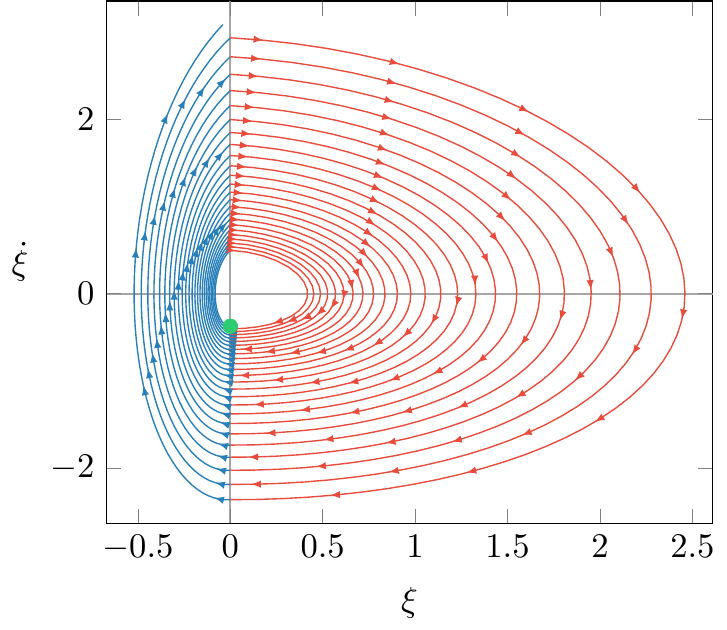}} \,
	{\includegraphics[]{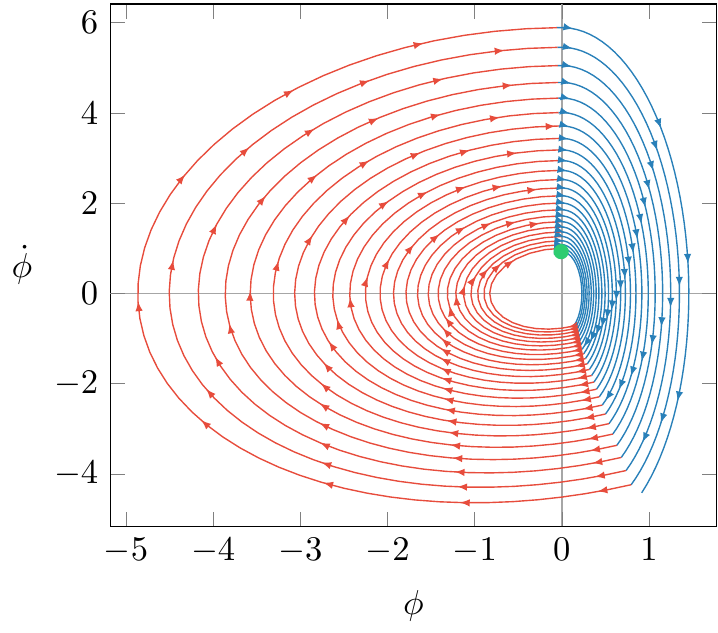}} \\
	{\includegraphics[]{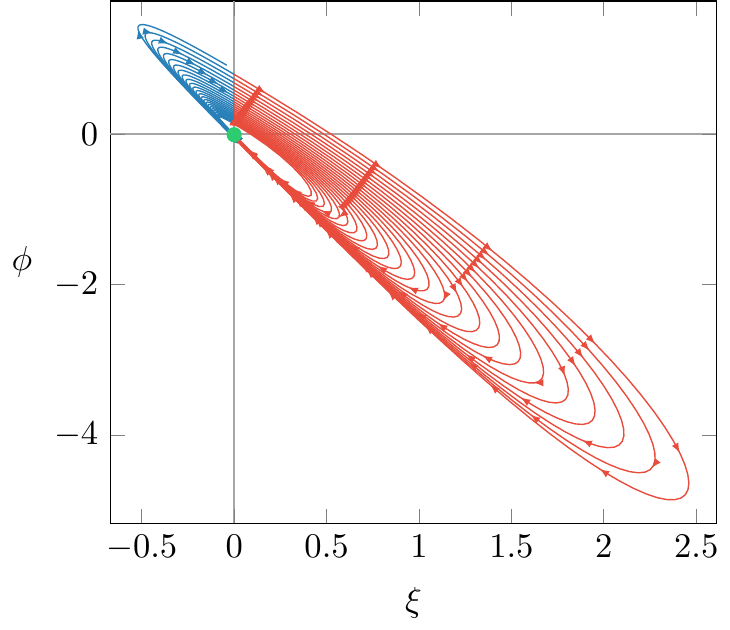}} \,
	{\includegraphics[]{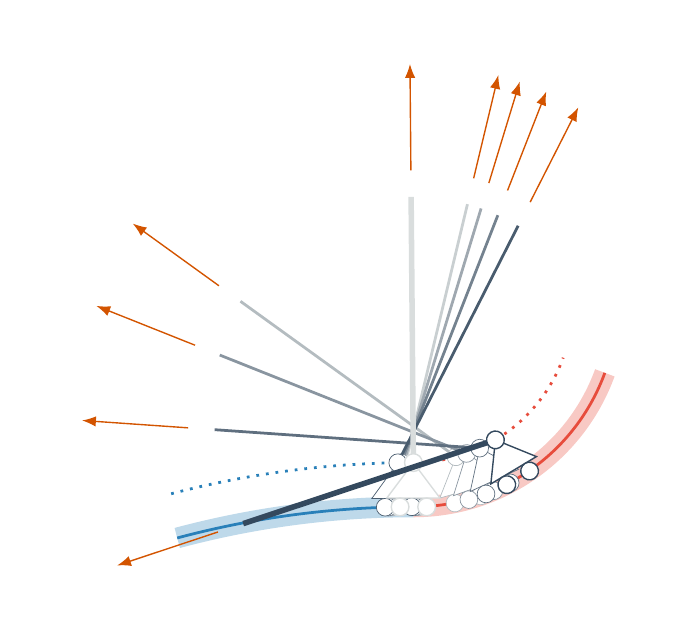}} \\
	\caption{Phase portraits for the non-smooth mechanical system analyzed in Fig. 
	\ref{fig:example1timePOSNEG} 
	(and shown on the lower part, right), evidencing instability although the two component substructures are stable.
	}
	\label{fig:example1phasePOSNEG}
\end{figure}

Although the two subsystems forming the piecewise linear structure are stable when considered separately, the combination of them is unstable, as can be appreciated in Figs.~\ref{fig:example1timePOSNEG} and \ref{fig:example1phasePOSNEG}, reporting a solution of the linearized equations of motion belonging to the unstable invariant cone. The evolution in time of the Lagrangian coordinates $\xi$ and $\phi$ shows an exponential increase in the amplitude of motion, analogous to the behaviour of a smooth mechanical system when flutter instability occurs. The phase portraits show an orbit lying on the invariant cone so that each phase portrait is a projection of this 4-dimensional invariant manifold onto a 2-dimensional plane. The orbit starting close to the origin (representing the trivial equilibrium configuration) evolves following a spiralling out motion, corresponding to an unstable behaviour. 


\subsubsection{Instability of the structure in the nonlinear range}
\label{sec:instnonlin}

All the theoretical and numerical results obtained in the previous sections are based on the linearization of the equations of motion describing the system and, in particular, on the piecewise linear response resulting from the combination of the two linearized responses for each subsystem forming the non-smooth structure. For a smooth dynamical system, the linearization of the equations of motion near an equilibrium configuration is a classical strategy for determining whether or not the considered configuration is stable. 
When the equilibrium of the linearized system is not \emph{marginally stable} (i.e. the real part of the eigenvalues of the Jacobian matrix vanishes), the Lyapunov theorem assures that the results obtained for the linearized case can be extended to the original nonlinear one.

A structure with a piecewise-linear behaviour cannot be further linearized so the Lyapunov theorem cannot be applied. An extension of this theorem to the nonlinear case of piecewise smooth dynamical system has been provided, see \cite{weissInvariantManifoldsNonsmooth2012a}, under regularity assumptions which are satisfied for the structures under examination.

Following this extension, the instability of the structures in a fully nonlinear range is expected, and indeed a direct integration of the nonlinear equations of motion for the previously analyzed structures confirms the presence of unstable behaviour. 

\begin{figure}[H]
	\centering
	{\includegraphics[]{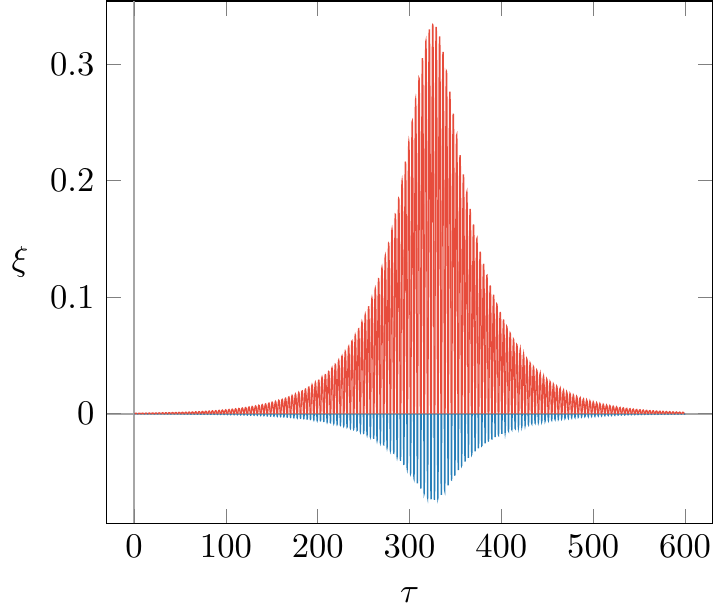}} \,
	{\includegraphics[]{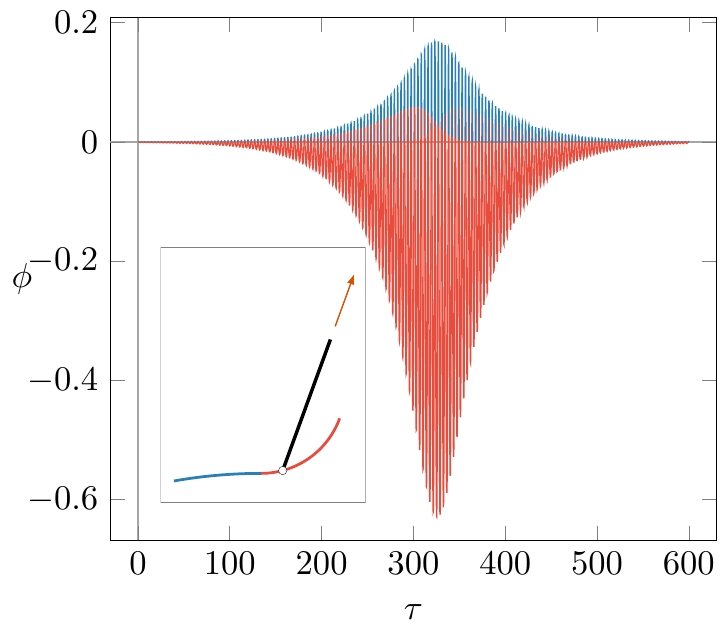}} \\
	{\includegraphics[]{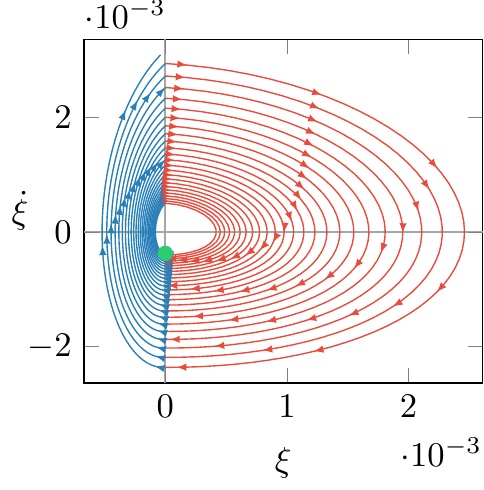}} \,
	{\includegraphics[]{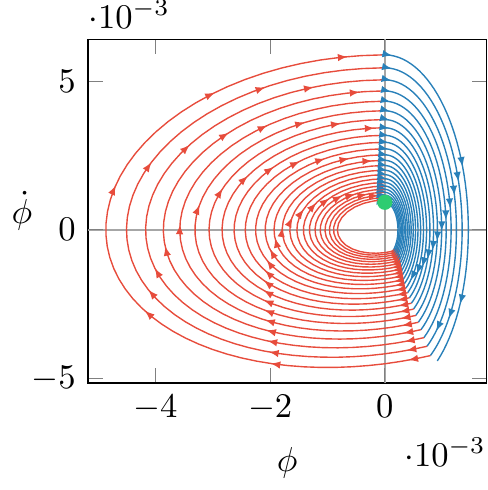}} \,
	{\includegraphics[]{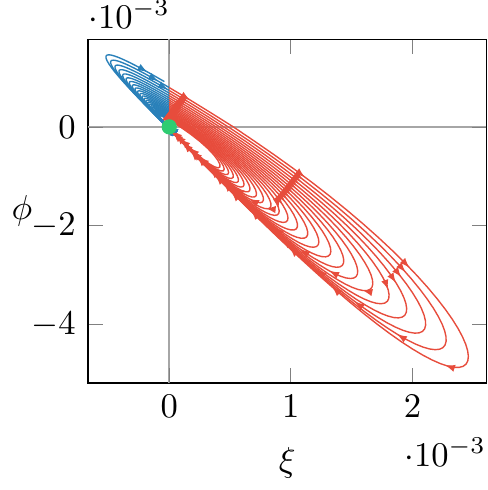}} \\
	\caption{
	Phase portrait and evolution diagrams for the non-smooth structure (its linearized analysis is reported in Figs. \ref{fig:example1timePOSNEG} and \ref{fig:example1phasePOSNEG} nonlinear case
	}
	\label{fig:example1NONLIN}
\end{figure}

In particular, the numerical solution of the nonlinear dynamics describing the reference structure is depicted in Fig.~\ref{fig:example1NONLIN}, for the same initial conditions used in the piecewise linear analysis, namely, vector $\by_0$, is selected 
now 
with a sufficiently small modulus to start from a neighbourhood of the equilibrium configuration, thus $\by_0$ has been scaled as 
\begin{equation}
    \by_0 = \left[\xi(0), \phi(0), \dot{\xi}(0), \dot{\phi}(0) \right] = f_s \times \left[ 0, -0.00838564, -0.372424, 0.928025 \right],
\end{equation} 
with the scaling factor $f_s = 0.001$.

The evolution of the Lagrangian generalized coordinates presents an exponential growth for small values of $\tau$ that can be associated to the presence of a quasi-invariant cone, as defined in \cite{weissInvariantManifoldsNonsmooth2012a}. The orbits do not reach a limit cycle as for the Hopf bifurcation in smooth structures, but evidence blowing-up oscillations which reach a peak and then decrease in amplitude until near the initial amplitude. This motion repeats itself several times (in a way similar to beats) and the peak values are found to be almost constant and independent of the modulus of the initial condition $\by_0$ (the cases $f_s = 10^{-4}$ and $f_s = 10^{-5}$ have also been tested and for all cases, a peak value of approximately $\xi = 0.3$ and $\phi = -0.6$ have been found,  
Fig. \ref{fig:example1NONLIN}~a-b). These features of the structural dynamics denote complex unstable behaviour.

\subsection{A non-smooth structure evidencing flutter in both tension and compression}

\subsubsection{Existence of an unstable invariant cone in both tension and compression}

\begin{figure}[H]
	\centering
	{\includegraphics[]{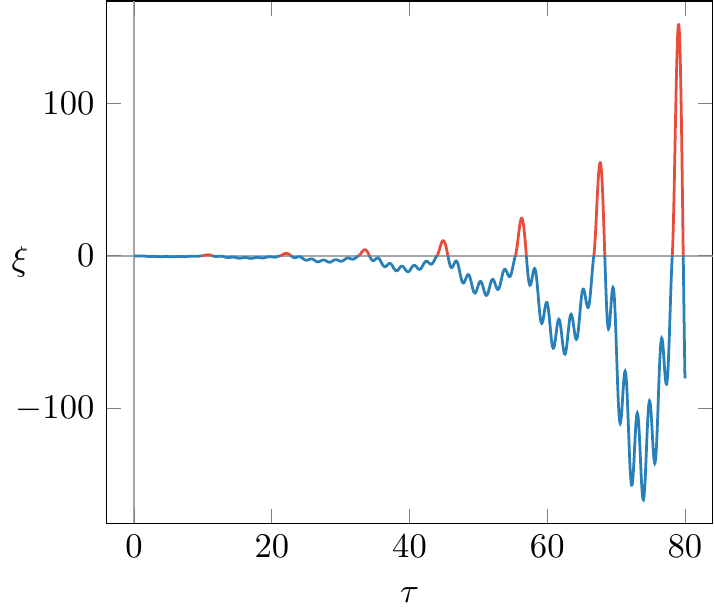}} \,
	{\includegraphics[]{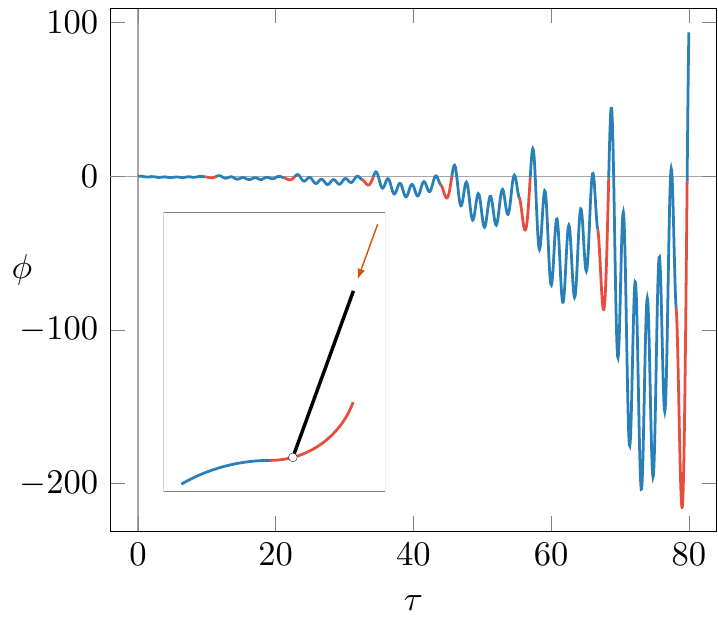}} \\
	{\includegraphics[]{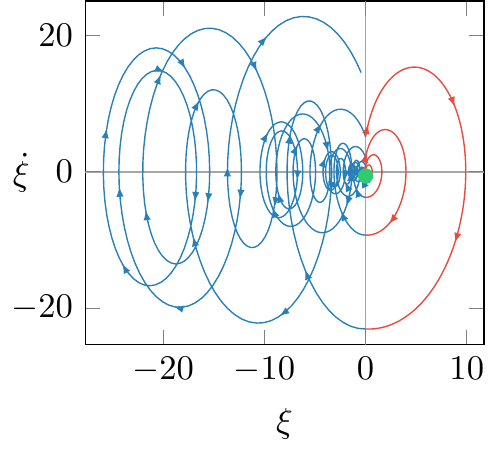}} \,
	{\includegraphics[]{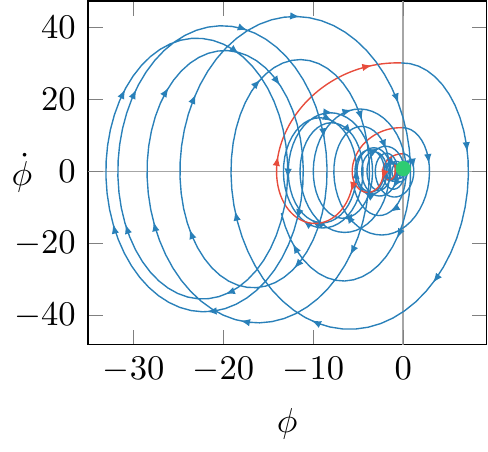}} \,
	{\includegraphics[]{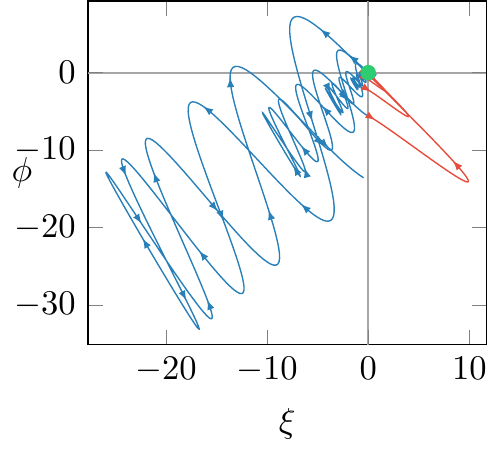}} \\
	\caption{Phase portrait and evolution diagrams for a non-smooth elastic structure displaying instability in compression (although the component substructures are stable).}
	\label{fig:example2A}
\end{figure}

The reference solution considered in the previous section is just an example of instability related to the non-smoothness of the structure. Considering other combinations of parameters leading to instability, the topological structure of the solution in the phase space has been found to remain similar, so that an orbit is found that spirals out from the origin. In certain cases, the evolution of the solution can be more or less irregular, as can be seen in the example reported in Fig.~\ref{fig:example2A}, where the wide difference in the values of $\Delta{t}^-$ and $\Delta{t}^+$ leads to an orbit that remains for a longer time in the negative subsystem. 

It is interesting to note that non-smooth structures presenting this kind of instability can be found both for tensile and compressive follower forces. For instance, consider the following set of design parameters
\begin{equation}
    \zeta_+  = 0.5,  \quad   k = 0.1, \quad \sigma = 0, \quad \Theta = 1, \quad \chi = 2, 
\end{equation}
together with the two cases of tensile or compressive follower force
\begin{equation}
    \label{eq:loadsGammaAB}
	\gamma^A = \frac{F l}{k_2}= -1.5, \qquad \qquad \gamma^B = \frac{F l}{k_2}= 0.75.
\end{equation}
Note that at loads \eqref{eq:loadsGammaAB} the substructures are both stable, as their critical loads for flutter are
\begin{equation}
    \gamma_{\text{flu}}^- = -2.62091, \qquad \qquad \gamma_{\text{flu}}^+ = 1.10455. 
\end{equation}
In the case of compressive load $\gamma^A = -1.5$, the initial condition defining the invariant cone is
\begin{equation}
    \bx^A = \by_0^A  = \left[\xi(0), \phi(0), \dot{\xi}(0), \dot{\phi}(0) \right] = \left[0, -0.00594364, -0.608652, 0.793415\right] ,
\end{equation} 
and the eigenvalue is equal to $\mu = 2.481844$, a value higher than the reference structure, in which the growth rate of the phase vector after each period $\Delta{t}^- + \Delta{t}^+$ was only $7.99\%$. The intersection time intervals can be calculated to be $\Delta{t}^- = 9.797295$  and $\Delta{t}^+ = 1.595396$.

In the case of tensile load $\gamma^B = 0.75$, the initial condition defining the invariant cone is
\begin{equation}
    \bx^B = \by_0^B = \left[\xi(0), \phi(0), \dot{\xi}(0), \dot{\phi}(0) \right] =  \left[0, 0.086944, -0.360442, 0.928721 \right] 
\end{equation} 
and the eigenvalue is equal to $\mu = 2.486877$, very close to the value for the compressive load, so that the growth rate of the solution on the invariant cone after each cycle is almost the same for both cases. The intersection time intervals can be calculated to be $\Delta{t}^- = 0.311784$ and $\Delta{t}^+ = 4.132277$.

\begin{figure}[H]
	\centering
	{\includegraphics[]{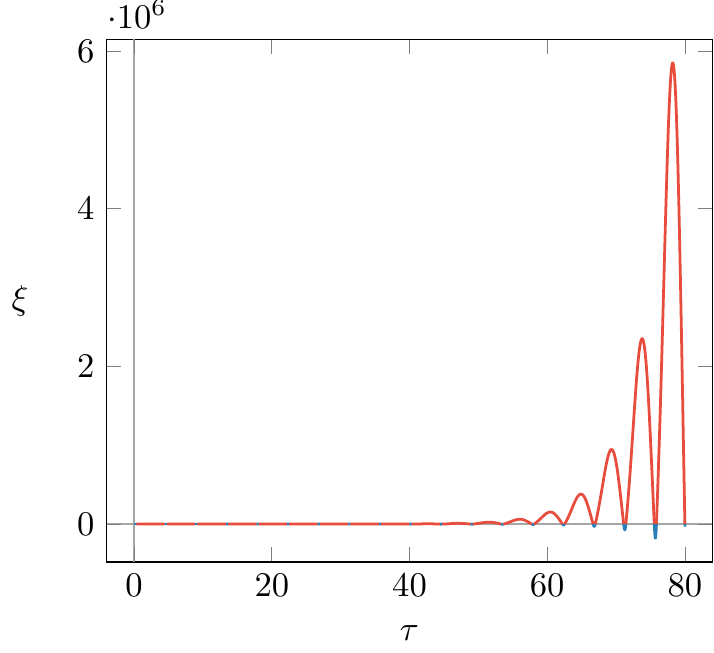}} \,
	{\includegraphics[]{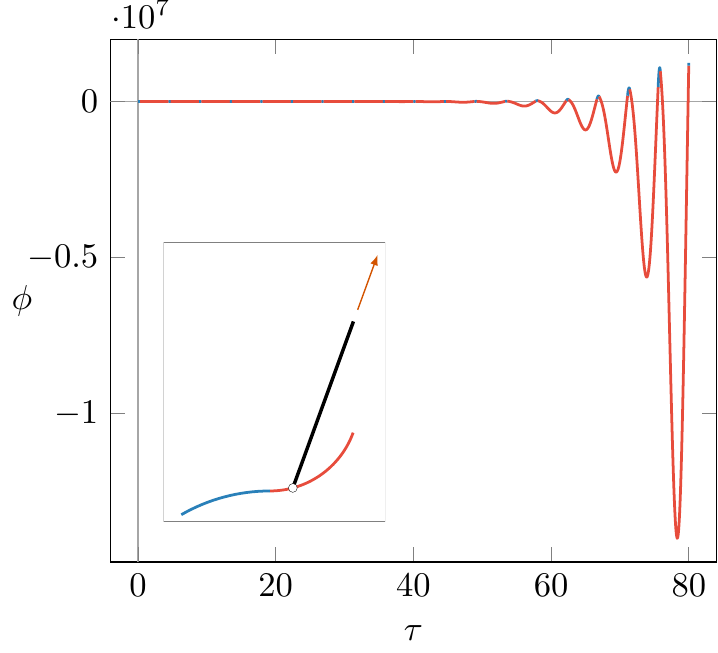}} \\
	{\includegraphics[]{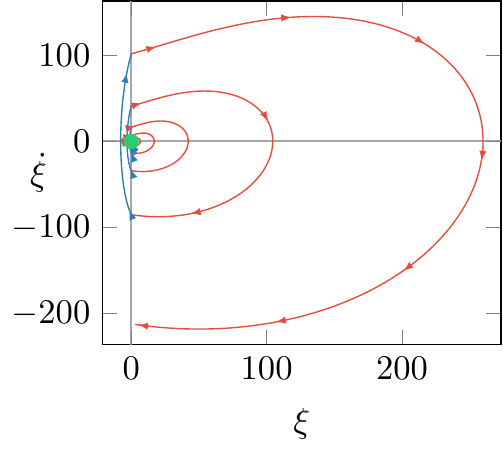}} \,
	{\includegraphics[]{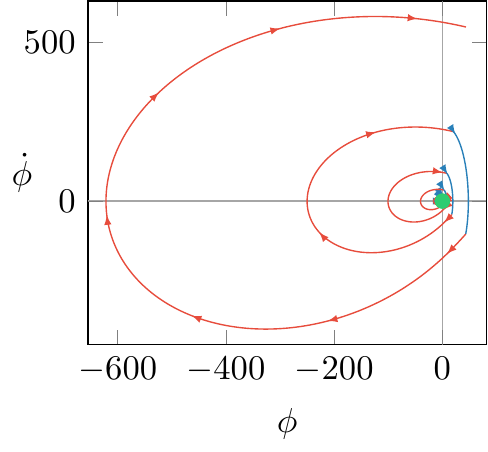}} \,
	{\includegraphics[]{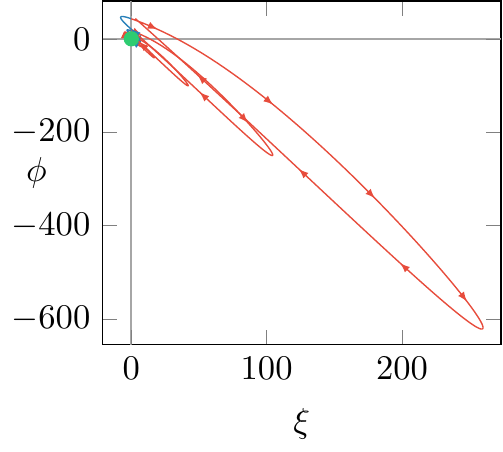}} \\
	\caption{
	Phase portrait and evolution diagrams for a non-smooth elastic structure displaying instability in tension  (although the component substructures are stable).
	}
	\label{fig:example2B}
\end{figure}

\subsubsection{Instability of the nonlinear structure}

A nonlinear analysis has been performed for the structure
linearly analyzed in Figs. \ref{fig:example2A} and \ref{fig:example2B}, 
evidencing unstable behaviour  both in tension and compression, to confirm the instability detected from the linearized analysis. The initial condition has been scaled as  
\begin{align}
    \by_0^A  &= \left[\xi(0), \phi(0), \dot{\xi}(0), \dot{\phi}(0) \right] = f_s \times \left[0, -0.00594364, -0.608652, 0.793415\right], \\
    \by_0^B &= \left[\xi(0), \phi(0), \dot{\xi}(0), \dot{\phi}(0) \right] =  f_s \times \left[0, 0.086944, -0.360442, 0.928721 \right],
\end{align} 
for tensile and compressive loads, respectively, with the scaling factor set equal to $f_s = 10^{-6}$.

\begin{figure}[H]
	\centering
	{\includegraphics[]{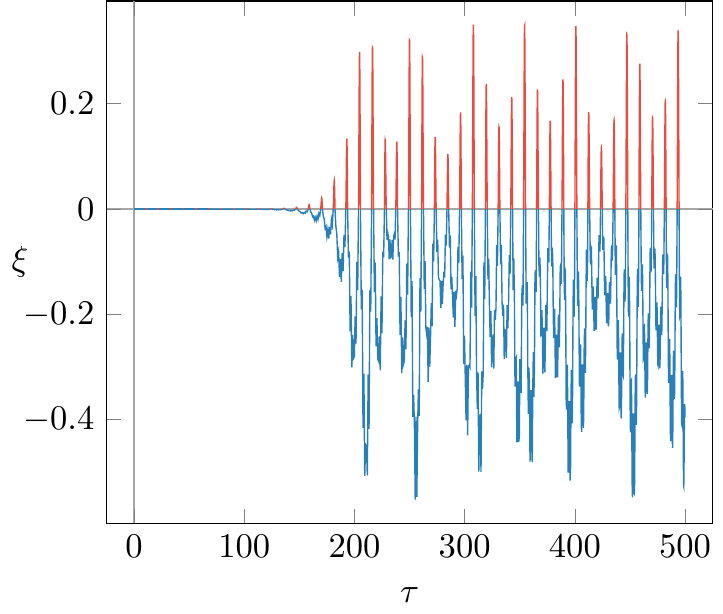}} \,
	{\includegraphics[]{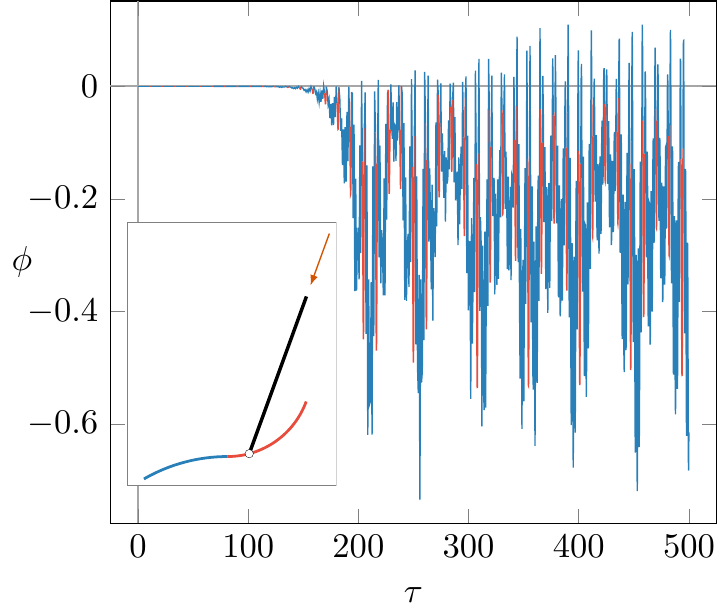}} \\
	{\includegraphics[]{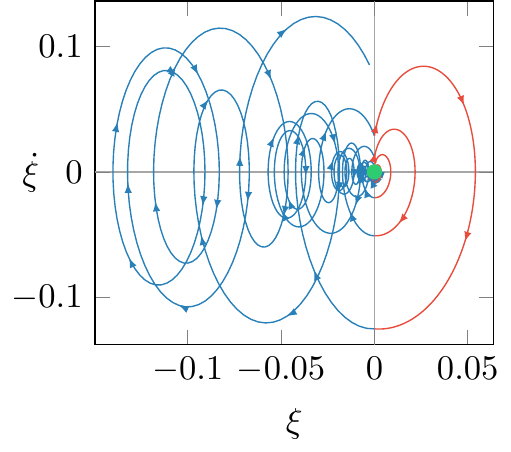}} \,
	{\includegraphics[]{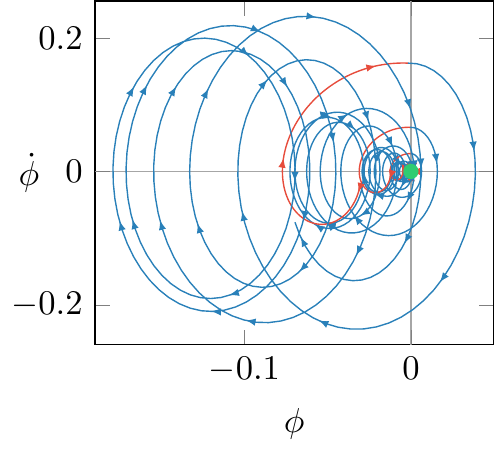}} \,
	{\includegraphics[]{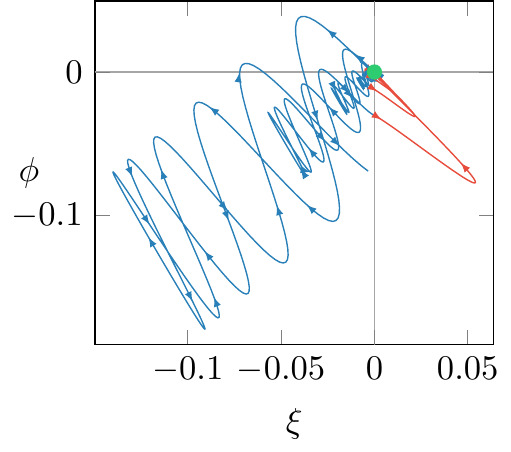}} \\
	\caption{Fully nonlinear behaviour of the structure 
	linearly analyzed in Fig. \ref{fig:example2A}. 
	The phase portrait and evolution diagrams for the case confirm instability in tension, resulting in a highly irregular motion.}
	\label{fig:example2NONLIN}
\end{figure}

\begin{figure}[H]
	\centering
	{\includegraphics[]{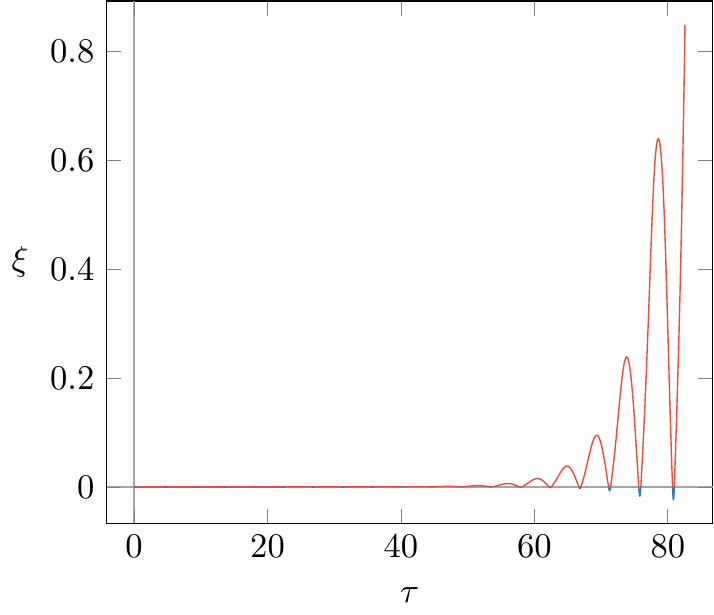}} \,
	{\includegraphics[]{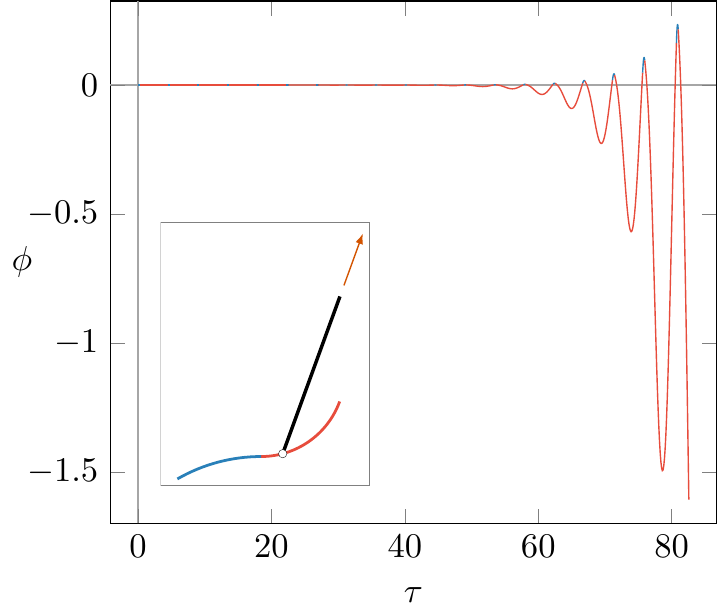}} \\
	{\includegraphics[]{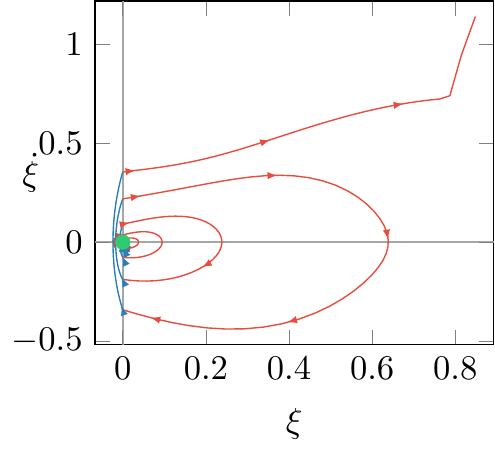}} \,
	{\includegraphics[]{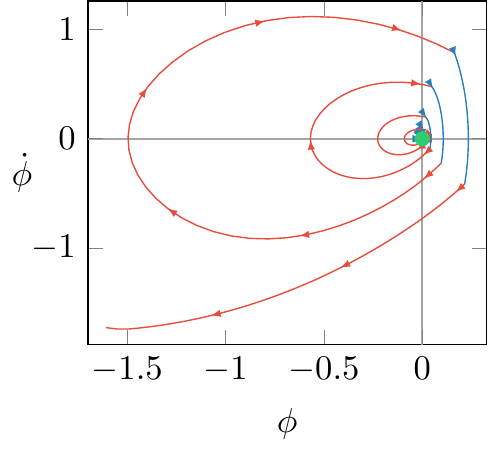}} \,
	{\includegraphics[]{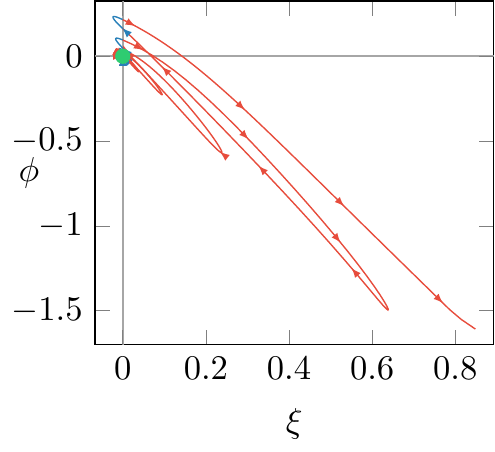}} \\
	\caption{Fully nonlinear behaviour of the structure 
	linearly analyzed in Fig. \ref{fig:example2B}. 
	The phase portrait and evolution diagrams for the case confirm instability in tension, resulting in a blowing up motion.}
	\label{fig:example3NONLIN}
\end{figure}

The behaviour of these cases, shown in Fig.~\ref{fig:example2NONLIN} and Fig.~\ref{fig:example3NONLIN}, is qualitatively different from that of the nonlinear reference structure reported in Sec.~\ref{sec:instnonlin}, because \lq beats' are not present. However, the orbits are more irregular, showing erratic behaviour. However, the instability of the system is evident for both tensile and compressive loads, since the orbits evolve along the invariant cone spiraling away from the equilibrium configuration.

\section{Conclusions}

A class of elastic structures has been disclosed, exposed to a new kind of instability, which, although already elaborated from a mathematical point of view, was never {\it directly} related to elastic structural systems. The instability results from the combination of both non-conservative follower load and non-smoothness of the equations governing the dynamics of the mechanical system. From this point of view, the proposed model is a discrete and simplified prototype of nonassociative elastoplasticity or frictional sliding, as it shares with these theories both the lack of self-adjointness of the governing differential operator and the piecewise linearity. 

We provide the first example of the use of the invariant cone theory as an instability criterion for elastic structures, permitting the design of an unstable structure as a fusion of two stable structures. The instability is fully explained and motivated from a mechanical perspective and is shown to be similar to the flutter instability occurring in smooth systems under follower loads. 

From the point of view of applications, keeping into account that nonconservative follower forces are fully feasible \cite{bigoniFlutterDivergenceInstability2018b}, we introduce a new design paradigm to avoid previously unknown structural instabilities or to design extremely deformable structures to be employed as sensors, or for energy harvesting, or as building blocks for archtected materials.

\section{Acknowledgments}
MR acknowledges support from Fondazione Carigo and University of Trieste (grant MetaSisGo). AP and DB acknowledge financial support from ERC-ADG-2021-101052956-BEYOND. This work was carried out under the auspices of the GNFM (Gruppo Nazionale per la Fisica Matematica) of the INDAM (Istituto Nazionale di Alta Matematica).

\appendix

\section{Flutter and divergence critical loads for smooth constraints}
\label{appendix1}

For a smooth profile with continuous curvature, see Fig.~\ref{fig:structure}, the flutter and divergence critical loads can be computed by solving the linearized equations \eqref{eq:linearisedEquationsMotion}, as described in Sec.~\ref{sec:smooth}. Consequently, any profile can be approximated by its osculating circle, since only the local curvature enters the linearized equations.

\begin{figure}[H]
	\centering
	{\includegraphics[width=0.38\linewidth]{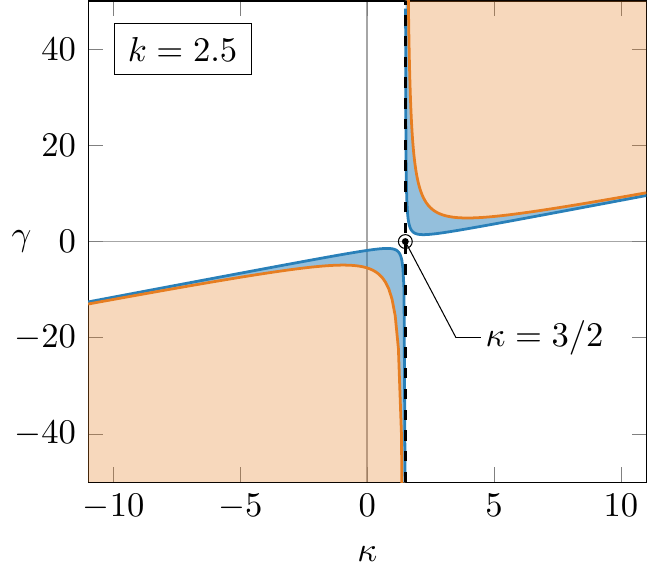}} \,
	{\includegraphics[width=0.38\linewidth]{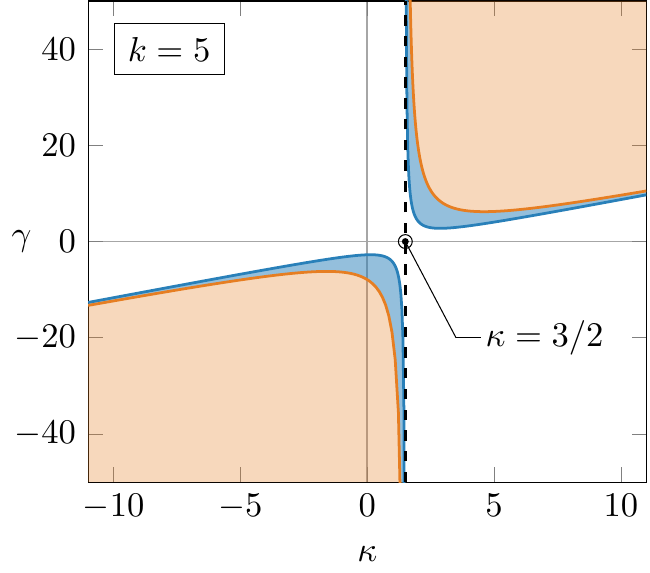}} \\
	{\includegraphics[width=0.38\linewidth]{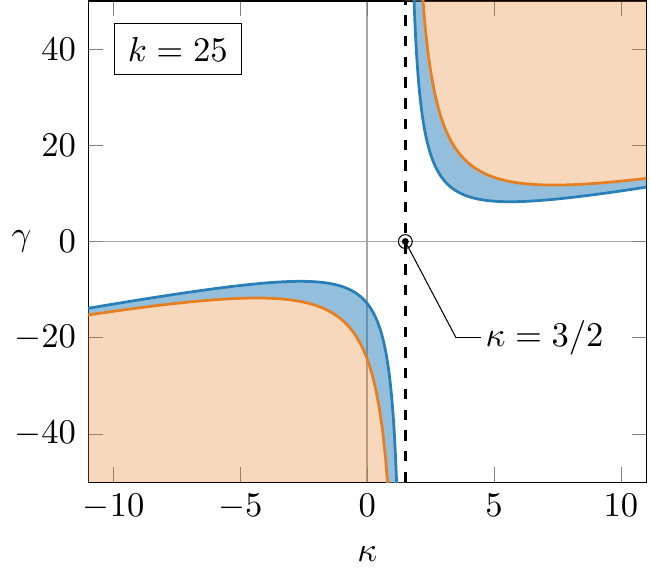}} \,
    {\includegraphics[width=0.38\linewidth]{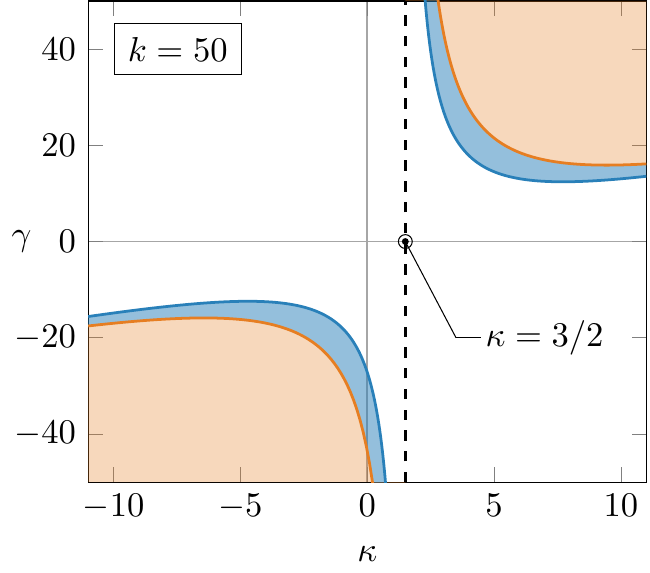}} \\
	{\includegraphics[]{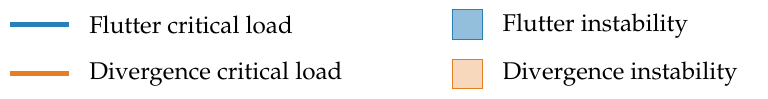}} 
	\caption{Flutter and divergence instabilities for smooth circular constraints: the flutter (blue curve) and divergence (yellow curve) critical loads are reported as a function of the dimensionless signed curvature $\kappa$ and four values of the dimensionless stiffness $k=\{2.5, 5, 25, 50\}$.}
	\label{fig:figloads}
\end{figure}

Therefore, assuming a circular profile defined by the dimensionless signed curvature $\kappa = \pm l/R_\pm = \pm 1/\zeta_\pm$, where $\zeta_\pm = R_\pm/l > 0$ is the normalized radius of curvature, the first and second invariants $I_1$ and $I_2$ of the matrix $\bGamma = -\bM^{-1} \bK$, see eq.~\eqref{eq:invariants}, are given by
\begin{equation}
    I_1 = \frac{2}{\Theta} \left\{ \gamma(2\kappa - 3) - 2(3 + k + \kappa^2 - 3\kappa) + 2k\kappa\sigma \right\}, \quad
    I_2 = \frac{12k}{\Theta^2} (1 - \kappa\sigma),
\end{equation}
in terms of the dimensionless quantities defined in \eqref{eq:nonDimensionalParameter}.
The flutter and divergence critical loads can then be computed from the equation 
\begin{equation}
    I_1^2 - 4I_2 = 0,
\end{equation}
of the parabola defining the critical condition in the $I_1 - I_2$ plane, see Fig.~\ref{fig:parabolaEigs}. The solution of this equation provides the critical loads
\begin{equation}
    \gamma_{\text{flu},\text{div}} = \frac{2}{2\kappa - 3} \left\{ 3 + k -3\kappa + \kappa^2 - k\kappa\sigma \pm \sqrt{3k(1-\kappa\sigma)} \right\},
\end{equation}
where it is understood that the critical load for flutter (divergence) corresponds to the minimum (maximum) absolute value. Since the term inside the braces is always positive, for any $k > 0$, $\kappa \in \Reals$ and $\sigma < 1/\kappa$, it follows that the follower and divergence loads are  compressive for $\kappa < 3/2$ and tensile for $\kappa > 3/2$, see Fig.~\ref{fig:figloads}.

\clearpage

\printbibliography

\end{document}